\documentclass[aps,prr,twocolumn,floatfix]{revtex4-2}

\usepackage[percent]{overpic}

\usepackage{graphicx}
\usepackage{epstopdf}
\graphicspath{{./}{./Figs/}} 

\usepackage{braket}
\usepackage{latexsym}
\usepackage{amsmath}
\usepackage{amssymb}
\usepackage{bm}
\usepackage{wasysym}
\usepackage{ifthen}
\usepackage{color}
\usepackage[colorlinks,allcolors=blue,bookmarks=true]{hyperref}
\usepackage{orcidlink}

\newcommand{\im}{\mbox{Im}}

\newcommand{\eexp}[1]{\mathrm{e}^{#1}}

\renewcommand{\ket}[1]{\left| #1 \right\rangle}
\renewcommand{\braket}[1]{\left\langle #1 \right\rangle }

\renewcommand{\Braket}[2]{\left\langle #1 \middle| #2 \right\rangle}

\newcommand{\beq}{\begin{eqnarray}}
\newcommand{\eeq}{\end{eqnarray}}

\newcommand{\hide}[1]{}  

\newcommand{\Eq}[1]{{\textcolor{blue}{Eq.}}~\!\!(\ref{#1})} 
\newcommand{\Sec}[1]{{\textcolor{blue}{Sec.}}~(\ref{#1})} 
\newcommand{\App}[1]{{\textcolor{blue}{Appendix}}~(\ref{#1})} 
\newcommand{\Fig}[1] {{\textcolor{blue}{Fig.}}~\!\!\ref{#1}}
\newcommand{\sect}[1]{{\bf #1.-- }}

\newcommand{\hrefl}[2]{\href{#2}{(#1)}}

\begin{document}

\title{Metastability, chaos, and spectrum tomography \\ for Bose-Hubbard rings and chains}

\author{Rajat, Doron Cohen}

\affiliation{
\mbox{Department of Physics, Ben-Gurion University of the Negev, Beer-Sheva 84105, Israel} 
}


\begin{abstract}
We analyze the metastability of Bose-Hubbard condensates for finite-size one-dimensional ring lattices and open chains, using a semiclassical tomographic perspective that emphasizes the relation of the many-body spectrum to the underlying classical phase-space structures. {In order to address quantum ergodicity in far-from-equilibrium scenarios of experimental interest, we inspect both local aspects (via Bogoliubov analysis) and global aspects (mixed regular-chaotic dynamics). In particular, we highlight the roles of the two parameters that control metastability, clarify the essential differences between low- and high-dimensional chaos, and show how the dynamical instabilities diminish in the limit of the Gross–Pitaevskii equation. It is somewhat frustrating that with more degrees of freedom, the dynamically metastable islands become better distinct from the ergodic sea, while their borders become ill-defined topologically. This stands in opposition to the very structured phase-space of two-degree-of-freedom systems, as reflected in the tomographic quantum spectrum.}
\end{abstract}

\maketitle

\section{Introduction} 
\label{sec:intro}

Condensation is a central theme in the theory of superfluidity (SF), and in some sense also in the theory of superconductivity \cite{Leggett,Leggett2,Leggett3}. The essence of SF is the feasibility of observing a metastable persistent current. An appropriate geometry for demonstrating SF is an $L_s$-site ring with Bosons. The standard version is described by the Bose-Hubbard Hamiltonian (BHH) \cite{exprBHH1,exprBHH2}, which generates the dynamics of $N$ bosons, with inter-site hopping $K$, and on-site repulsive interaction~$U$.
The 2~site version, aka Bosonic Josephson Junction, is well studied \cite{Oberthaler,Steinhauer,DimerQPT,exprDimerHys}. The minimal version for a ring requires $L_s{=}3$ sites, aka trimer \cite{trimer2,trimer3,trimer4,trimer6,trimer15,trimer7,trimer19,trimer20,trimer18,trimer12,trimer13,trimerSREP2,gallemi}.

The {\em classical} equations of motion for an $L_s$-site ring or chain, that are generated by the BHH, are known as the Discrete Non-Linear Schrodinger Equation (DNLSE) \cite{NLS1, NLS2}. In the continuum limit, one obtains the Gross-Pitaevskii Equation (GPE).  From a classical perspective, 
the dimer is integrable, while the trimer features a phase-space with a mix of quasi-regular and chaotic dynamics. Its analysis provides many insights regarding the feasibility of observing a metastable condensate \cite{sfs,sfc,sfa}. However, the trimer is non-generic in many respects, as clarified below. Our purpose is to extend the perspective. In particular, we would like to place the emphasis on the questionable metastability of condensates in Bose-Hubbard chains with ${L_s>3}$ sites. We also intend to clarify the GPE limit (${L_s \rightarrow \infty}$) where chaos is diminished, and integrability is recovered.

The BHH has inspired an extensive experimental work. The formation of lattices for trapping Bosons is state of the art; see review \cite{atomtronics} and references therein. The bosons can be trapped by a painted optical potential (see e.g. Fig.11 there). Optionally, the ring can be rotated {(control over the scaled rotation velocity $\Phi$, aka Sagnac phase)}. The occupation of the momentum orbitals or the current can be measured from time-of-flight images (see e.g. Fig.15 there). See, for example, experiments reported in \cite{exprRingRev,exprRingNIST}. The questions that we ask regarding metastability concern far-from-equilibrium scenarios of experimental interest, in particular in box geometry \cite{Box1,Box2,Box3,Box4,Box5,Box6} or possibly after invoking turbulence \cite{turbPRA,turbPRL,NTFP,WWT1,WWT2}.

\sect{Phase-space}
Formally, the BHH describes coupled non-linear oscillators and therefore provides an ideal arena for the study of ``quantum chaos" themes \cite{KolovskyReview,sfc,BHHchainChaos1,BHHchainChaos2,Swallow1,Swallow2,Swallow3,Swallow4,Swallow5}.
Taking into account that $N$ is a constant of motion, the BHH has effectively ${d=L_s{-}1}$ degrees-of-freedom (dof). For $L_s=2$, the phase space is the Bloch sphere, while for ${L_s=3}$ it is a 4-dimensional space. In the latter case, the energy surface is 3-dimensional, whereas the Kolmogorov-Arnold-Moser (KAM) tori are 2-dimensional surfaces \cite{LLbook}. It follows that chaotic regions are strictly separated by the KAM tori. If $L_s$ is larger, the $d$-dimensional KAM tori cannot divide the $2d{-}1$ energy surface into distinct regions: even if the non-linearity is small, we have in principle Arnold diffusion \cite{Chirikov1,Chirikov2}. Thus, ${L_{s}=3}$ is, in fact, non-generic. Generic chaos requires ${L_s>3}$.

\sect{Ground state}
Semiclassically, the ground state of the BHH can be visualized as a squeezed Gaussian located at the minimum of the energy landscape. Formally, the minimum is a stationary point (SP) of the Hamiltonian flow. If $U$ is large (compared to $N^2$), the squeezed Gaussian cannot be accommodated within the island that surrounds this SP, leading to a transition to a Mott Insulator (MI) state.  

\sect{Energetic Metastability (ES)}
The $L_s$ site BHH has $L_s$ single-particle orbitals. Each orbital is formally a trivial SP in phase space. For example, for ${L_s{=}2}$, these orbitals are the West and East poles of the Bloch sphere (the North and South poles correspond to the two sites of the system). If the SP is a local minimum, its island possibly can accommodate a squeezed Gaussian, and then we say that we get an energetically stable (ES) condensate. This is the type of stability that has been discussed by Landau (the ``Landau criterion"), i.e., thermodynamically stable SP. As pointed out by Leggett \cite{Leggett}, this type of metastability is feasible for condensate in a ring, but {\em not} for a condensate in a box. We shall inquire further this observation in the present work.

\sect{Dynamical Metastability (DS)}
If the SP is neither a local minimum nor a local maximum, then it is a saddle point of the energy landscape. There is a simple way to diagnose such a case: the Bogoliubov frequencies of the SP do not have the same sign or become complex. The latter option implies that the SP is dynamically unstable, whereas the former option opens the possibility to observe dynamical stability (DS). 

\sect{Quantum DS}
The question arises whether a dynamically stable SP can support a metastable condensate. The reasoning is based on the idea that the SP should be surrounded by a large enough stability island that can accommodate a squeezed Gaussian. The role of the Planck constant is played here by~$1/N$. Then, the feasibility of a DS condensate is based on the prevailing Semiclassical eigenfunction hypothesis of Percival \cite{Percival, Berry}, who suggested a classification of quantum mechanical spectra into 'regular' and 'irregular' eigenstates, supported respectively by quasi-regular islands and chaotic seas. 
This conjecture, that eigenstates can be associated with phase space regions, is complementary to Berry's conjecture \cite{Berry,Heller}, and to the Eigenstate Thermalization Hypothesis (ETH) \cite{ETH1,ETH2,ETH3,ETH4,ETH5,ETH6,ETH7,bht,KottosVardi}.

\sect{Quantum Hybrids}
In practice, it is likely to observe hybrid eigenstates that do not respect Percival's conjecture. Rather, they form a superposition of pieces that are localized in different regions in phase-space. This has been demonstrated in \cite{sst}. It possibly can be regarded as a variant of chaos-assisted tunneling \cite{CAT}, or as amphibious eigenstates that ignore the classical structures \cite{HybridK}. 

\sect{Quantum Scarring}
Even if we do not have a stability island, there might be a possibility for the SP to support a condensate. In particular, it has been demonstrated \cite{sst} that an unstable SP of an $L_s{=}3$ chain can localize a wavepacket and, hence, support a metastable condensate. But the latter option is rather exotic, and fragile, and better be regarded as a ``quantum scarring" effect \cite{scarsH,scarsKH,scarsK,scarsDag,ckt}.


\sect{Tomography}
Rather than asking yes/no questions about metastability versus ergodization, we prefer to use the
traditional ``quantum chaos" perspective, namely, we look for the quantum {\em signatures} of ES / DS and chaos. 
The common practice is to look on the energy level statistics. But such an approach has two issues: {\bf (i)}~It provides numerically clear results only for rather simple systems with large $N$, such that the spectrum in the range of interest is dense enough; {\bf (ii)}~It does not allow classification of the eigenstates in the typical situation of underlying mixed chaotic and quasi-regular dynamics. 
As opposed to that, our approach \cite{bem} is {\em numerically cheap} and flexible enough to deal with small complex systems. We use the term {\em quantum phase-space tomography} in order to emphasize that the inspection of the spectrum is not limited to ``level statistics", but rather oriented to reveal the detailed relation to the underlying phase space structures, that serve as a classical skeleton for the quantum eigenstates.   
A useful practice is to provide 3D images of the spectrum. Each point in such an image represents an eigenstate, whose vertical position is the energy. The extra dimensions (horizontal axis/axes and/or color-code) are exploited to reveal properties of the eigenstates. Such a technique has been exploited, e.g., to highlight Monodromy-related features in the spectrum \cite{bhm}.

\sect{Outline}
%
We introduce the model, its semiclassical limit, and its continuum GPE limit in \Sec{sec:model}, \Sec{sec:classical}, and \Sec{sec:GPE}, respectively. The energy landscape discussed in \Sec{sec:landscape} motivates the Bogoliubov stability analysis presented in \Sec{sec:zbogo}, \Sec{sec:SP}, and \Sec{sec:ebogo}, leading to the big picture of stability regimes in \Sec{sec:stability}. After inspection of representative trajectories in \Sec{sec:traj}, we present both quantum and classical tomography of the spectrum in \Sec{sec:tomo}. Furthermore, we show in \Sec{sec:spectral} how the transitions between the stability regimes are reflected. Quantum ergodicity is then carefully inspected in \Sec{sec:erg}. We conclude our findings in \Sec{sec:end}.

\section{Model Hamiltonian}
\label{sec:model}

Consider a system with $L_s$ equally-spaced sites $x_j$, labeled by~$j$. The BHH is expressed in terms of Fock operators $\bm{a}_j$ and associated site occupation operators 
${\bm{n}_j = \bm{a}_j^{\dagger}\bm{a}_j}$. 
Given on-site potential $V_j$, inter-site hopping $K_{ji}$,  and on-site interaction~$U$, it takes the form:
\beq \label{eBHH}
\mathcal{H} = \sum_{j=1}^{L_s} \left[ V_j\bm{a}_j^{\dag}\bm{a}_j + \frac{U}{2} \bm{a}_j^{\dag}\bm{a}_j^{\dag}\bm{a}_j\bm{a}_j \right]  
\ - \  \sum_{\langle ij \rangle} \frac{K_{ji}}{2} \bm{a}_j^{\dag} \bm{a}_i.
\ \ \ \ 
\eeq
We assume below that the on-site potential is zero. 
For a chain, the non-zero couplings are 
${ K_{j{+}1,j} = K_{j,j{+}1} = K }$,
where ${1 \le j < L_s}$. 
For a ring, periodic boundary conditions are imposed, 
such that the non-zero couplings are 
${K_{j{+}1,j} = K \exp(i\Phi/L_s)}$, and their conjugates, 
where $j$ is defined mod($L_s$). 
In the latter case, the Sagnac phase~$\Phi$ is proportional to the rotation frequency of the device, and can be regarded as a control parameter, as in recent experiments \cite{exprRingRev,exprRingNIST}.  The number of particles~$N$ is a constant of motion that plays the role of inverse Planck constant. 

\subsection{Phase-space}

We can define optional creation operators via a unitary transformation as follows:
${ \bm{a}_{\alpha}^{\dag}  = \sum_j \Braket{x_j}{\alpha} \, \bm{a}_{j}^{\dag} }$. The associated single-particle states,
aka orbitals, are 
${\ket{\alpha}  = \bm{a}_{\alpha}^{\dag} \ket{\text{vacuum}} }$. 
We can regard each orbital as a point in phase-space. For example, for 2~sites the ${ \{ \alpha \} }$ phase-space is
the two-dimensional Bloch sphere. 
Condensation of $N$ particles in the $\alpha$ orbital means 
${\left[ \bm{a}_{\alpha}^{\dag} \right]^N \ket{\text{vacuum}} }$. 
The $\alpha$-condensates can be regarded as a set of coherent states in the sense of Perelomov and Gilmore \cite{CSP,CSG}.

\subsection{The $k$-orbitals}

The stationary single-particle states of the Hamiltonian 
are the $k$-orbitals that are labeled by~$k$. 
For a ring, these are the momentum orbitals with 
${k=k_m=[2\pi/L]m}$, where $m$ is an integer mod($L$), 
and ${L=L_s}$ is the length of the ring. For a box, the labeling is 
\beq
k =&& k_m [\text{box}] = \frac{\pi}{L}m, 
\ \  \text{with} \ m=1,2,... 
\nonumber \\ 
&& \ \ \text{where} \  L=L_s{+}1. 
\eeq
Here, $L$ rather than $L_s$ should be regarded as the length of the box. The associated Fock occupation operators are labeled as $\bm{n}_k$, 
with subscript~$k$ instead of site-index~$j$. The respective unperturbed ($u{=}0$) Fock-basis states are indexed by $\{ n_k \}$. For the associated Fock creation operators, we used the dedicated notation 
\beq \label{eFock}
\bm{b}_{k}^{\dag}  = \sum_j \Braket{x_j}{k} \, \bm{a}_{j}^{\dag}.
\eeq  
For a ring the coefficients are $\sqrt{1/L}e^{ikx_j}$, 
while for a chain they are $\sqrt{2/L}\sin{kx_j}$. 

\subsection{Eigenstates}

After diagonalization of the Hamiltonian for $N$ particles, we get the many-body eigenstates $\ket{E_{\nu}}$ that are indexed by $\nu$. The ring Hamiltonian has translational symmetry, hence it decomposes into $L$ blocks that differ by the total momentum $P$ mod($2\pi$).
For the sake of numerical efficiency, we consider only the $P{=}0$ block. Likewise, in the case of the chain, we have mirror symmetry, and therefore, for the sake of numerical efficiency, we exclude the odd-parity states.   

From each eigenstate, we can extract an associated distribution in $\{ n_k \}$ space. In particular, we shall focus our attention on a selected orbital $k_{o}$, and will use the notation $P^{\nu}(n_o)$ for the associated probability distribution of its occupation and $\braket{n_o}$ for the respective expectation value. Furthermore, for a given eigenstate, we calculate the one-body reduced probability matrix 
\beq
\rho_{ij}^{(\nu)} = \frac{1}{N} \langle \bm{a}_{i}^{\dagger} \bm{a}_j \rangle_{\nu}.
\eeq
The {(basis independent)} purity of the state is
\beq
\mathcal{S} = \mathrm{Tr} \left(\rho^{2}\right).
\eeq
The inverse purity $1/\mathcal{S}$ reflects the number of participating orbitals. In particular, $\mathcal{S} =1$ implies that all the particles are condensed in a single orbital. A major objective is to identify the conditions for the appearance of such states, to characterize their stability, and to identify the associated excitations.  

\section{Semiclassical limit}
\label{sec:classical}

Formally, the BHH describes coupled non-linear oscillators, and therefore provides an ideal arena for the study of ``quantum chaos" themes \cite{KolovskyReview,sfc}.  
In a semiclassical context, one defines action-angle coordinates via 
\beq \label{act-ang}
\bm{a}_j = \sqrt{\bm{n}_j} \eexp{i\bm{\varphi}_j},
\eeq
and the Hamiltonian expressed as ${\mathcal{H}=H(\bm{\varphi},\bm{n})}$, becomes similar (but not identical) to that of a superconducting Josephson circuit.

In a classical context, we regard the $\bm{a}_j$ as phase-space coordinates of ``oscillators", with $\bar{\bm{a}}_j$ interpreted as the conjugate coordinates. The classical equation of motion for $(d/dt)\bm{a}_j$ is the DNLSE. For the generated trajectory, we can calculate $\langle \bar{\bm{a}}_{i} \bm{a}_j \rangle$ as a temporal average, and hence calculate the associated classical purity $\mathcal{S}$.


In the absence of interaction, condensation in the $k_o$ orbital, (${n_k = N\delta_{k,k_o}}$), is a stationary state. We already pointed out that such condensate can be regarded as a coherent state in the sense of Perelomov and Gilmore \cite{CSP,CSG}. In the Wigner-function picture, it is represented in phase-space by a minimal Gaussian-like distribution of uncertainty width~$2/N$ in $n_o$.   
With interaction, assuming that the condensate is {\em stable} (in the ES or in the DS sense), we expect to find an eigenstate with ${\mathcal{S} \sim 1}$, reflecting occupation ${n_o\sim N}$. Such an eigenstate possibly can be described as a {\em squeezed} coherent state. For very large interaction~$u$, such states get fragmented, as discussed in detail in \cite{bem}. The discussion of this SF-MI transition is outside the scope of the present work.

\section{GPE limit}
\label{sec:GPE}

The single-particle energies and the associated velocities 
in a tight-binding chain that has lattice constant $\ell$ 
are ${\varepsilon_k = -K\cos(\ell k)}$ 
and ${v_k = K\ell \sin(\ell k)}$. 
In  the continuum limit $v_k=(1/M)k$ with mass $M=(K\ell^2)^{-1}$. %
The dimensionless semiclassical interaction parameter of the BHH, that controls the local non-linearity and the manifestation of chaos, is
\beq
u = \frac{NU}{K} = \ell \, N M g,
\eeq
where ${g=U\ell}$. In the continuum limit, the DNLSE becomes the GPE with interaction~$g$. The only dimensionless parameter that survives in the GPE limit is 
\beq
u_L = Lu = L\frac{NU}{K} 
= \mathcal{L} \, N M g  
\equiv \frac{1}{2}\left(\frac{\mathcal{L}}{\xi}\right)^2,
\eeq
where $\mathcal{L}=L\ell$, while $\xi$ is known as the healing length. Fixing $u_L$, the parameter ${u=u_L/L}$ that controls the local chaos, diminishes in the GPE limit (no chaos).  
In the numerics, we treat the GPE and the DNLSE on equal footing. 
The units of length are chosen such that the lattice constant is $\ell=1$, and the units of time are chosen such that ${K=1}$.

A particle that is placed in the orbital $k_o$, has in the GPE limit kinetic energy ${\mathcal{E} = (K/2)(([2]\pi/L)m_o)^2 }$. Condensing $N$ particles in the same orbital implies that there is a shift in the potential floor, which is characterized by the parameter   
\beq
\Delta = \frac{N}{\mathcal{L}}g = \frac{NU}{L}.
\eeq
Accordingly, we identify two dimensionless parameters that are going to play an important role in the stability analysis of such condensate:
\beq \label{eGPE}
\text{GPE parameter} &=& \frac{\Delta}{\mathcal{E}} \ \sim \ \frac{u_L}{\pi^2 m_o^2},
\\ \label{eDNLSE}
\text{DNLSE parameter} &=& \frac{\Delta}{K} \ = \ \frac{u_L}{L^2}. 
\eeq
The DNLSE parameter reflects the existence of the lattice. In other words, the DNLSE, unlike the GPE, features a finite bandwidth because the kinetic energy is bounded ${\mathcal{E}=K [1-\cos(k)]}$. Once the DNLSE parameter becomes less than unity, it is implied that the healing length $\xi$ becomes smaller than the lattice constant $\ell$, and hence outside of the GPE regime.

\section{The energy landscape} 
\label{sec:landscape}

The standard textbook procedure is to transform the BHH to the momentum basis using \Eq{eFock}. Here, we treat the chain and the ring on equal footing and transform the finite-size BHH to the appropriate orbital basis. The Hamiltonian takes the form 
\beq 
&& \hspace*{-5mm} \mathcal{H} =  \sum_{k} \varepsilon_k \bm{b}_k^{\dag}\bm{b}_k 
\ + \ \frac{U}{2L} \sum 
C[k] \  \bm{b}_{k''''}^{\dag}\bm{b}_{k'''}^{\dag}\bm{b}_{k''}\bm{b}_{k'}  
\\ \label{eE0}
&&  \hspace*{-5mm} \ = 
\sum_{k} \varepsilon_k \bm{n}_k
+ \frac{U}{L}N^2
-\left[\frac{1}{2}\right]_{c} \frac{U}{2L} \sum_k \bm{n}_k^2 
+\text{transitions}
\ \ \ \ \ \ 
\eeq
where $C[k] \equiv C[k'''',k''',k'',k']$ are combinatorial coefficients, and $[1/2]_c$ is an extra prefactor in the case of a chain. 
{In the second line, for pedagogical purposes, the same expression is written as a function of the occupations $\bm{n}_k$, and therefore the terms in the $C[k]$ sum that involve {\em changes} in the occupations (i.e. ``transitions" of particles between orbitals) have been omitted.}
The energies of the orbitals are 
\beq 
\varepsilon_k [\text{ring}] &=& -K\cos(k-(\Phi/L)),
\\
\varepsilon_k [\text{chain}] &=& -K\cos(k), 
\\ 
\varepsilon_{\text{floor}} &\equiv& -K,
\\ \label{eEkin}
\mathcal{E}_k &\equiv& \varepsilon_k - \varepsilon_{\text{floor}}. 
\eeq
In the case of a ring, the $C[k]$ coefficients restrict the 
summation such that ${C[k]=1}$ is non-zero only if the condition 
${k''''+k'''- k''-k' = 0}$ is satisfied mod($L$).  
In the case of a chain, all possible variations of $\pm$ signs are allowed, and $C[k]$ depends on the combination. 
Specifically, $C_{chain}=3/2$ if all the $k$-s are identical, 
and $C_{chain}=1/2$ if only two $k$-s are identical,   
and $C_{chain}=1$ if there are two different pairs. 
Hence, an extra $1/2$ pre-factor is implied in front of the second term in \Eq{eE0}.    

Ignoring `transitions' that change the occupations of the orbitals, one concludes by inspection of \Eq{eE0} that for the assumed repulsion ($U>0$) there is an energy gain due to condensation. The impression is that a finite $U$ leads to metastability: the energy is reduced if all the particles are condensed in a single orbital. So, for a large enough $U$, any orbital can support an ES condensate. {\em This conclusion is wrong}, because the transitions are formally a ``kinetic term" in the Hamiltonian that deforms the potential floor, as implied by the Bogoliubov analysis (see subsequent section).  Furthermore, the implications are different for a ring and for a chain.

\section{Zero-order Bogoliubov analysis}
\label{sec:zbogo}

Assuming condensation in orbital ``o", and using that the number of particles is a constant of motion, we can substitute in the first term of the Hamiltonian 
${ \bm{b}_{k_o}^{\dag}\bm{b}_{k_o} = N - \sum \bm{b}_k^{\dag}\bm{b}_k }$.
The zero-order Bogoliubov approximation is obtained if we furthermore keep only the leading terms in the interaction that transfer {\em pairs} from the condensate $k_o$ to different orbitals.
For a ring, we get
\beq
&& \mathcal{H}_0 [\text{ring}] = 
\sum_{k \neq k_o} 
\left(\varepsilon_k-\varepsilon_{k_o} + \Delta  \right)
\bm{b}_k^{\dag} \bm{b}_k 
\nonumber \\ 
&&+ \Delta \sum_{q>0} 
\left( 
\bm{b}_{k_o-q}^{\dag}\bm{b}_{k_o+q}^{\dag} 
+ \bm{b}_{k_o-q}\bm{b}_{k_o+q}
\right).
\eeq 
For a chain, we have additional terms, notably those that create pairs in the same orbital,   
\beq 
&& \mathcal{H}_0 [\text{chain}] =
\sum_{k \neq k_o} 
\left(\varepsilon_k-\varepsilon_{k_o} + \Delta  \right)
\bm{b}_k^{\dag} \bm{b}_k  
\nonumber \\
&& + \ \ \sum_{k \neq k_o} \ \ 
\Delta 
\left( \bm{b}_{k}^{\dag}\bm{b}_{k}^{\dag} + \bm{b}_{k}\bm{b}_{k} \right)
\nonumber \\ 
&& + \!\!\! \sum_{|k'{\pm}k''|=2k_o} \!\!
\Delta  
\left( 
\bm{b}_{k'}^{\dag}\bm{b}_{k''}^{\dag} 
+ \bm{b}_{k'}\bm{b}_{k''}
+ 2\bm{b}_{k'}^{\dag}\bm{b}_{k''} 
\right)
\nonumber \\ 
&& - \ \ \sqrt{N}\frac{\Delta}{2} (\bm{b}_{3k_o}+\bm{b}_{3k_o}^{\dagger}). 
\label{eH0chain}
\eeq 
The last term shifts the SP from the origin and is ignored in what we call the {\em zero-order} Bogoliubov analysis.

\subsection{Ring}

For a ring, the diagonalization decomposes into `blocks', such that each block consists of two opposing orbitals, namely, ${k_o \pm q}$. We define the excitation energy per particle for a paired transition and the transverse excitation energy as follows:
\beq 
\varepsilon_q^{\parallel} 
&=& \frac{1}{2}(\varepsilon_{k_o+q}+\varepsilon_{k_o-q}) - \varepsilon_{k_o}
\ = 2K \cos\frac{\phi}{L} \sin^2\frac{q}{2}, 
\nonumber
\\ 
\varepsilon_q^{\perp} 
&=& \frac{1}{2}(\varepsilon_{k_o+q}-\varepsilon_{k_o-q}) 
\ = K\sin\frac{\phi}{L}\sin q.
\nonumber
\eeq
These frequencies are conveniently expressed as a function of the unfolded phase ${\phi = \Phi - 2\pi m_o }$. Accordingly, without loss of generality, one assumes $m_o{=}0$, gaining the option to address all the flow states in a single ${(\phi,u)}$ stability diagram. Namely, the associate Bogoliubov frequencies for a ring can be written as 
\beq \label{eBogoRing}
\omega_q [\text{ring}] \ \ = \ \ \varepsilon_q^{\perp} 
+ \sqrt{\left[\varepsilon_q^{\parallel} + 2\Delta\right] \varepsilon_q^{\parallel}}.
\eeq
See \Fig{fBogoRing} for illustration. 
Let us focus on a condensate that is prepared in an excited orbital {\em in the lower half of the spectrum}. In the absence of interactions, the Bogoliubov frequencies are {\em real},  but some of them are {\em negative}, reflecting the possibility of downwards transitions. But the excitation energy $\varepsilon_q^{\parallel}$ for a paired transition is always {\em positive}. Therefore, as $U$ increases, negative $\omega_q$ becomes positive, which implies that the condensate gets stabilized.  It becomes energetically stable.  An illustration for that is provided in \Fig{fBogoFiveSites}a.

\subsection{Chain}

For a chain, the diagonalization is more complicated, because we have to distinguish between several possibilities, and we end up with larger `blocks' in the diagonalization procedure. We shall provide a detailed account in the next section.
For pedagogical purpose, let us make the (generally incorrect) assumption that transitions between different orbitals can be ignored, keeping only the first two lines in \Eq{eH0chain}. Then the diagonalization of the Bogoliubov frequencies are 
\beq \label{eWc}
\omega_q [\text{chain}] \ \ = \ \ \sqrt{\left[\varepsilon_q^{\parallel} + 2\Delta \right] \varepsilon_q^{\parallel}},
\eeq
where the excitation energy per particle for a paired transition to the same orbital is simply 
\beq \label{eBogoChain}
\varepsilon_k^{\parallel} [\text{chain}] 
&=& \varepsilon_{k}-\varepsilon_{k_o}
\ = K[\cos(k_o)-\cos(k)].
\nonumber
\eeq
This expression, in spite of its questionable validity, illuminates a key difference between rings and chains. For a chain,  $\varepsilon_q^{\parallel}$ to the lower orbitals are negative. Consequently, as $U$ increases, the Bogoliubov frequencies become {\em complex}, implying that the condensate becomes dynamically unstable rather than ES.    

\begin{figure*} 
\centering 
\begin{overpic}[width=5cm]{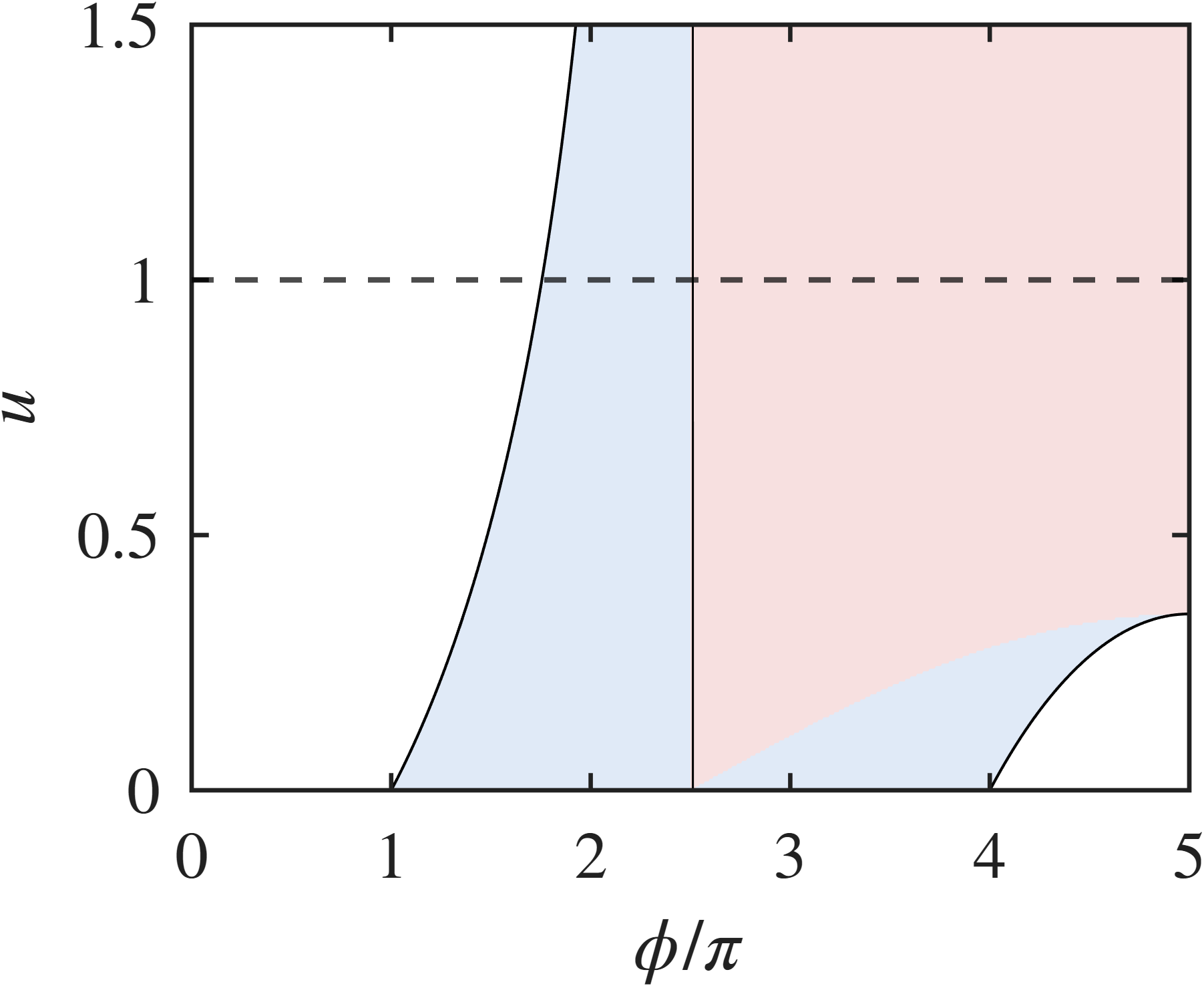} 
\put(-3.5,77){(a)}
\end{overpic}
\ \ \ 
\begin{overpic}[width=5cm]{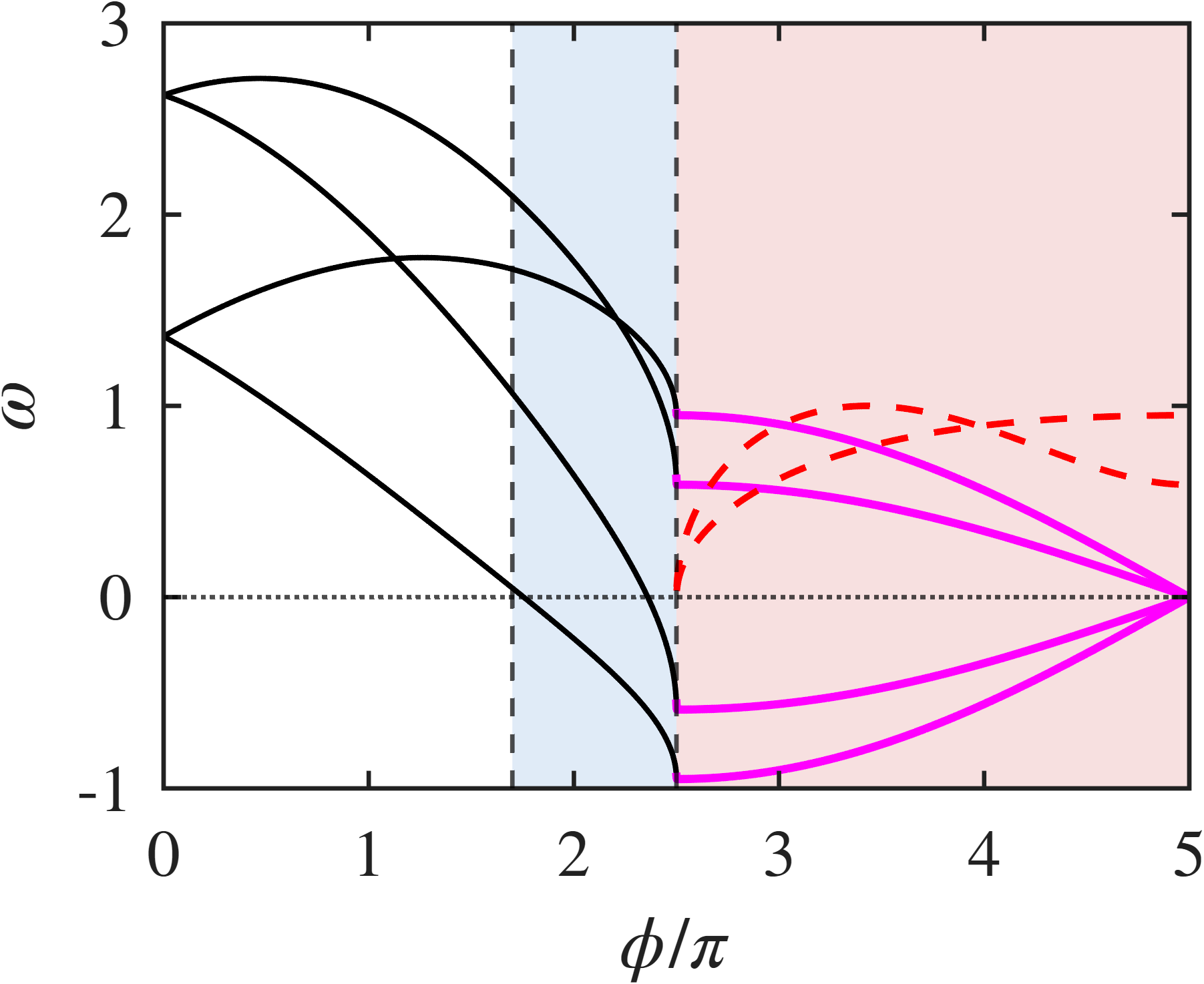} 
\put(-3.5,77){(b)}
\end{overpic}
\caption{{\bf Stability regimes for 5-site ring.}
(a)~The $(\phi,u)$  regime diagram. White regions indicate Landau energetic stability (all Bogoliubov frequencies are real and positive). Blue region indicates linear dynamical stability (Bogoliubov frequencies are real, but some become negative). Orange region indicates instability (Bogoliubov frequency becomes complex). 
(b)~Bogoliubov frequencies along the indicated ${u=1}$ section. In the region where they become complex, the real part is plotted in magenta, while the imaginary part is plotted as a red dashed line.    
}
\label{fBogoRing}
%
%
\ \\ \ \\ 
%
\centering
\begin{overpic}[width=5cm]{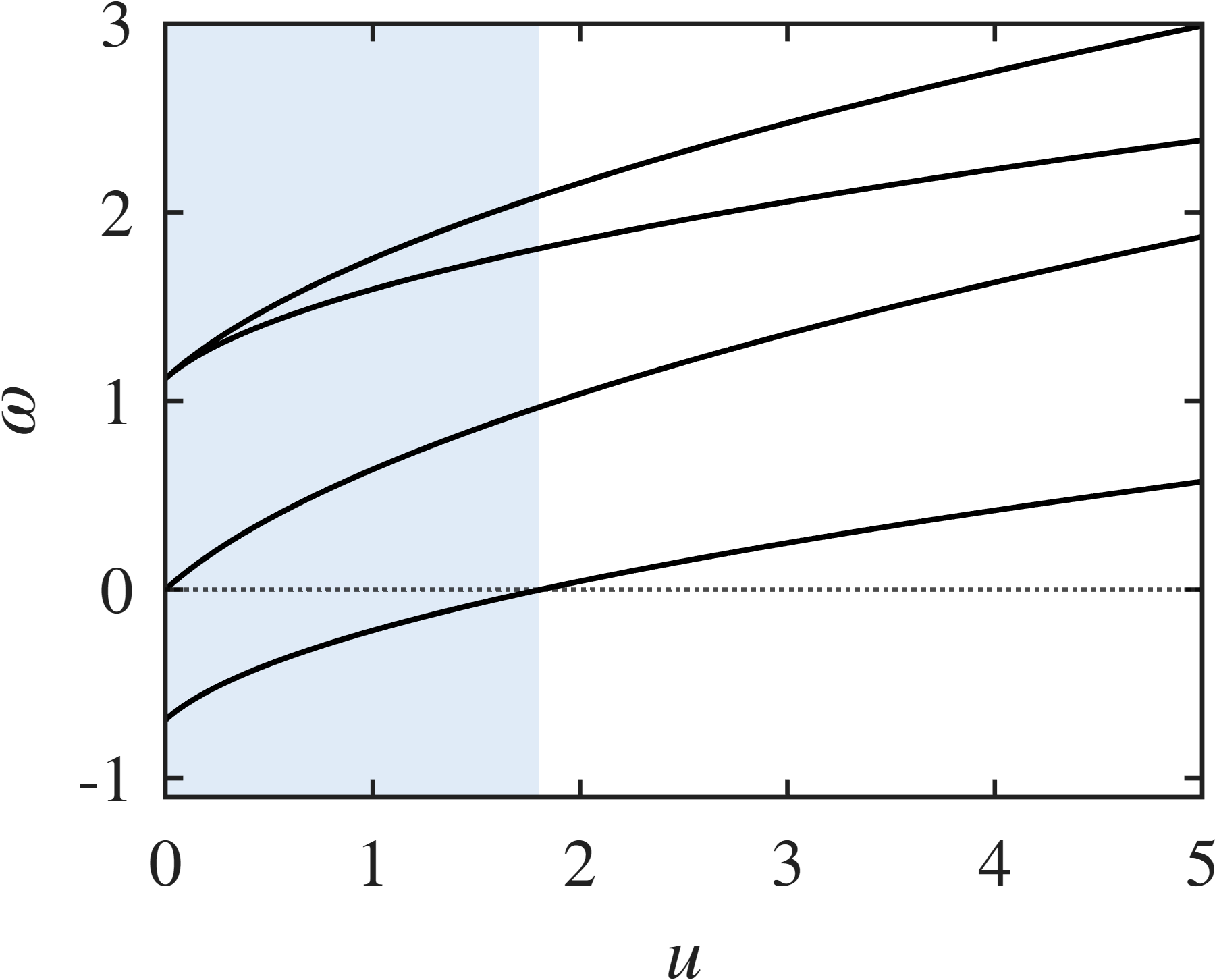}  
\put(-3.5,75){(a)}
\end{overpic}
\ \ \
\begin{overpic}[width=5cm]{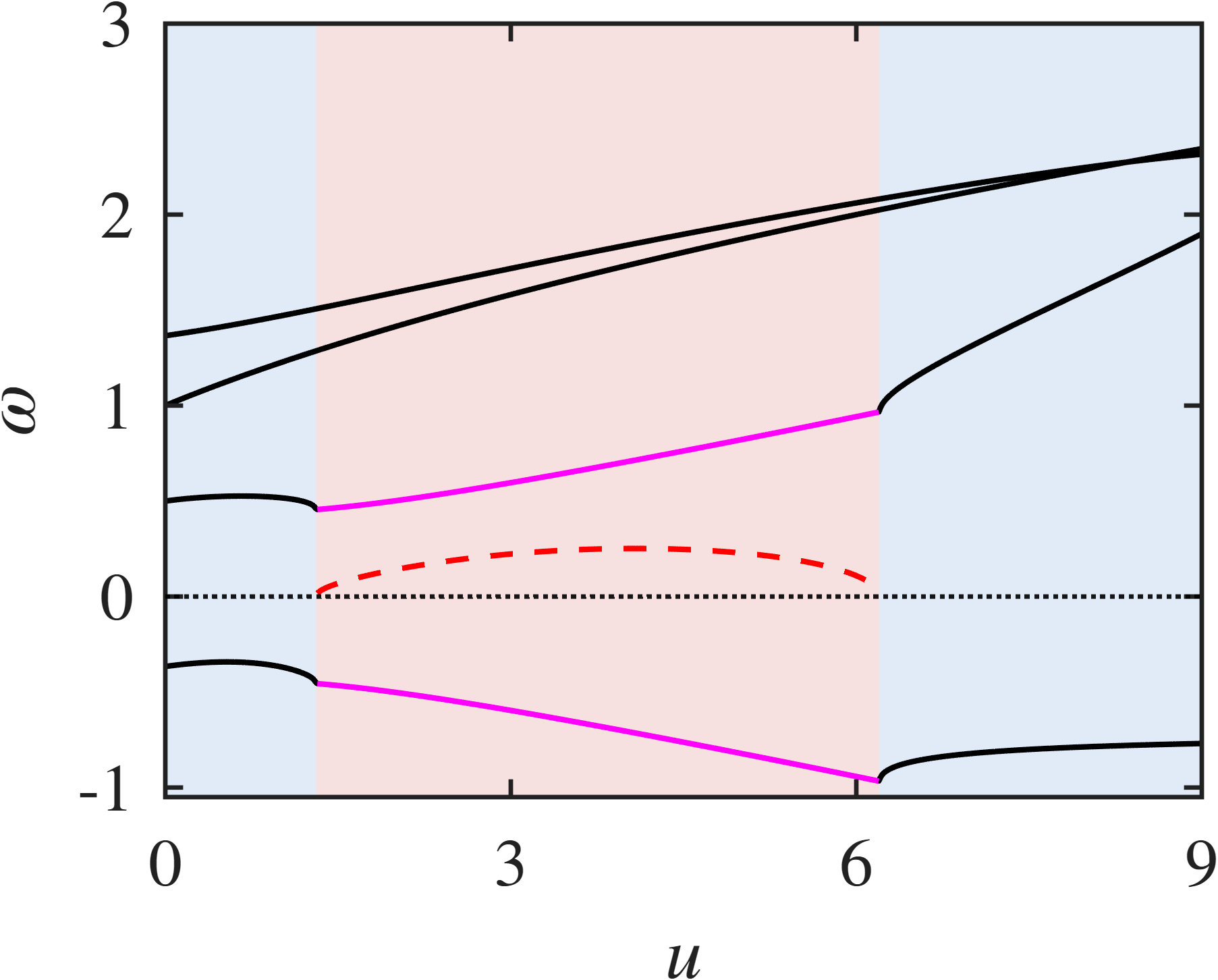} 
\put(-3.5,75){(b)}
\end{overpic}
\caption{{\bf Bogoliubov frequencies versus $u$, Ring vs Chain.} 
We compare the dependence of $\omega_q$ on $u$ for Ring and Chain with 5 sites. We inspect the SP that supports condensation in the first excited orbital.
(a) The Bogoliubov frequencies $\omega_q$ for $L_s{=}5$ ring. 
The DS to ES transition is illustrated, 
meaning that all frequencies become positive.
(b)The Bogoliubov frequencies $\omega_q$ for $L_s{=}5$ chain. 
In the region where they become complex, the real part is plotted in magenta, while the imaginary part is plotted as a red dashed line. 
The transition from DS to instability is illustrated,
meaning that some frequencies become complex.
}
\label{fBogoFiveSites}
%
\ \\ \ \\ 
%
%
\centering
\begin{overpic}[width=5cm]{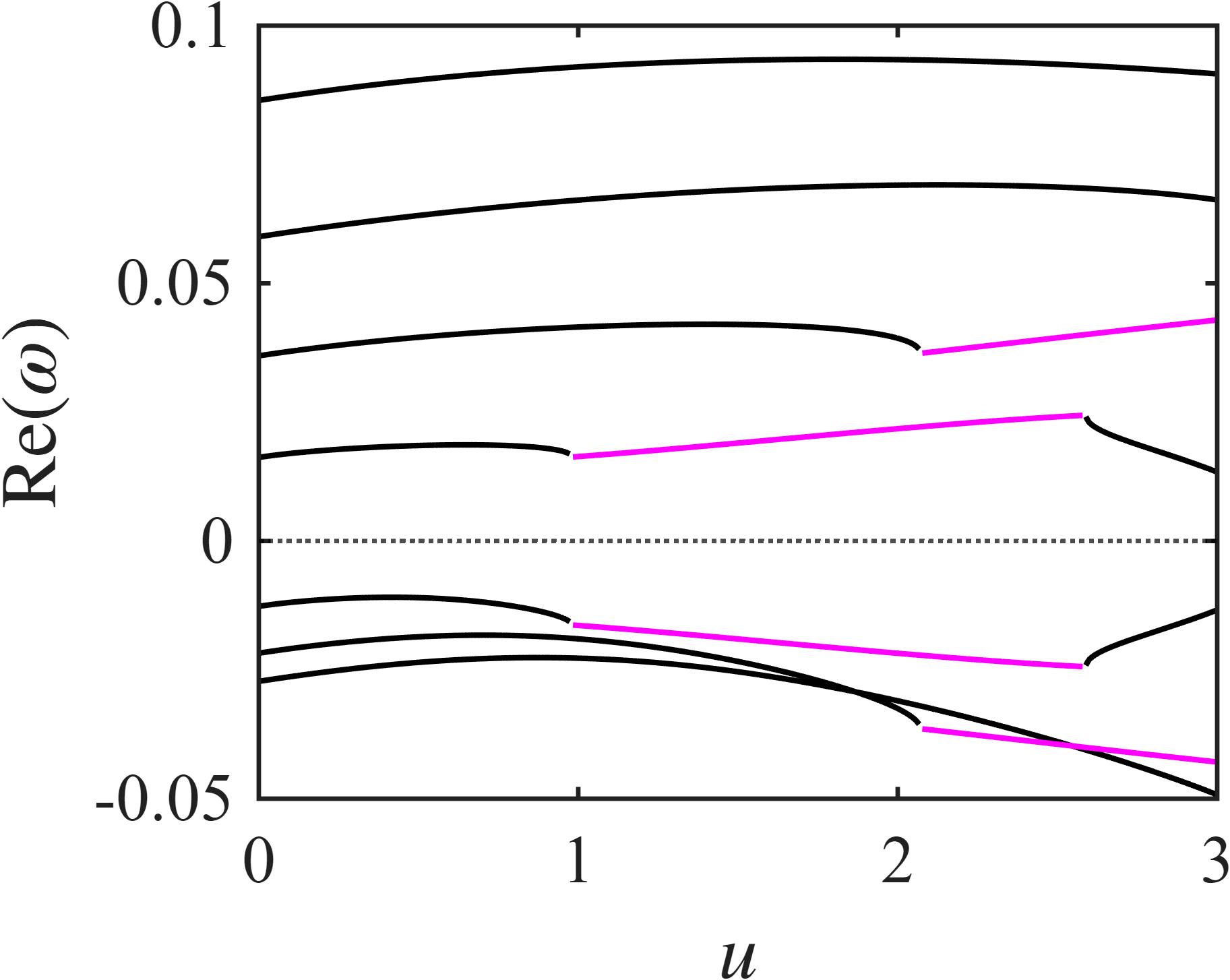}
\put(-3.0,74){(a)}
\end{overpic}
\ \ \
\begin{overpic}[width=5cm]{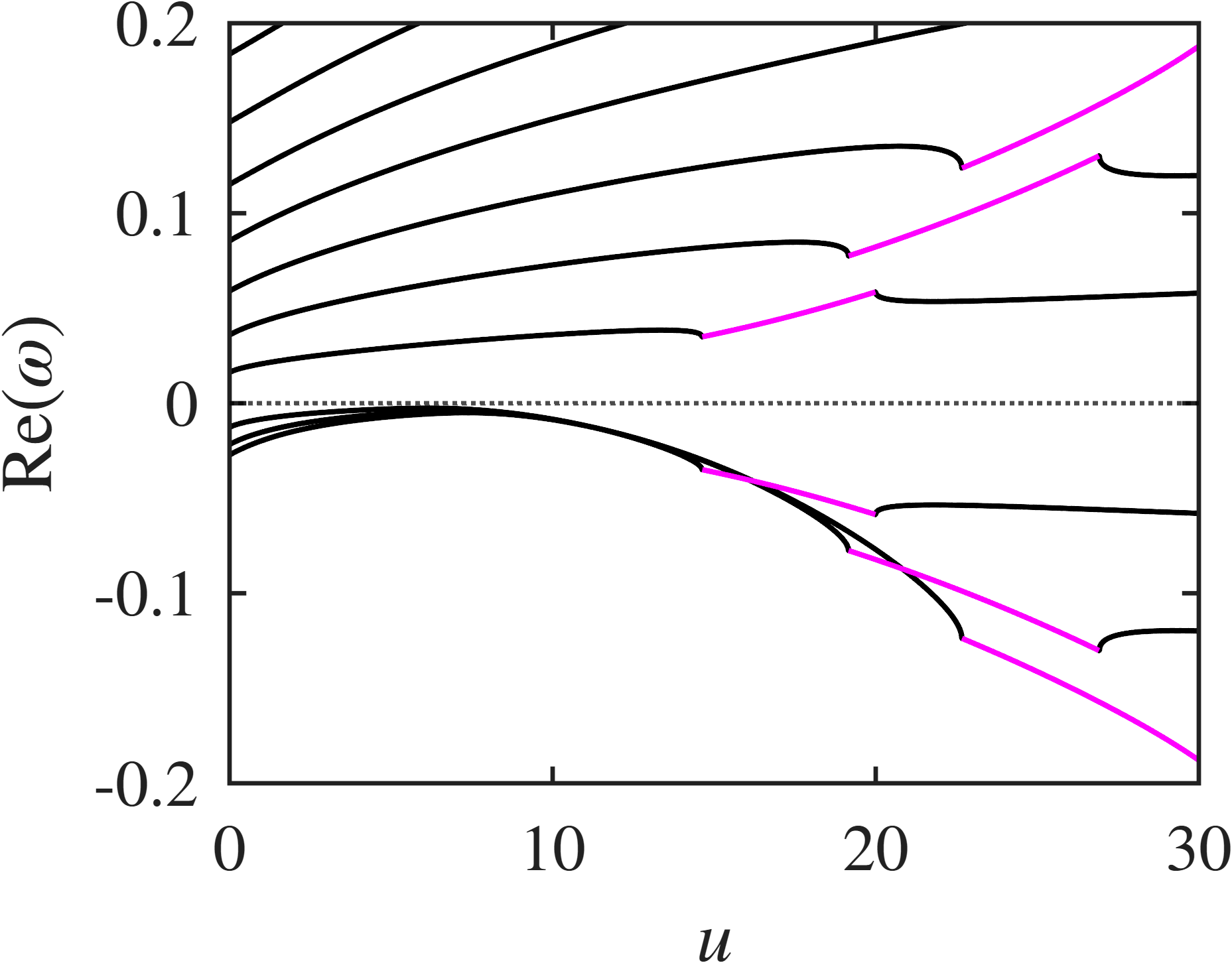}
\put(-3.0,73){(b)}
\end{overpic}
\ \ \
\begin{overpic}[width=5cm]{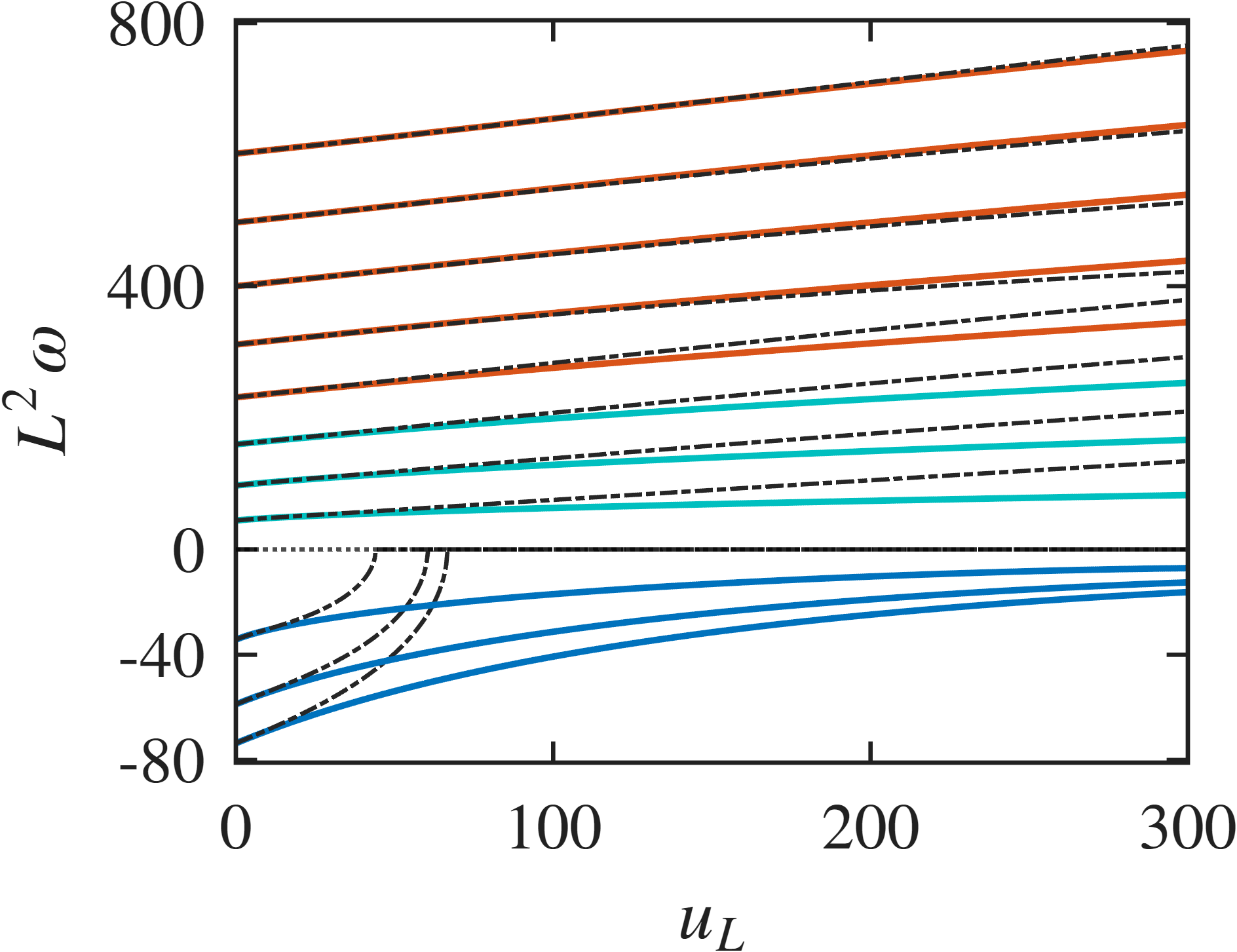}
\put(-3.0,72){(c)}
\end{overpic}
\ \ \
\caption{{\bf Bogoliubov freq versus $u$, large  Chain.} 
We compare the prediction of the Bogoliubov analysis 
to the exact numerical calculation for a chain with 51~sites, 
and clarify the GPE limit for chains with many sites. 
We inspect the SP that supports condensation in the 3rd excited orbital; therefore, for $u{=}0$, there are 3~negative Bogoliubov frequencies.   
(a) Zero-order result.
Complexity is indicated by the magenta color, as in \Fig{fBogoFiveSites}, but the imaginary part is not plotted. 
Note that instability shows up for a relatively small value of~$u$. (b) Exact result. 
Note that the transition to complexity is pushed to a large $u$ regime, which gets excluded in the GPE limit. 
(c) Zoom over the GPE regime, where the frequencies are real. One observes that the negative frequencies do not change sign but rather accumulate to zero. 
The negative frequencies are colored in blue, 
and the associated low positive frequencies are colored in cyan. 
Additional high positive frequencies are colored in red. 
The thin dashed-dotted lines are the outcome of a simple-minded approximation that is explained in \App{sec:BogoApp}.
}
\label{fBogoManySites}
\end{figure*}

\begin{figure}[b!]
\centering
\begin{overpic}[width=6cm]{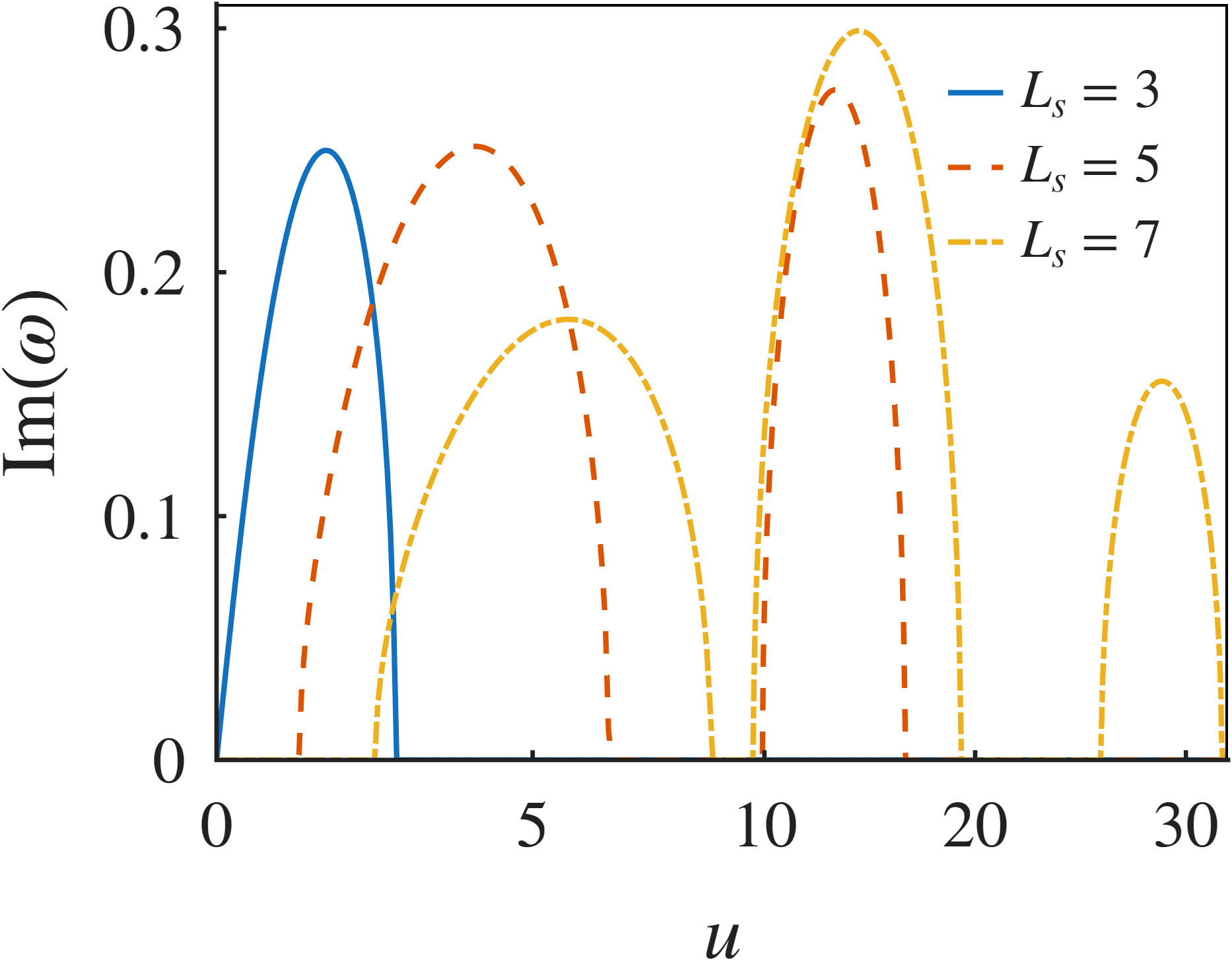}
\put(-3.5,74){(a)}
\end{overpic}
\ \ \ \
\begin{overpic}[width=6cm]{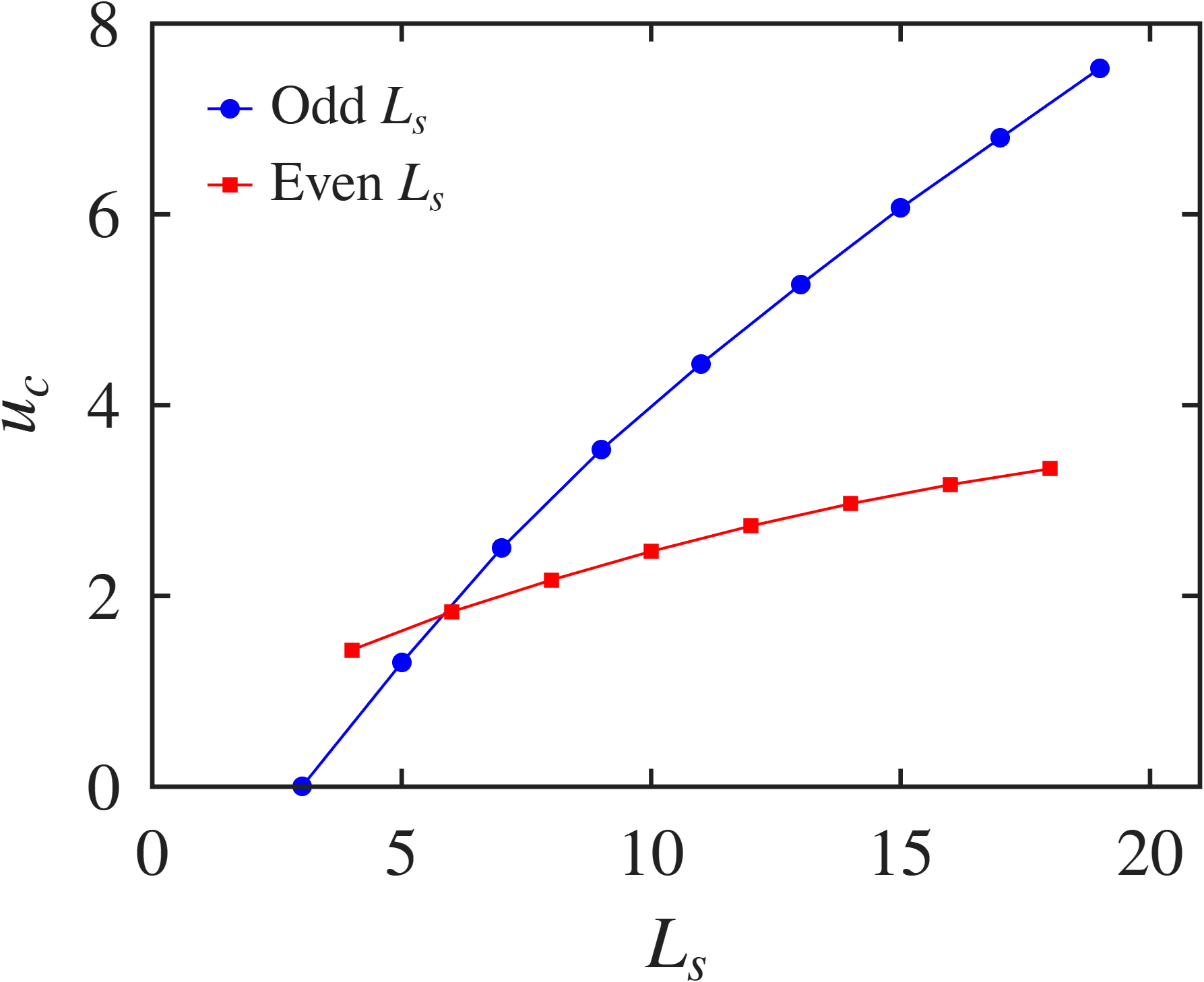}
\put(-3.5,76){(b)}
\end{overpic}
\caption{
{\bf Bogoliubov freq versus $u$, complex frequencies.}  
(a) The imaginary part $\im[\omega_q]$ of the Bogoliubov frequency is plotted versus $u$ for a chain with 3~(blue), 5~(orange), and 7~(yellow) sites. We inspect the SP that supports condensation in the first excited orbital.   
(b) The critical value $u_c$ above which complex frequencies appear as a function of $L_s$. Note that for 3~sites ${u_c=0}$. 
For ${u_L > u_cL}$, the SP becomes unstable; hence, in the GPE limit, the instability regime gets excluded, and chaos is globally diminished.  
}
\label{fBogoComplex}
\end{figure}

More precisely, the expression in \Eq{eWc} is dominated by two dimensionless parameters: the GPE parameter \Eq{eGPE} and the DNLSE parameter \Eq{eDNLSE}. For a large chain, we first hit the GPE border and later the DNLSE border.
This is illustrated in \Fig{fBogoManySites} and will be further discussed in the next section. Namely, one observes accumulation of negative frequencies at zero, after the GPE border is crossed, and they become complex only later, after the DNLSE border is crossed. To understand the GPE accumulation, we have to go beyond the zero-order analysis.   
As opposed to that, for a short chain, we encounter a premature DNLSE transition to complex frequencies, as seen in \Fig{fBogoFiveSites}b. In \App{sec:BogoApp} we provide the $u$ dependence for chains with ${L_s= 7, 9, 11, 13, 15, 21, 51, 91}$ sites to show how the GPE limiting behavior is gradually exposed, while in  \Fig{fBogoComplex} we show the $u$-regions where the DNLSE instability comes into play.

\section{The displaced condensate} 
\label{sec:SP}

There is a one-to-one correspondence between quantum and the classical Bogoliubov procedure, and in practice, the latter is more transparent and more convenient. Using phase-space language, the first step is to identify the SP that supports the condensate. 
The DNLSE is derived from the BHH via canonical Hamilton's equations using the action-angle coordinates defined in \Eq{act-ang}, and can be written as
\begin{equation}
i\frac{d}{dt} \psi = (\bm{D} + u \bm{P}) \psi, 
\label{A1}  
\end{equation}
where ${ \psi = \{ \psi_j \} }$  
is a column vector of the rescaled coordinates, 
namely, ${ \psi_j = N^{-1/2} \bm{a}_j }$. 
Commonly, $\psi$ is called {\em the~SP~wavefunction}. 
The definitions of the $\bm{D}$ and $\bm{P}$ matrices are implied, 
e.g. for the ${L_s=5}$ chain 
\beq
\bm{D} {=} {-}\frac{1}{2} 
\begin{pmatrix}
 0 & 1  & 0  & 0 & 0 \\ 
1 & 0 &  1  & 0 & 0\\ 
0  & 1 & 0 &  1 & 0\\
0  & 0 & 1 & 0 & 1 \\
0  & 0 & 0 &  1 & 0  \\
\end{pmatrix},
\ \ 
\bm{P} {=} 
\begin{pmatrix}
p_{1} & 0 & 0 & 0 & 0\\ 
0 & p_{2} & 0 & 0 & 0\\  
0 & 0 & p_{3} & 0 & 0\\ 
0 & 0 & 0 & p_{4} & 0\\
0 & 0 & 0 & 0 & p_{5}
\end{pmatrix},
\ \ \ \ 
\label{eDP}   
\eeq
where ${p_{j}= |\psi_j|^{2} =n_j/N}$ are the rescaled occupations, such that ${\sum_j p_j =1}$. The SP of the Hamiltonian are the non-zero solutions of the equation  
\beq
(\bm{D} + u \bm{P}) \psi  \ = \ \mu  \psi. 
\eeq
The non-linear eigenvalues can be written as 
\beq \label{eMu}
\mu \ \ = \ \ \varepsilon_{floor} + \Delta + \mathcal{E}, 
\eeq
where $\varepsilon_{floor} + \Delta$ is the potential floor, and $\mathcal{E}$ is a positive ``kinetic energy". The kinetic energy in the absence of interaction is given by \Eq{eEkin}, such that ${\mu=\varepsilon_k}$. As further discussed below, we can, by continuity, keep the same indexing for the non-linear eigenvalues as the interaction is increased, though in general one can obtain additional or different non-trivial SPs via bifurcations.  

For a {\em ring}, the trivial SP wavefunctions are {\em not} affected by the interaction. On the technical level, $\bm{P}$ is proportional to the identity matrix, and therefore the trivial zero-order SP wavefunctions remain eigenvectors of ${ (\bm{D} + u \bm{P}) }$. The only effect of the interaction is the shift of the potential floor in \Eq{eMu}. Namely, all the $\mu$ eigenvalues are shifted by~$\Delta$, but otherwise not affected. Note that, for a ring, there are also non-trivial complex eigenfunctions \cite{NLS1}, whose discussion is beyond the scope of the present study. 

For a {\em chain}, the SP wavefunctions are, in general, affected by the interaction. {The simplest example is the SP wavefunction of an $L_{s}=7$ chain that supports the $m_o{=}2$ condensate. For zero interaction ${\psi \propto \sin(kx)}$ with ${k=(\pi/8)m_o}$, while for very large $u$ the wavefunction get flatten except the zero at the ${x=4}$ node.} The associated densities are 
\beq \label{eSP7}
\bm{p} \big|_{ \text{zero $u$} } &=& \{ |\psi_j|^2 \} \ = \frac{1}{8} \{1, 2, 1, 0, 1, 2, 1\}, \\
\bm{p} \big|_{ \text{large $u$} } &=& \{ |\psi_j|^2 \} \ \approx \frac{1}{6} \{1, 1, 1, 0, 1, 1, 1\}.
\eeq
More generally, for larger $L_s$, in the GPE limit, there is a nodal structure that does not change with the interaction. Namely, the stationary solutions deform smoothly as the interaction is increased. For small $u$ they look like sinusoids, while for larger $u$ they become elliptic functions. For very large $u$, the density ${p_j=|\psi_j|^2}$ becomes uniform, interrupted by dark solitons at the nodes. Nodes neither disappear nor are generated. Based on Eq(24) of \cite{NLS1}, one obtains 
\beq
\mathcal{E} \Big|_{ \text{zero $u$} }  \ \ &=& \ \  \frac{K}{2} \left(\frac{\pi}{L} m_o\right)^2, \\
\mathcal{E} \Big|_{ \text{large $u$} } \ \ &\approx& \ \  2K \frac{\sqrt{u_L}}{L^2}m_o.
\eeq
One observes that the crossover from the small to the large $u$ regime happens at the GPE border of \Eq{eGPE}. The equal spacing for large $u$ reflects the cost of each soliton notch.

Once we find the SP wavefunction, we can also calculate the optional $n_k$ coordinates from its $n_j$ coordinates. In particular, the SP wavefunctions of a chain in the GPE limit feature approximately uniform density for large $u$, hence we get  
\beq 
\frac{n_{k}}{N} \ \ \approx \ \ \frac{8}{\pi^2} \left(\frac{k_{o}}{k}\right)^2,
\eeq
where $k$ should be an odd harmonic of $k_{o}$. We use the notation ${n_{\text{SP}} \equiv n_{k_o}}$. For a non-depleted condensate ${n_{\text{SP}} {=} N}$. For strong interaction, we get from the above formula ${n_{\text{SP}} \approx 0.8N }$, meaning that even for very large $u$, the depletion of the condensate is no more than $20\%$. 
We can also calculate $E_{\text{SP}}$. Based on \Eq{eE0} one obtains 
\beq
E_{\text{SP}}=\varepsilon_{k_o} N + \frac{3U}{4L}N^2.
\eeq
If we go beyond zero-order, the SP might become shifted such that ${n_o {=} n_{\text{SP}}}$ is less than~$N$. If this is the case, it means that the stationary condensate is somewhat depleted. Accordingly, in the quantum treatment, the condensate is semi-classically represented by a Gaussian-like cloud that might be both {\em shifted} and {\em squeezed} relative to the zero-order solution.

\section{Exact Bogoliubov analysis} 
\label{sec:ebogo}

This section is rather technical and can be skipped in the first reading. In the next section, we discuss the implied stability regimes for rings and chains, which provide a bridge to the tomographic inspection of the spectrum.  

For the purpose of stability analysis, one finds the Bogoliubov frequencies $\omega_q$ of the small oscillations around the SP.  For that purpose, one has to carry out the diagonalization of the Bogoliubov matrix $\bm{W}$, which is obtained from the Hessian of the Hamiltonian ${\mathcal{H}(a,\bar{a})}$. For a chain and also for a non-rotating ring, the SP wavefunction $\psi$ is real, and $\bm{W}$ can be written as follows:
\beq \label{eW}
\bm{W} =
\begin{pmatrix}
\bm{D} + 2 u \bm{P}- \mu & -u \bm{P}   \\ 
u \bm{P}  & -(\bm{D} + 2 u \bm{P}- \mu)  
\end{pmatrix}.
\label{A3}
\eeq
The diagonalization results in $L_s$ pairs of characteristic frequencies $\pm\omega_{q}$, that are indexed by ${q}$. The trivial frequency ${\omega_{0}{=}0}$ is implied by conservation of particles.  

The zero-order Bogoliubov analysis assumes that the SP for ${u>0}$ is the same as the SP of the non-interacting system. We already pointed out that this assumption is correct for the trivial SP wavefunctions of the ring. Namely, the uniform density of the condensed particles is not affected, and therefore the $\omega_q$ of \Eq{eBogoRing} are exact. 

The zero-order Bogoliubov analysis is also exact for the middle orbital of a chain with an odd number of sites, see \App{sec:BogoExact}. But this is not very interesting because only the lower half of the spectrum survives in the GPE limit. The upper half of the spectrum formally corresponds to bosons with attractive interactions, which is not of our interest here. In the following, we discuss the results for $\omega_q$ for condensation in orbitals that belong to the lower half of the spectrum.

\subsection{Chain with 5 sites}

For a general chain, the SPs that support the $u{=}0$ condensates are displaced once the interaction is turned on. Conveniently, this is not the case with the SP wavefunction $\psi^{(1)}$ that supports the first excited orbital of the 5-site chain. This SP is not displaced by the interaction:
\beq
\psi_j^{(1)} = \frac{1}{2}(1, 1, 0, -1, -1).
\eeq
The 5~site chain is going to serve as our leading numerical illustration; therefore, we provide here explicit expressions for the Bogoliubov frequencies. They are indexed as ${q{=}-1,0,1,2,3}$. Two roots ($\omega_0$ and $\omega_2$) are rather trivial, while the other three are determined by a cubic equation, see \App{sec:BogoExact}. The result is 
\beq \label{eOmegaChain}
\omega_{\pm1}^2 &=& \frac{1}{3}\left(c_2 +  e^{\pm i\pi/3} C + e^{\mp i\pi/3} \frac{\Delta_0}{C}\right), \\
\omega_2^2 &=&  1+\frac{1}{2} u, \\
\omega_3^2 &=& \frac{1}{3}\left(c_2 + C + \frac{\Delta_0}{C}\right). 
\eeq
For small $u$, the $C$ comes out negative 
and consequently the frequencies come out real. 
The critical value ${u_c \approx 1.307}$ is the minimal~$u$
that makes $\omega_{\pm1}$ complex. 
The dependence of the Bogoliubov frequencies on $u$ is plotted in \Fig{fBogoFiveSites}b. As opposed to the ring, here, as $u$ is increased, we have a transition to instability instead of a transition to $ES$.

\subsection{Chain with many sites}

\begin{figure}
\centering
\begin{overpic}[width=4.2cm]{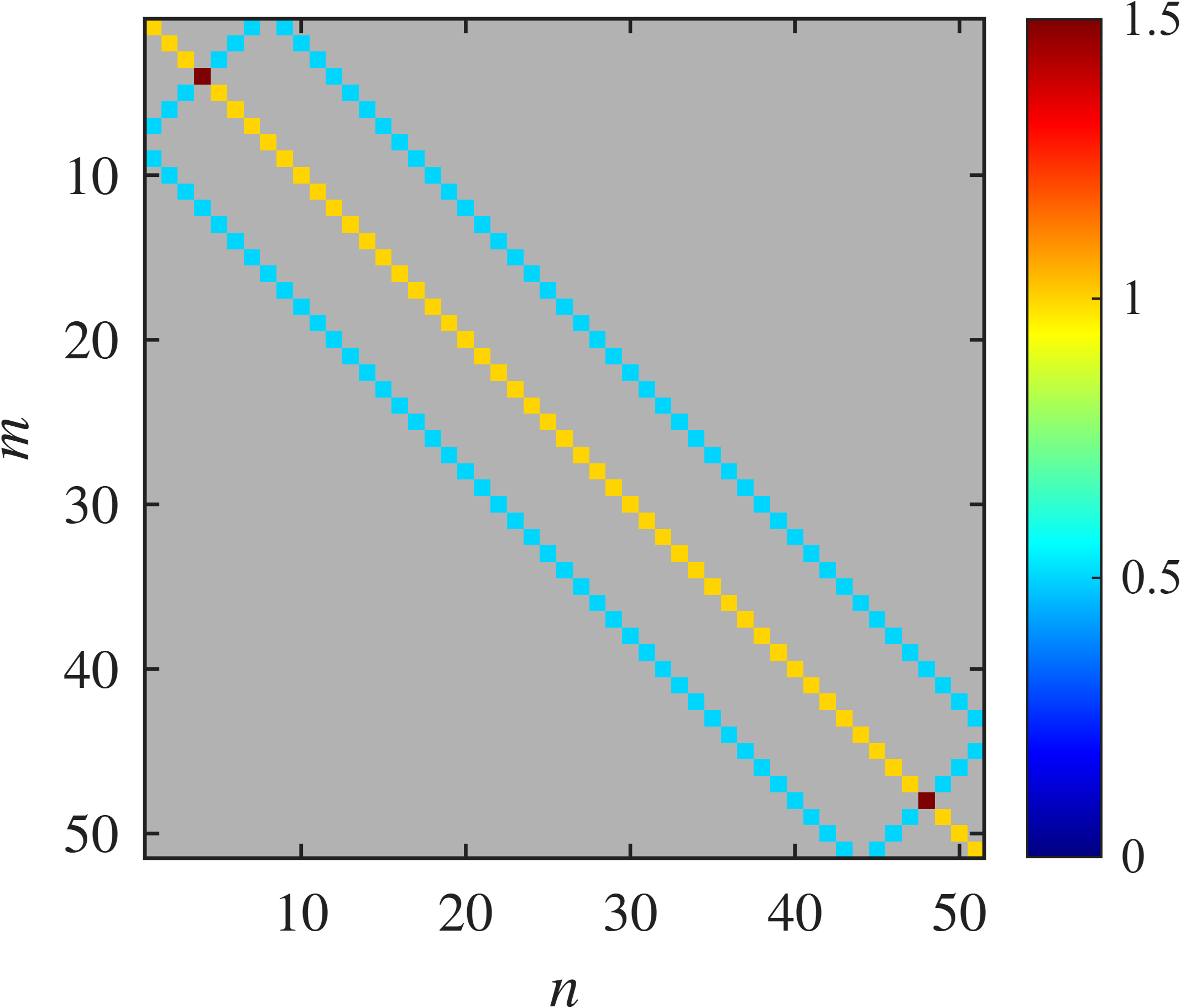}
\put(-3.0,79){(a)}
\end{overpic}
\begin{overpic}[width=4.2cm]{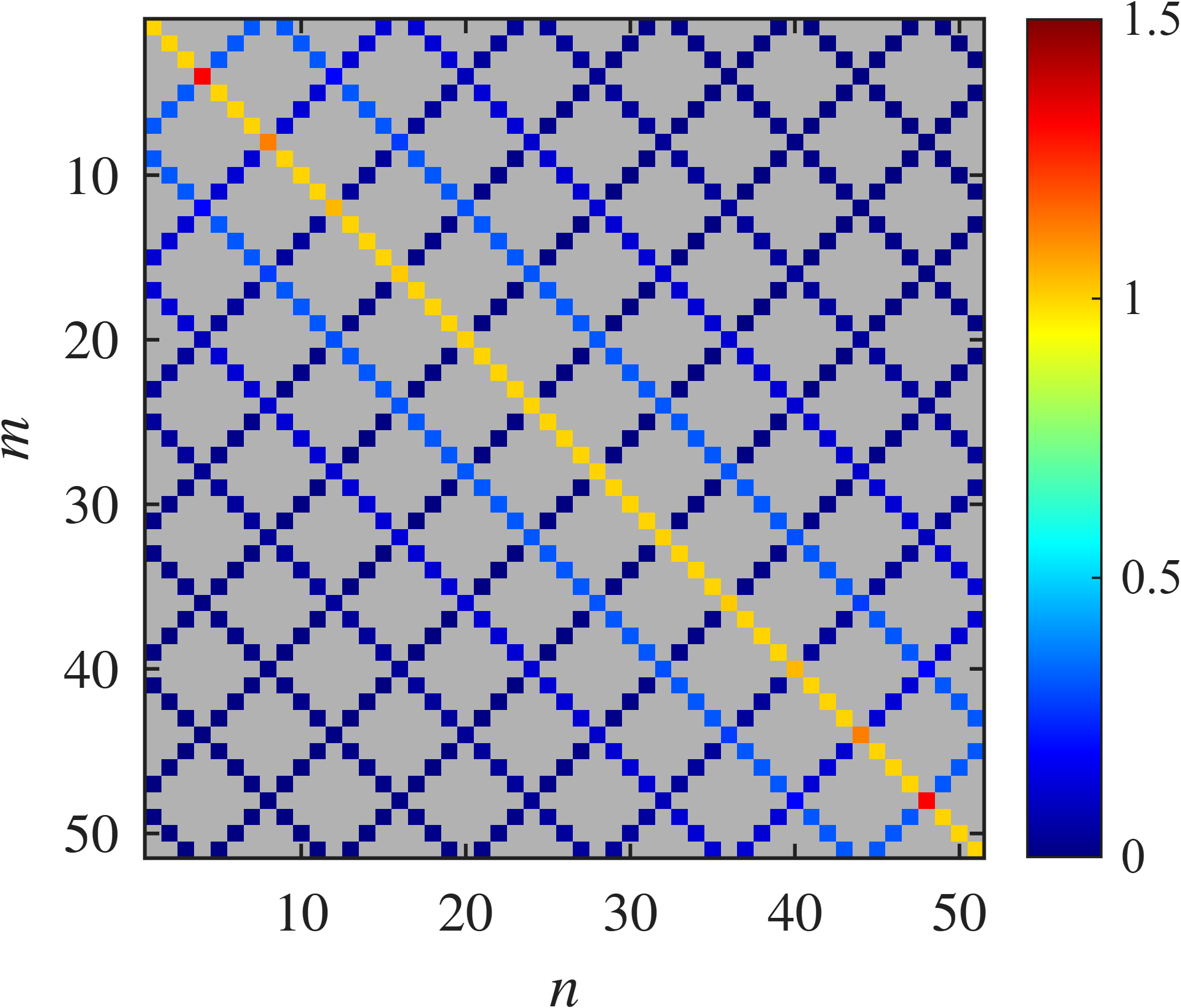}
\put(-3.0,79){(b)}
\end{overpic}
\caption{
{\bf The structure of the $\tilde{\bm{P}}$ matrix.}
Image of the $\tilde{\bm{P}}$ matrix (multiplied by $L=L_s{+}1$) in the orbital representation for the $m_o{=}4$ stationary state of an $L_s{=}51$ chain. Left and right panels are for $u{=}0$ and $u{=}8$. 
Exact zero values are colored in gray, while blue color indicates very small non-zero numerical values.}

\label{fPmatrix}
\end{figure}

In general, the SP of a chain is affected by~$u$. Consequently, the zero-order Bogoliubov results are flawed. The exact calculation requires first finding the SP wavefunction and then calculating ${ \bm{P}=\text{diag}\{ \bm{p} \} }$. It is illuminating to transform this matrix into the orbital representation. An illustration is provided in \Fig{fPmatrix}. For zero interaction, the non-zero elements are in one-to-one correspondence with the bi-linear terms of \Eq{eH0chain}. As $u$ is increased, the off-diagonal elements get smaller, but on the other hand, a pattern appears that reflects the comb of notches that is featured by the solitonic stationary distribution.

The failure of the zero-order approximation is demonstrated numerically in \Fig{fBogoManySites}, where we plot (panel~a) the prediction that is based on the zero-order Bogoliubov analysis. The $\omega_q$ becomes complex as $u$ becomes of order unity. But in the exact calculation (panel~b), this happens for a much larger value of~$u$. We zoom over the GPE regime where the frequencies are real (panel~c). The outcome of a simple-minded approximation that is based on block-wise diagonalization of $\bm{W}$ is indicated by thin dashed-dotted lines. Its reasoning is explained in \App{sec:BogoApp}. It provides expressions of the type \Eq{eBogoRing} and \Eq{eWc}. It fails to deliver a proper explanation for the accumulation of the negative frequencies.   

In order to explain the accumulation of the negative frequencies, it is better to take a second look at $\bm{W}$ of \Eq{eW} as written in the site representation. If we ignore pair destruction and creation, the eigen-frequencies are obtained from the diagonalization of the GPE-like block, where the mean-field potential is enhanced by a factor~$2$, and shifted by~$\mu$. If this potential were flat, we would get ${\omega_q(u) = \omega_q(0) + \Delta }$, reflecting the extra cost~$\Delta$ for relocating a particle from the $k_o$ orbital to a different orbital. But the mean-field potential has notches where the potential is $-\Delta$. Those notches can support eigenvectors that have negative frequencies. These frequencies become less negative as $u$ increases, because we have to add the kinetic energy that comes from $\bm{D}$. This kinetic energy is bounded by unity (which means $K$ in unscaled units). Hence, as we come close to the GPE border of \Eq{eGPE}, the kinetic energy of the eigenvectors compensates the $\Delta$, leading to close-to-zero frequencies. There are, of course, the pair destruction/creation terms that couple the two blocks of $\bm{W}$. Their effect is similar to what we have encountered in the calculation for a ring. They mainly affect the positive frequencies, and we can use the simple-minded approximation as discussed in \App{sec:BogoApp}.

Still, one wonders whether it is possible to obtain a tangible approximation for the negative frequencies. This goes beyond the scope of our interest here, but we point out the following hand-waving estimate. If we had an infinite chain, the notch would gain translational symmetry, meaning that the ``negative" frequencies are in fact Goldstone modes with zero frequency. But if we have a comb of notches distance $L/m_o$ apart, we expect negative frequencies of order ${\exp[-L/(m_o\xi)]}$ due to tunnel-overlap.

\section{Stability regimes}
\label{sec:stability}

On the basis of the previous section, let us discuss the implications with regard to the stability of a given SP. If all the Bogoliubov frequencies are real and positive, then the SP is a stable minimum (ES). If some are negative (but real), it indicates that the SP is a stable elliptic point that sits on a saddle in the energy landscape, suggesting DS.  If some of the frequencies become complex, it indicates that the SP is an unstable hyperbolic point, and possibly it is associated with the emergence of chaos in its vicinity.

\subsection{Ring}

For a ring, a given orbital, indexed as ${k_o = 2\pi m_o}$, constitutes a trivial SP in phase-space: its location is not affected by the interaction because the particles are distributed uniformly over the sites. 
The stability diagram for ${L=5}$ was displayed in \Fig{fBogoRing}. For pedagogical purpose we consider a section along ${u=1}$. As $\phi$ is increased the ${m_o=0}$ condensate becomes metastable once ${\phi>\pi}$. Then at ${\phi_c^{ES}}$ it goes from ES to DS, and after that at $\phi_c$ it goes from DS to instability. 

Let us expand on the determination of the borders between the stability regimes. 
If the ring is large (${L\gg 1}$) an approximation for \Eq{eBogoRing} is ${\omega_q \approx c_s |q| + v_{o}q }$, 
where ${c_s =\sqrt{NUK/L}}$ is the speed of sound 
and ${v_{o} \approx (K/L)\phi}$. 
Recall that ${\phi=\Phi-2\pi m_o}$, hence $v_o$ is the rotation velocity of the ring {\em relative} to the super-flow. Consequently, we deduce the Landau criterion for stability ${|v_o| <  c_s}$, 
leading to ${|\phi|<\phi_c^{ES}}$, where  
\beq
\phi_c^{ES} \ \ = \ \ \sqrt{u_L}.
\eeq
It ensures that all the excitations in the rotating frame are positive, and hence the SP is a local minimum.

More generally, if $L$ is not large, the generalized Landau criterion for ES involves the lowest excitation frequency $\omega_q$ that is obtained for ${ |q|=2\pi/L }$. Hence, we get
\beq
2 \cos\left(\frac{\phi_c^{ES}}{L}\right) = -\left(\frac{u}{L}\right) 
+ \sqrt{ \left(\frac{u}{L}\right)^2 + 4 \cos^2\left(\frac{q}{2}\right)}. 
\eeq     
Consistency with the large $L$ approximation is easily verified.
Given $L$, the critical value ${\phi_c^{ES}}$ increases with $u$ from the minimal value ${\phi_c^{ES} \sim \pi}$  to the limiting value ${\phi_c^{ES} \sim (L/2)\pi}$. 
Irrespective of that, the Bogoliubov frequencies become complex once the stability border $\phi_c$ is crossed.  
Namely, we get the following sequence of transitions: 
\beq
\pi \ \ < \ \ \phi_c^{ES}(u) \ \ < \ \ \phi_c \ \ = \ \ L\frac{\pi}{2}.
\eeq

Optionally, we can consider a non-rotating ring. In \Fig{fBogoFiveSites}a, we plot the $\omega_q$ of the ${m_o=1}$ excited condensate as a function of $u$. As the interaction increases, $\phi_c^{ES}(u)$ becomes larger than $\phi$, and consequently, the negative Bogoliubov frequency becomes positive, and the stability of the condensate changes from DS to ES.

\subsection{Chain}

We now ask what happens if we have a chain instead of a ring. The dimensionless parameters that control the stability analysis are \Eq{eGPE} and \Eq{eDNLSE}. As already discussed in previous subsections, the GPE and the DNLSE parameters determine, respectively, borders that indicate the accumulation of negative frequencies, and for getting instability. This was illustrated in \Fig{fBogoManySites}. 
For short chains, the DNLSE border comes before the GPE border, and then the GPE region is not exposed, and no accumulation is observed, as was demonstrated in \Fig{fBogoFiveSites}b.

For the purpose of comparison between chains and rings, let us consider a non-rotating ring ($\Phi{=}0$). Then, the relative velocity of the super-flow is 
${|v_o|=(K/L)2\pi |m_o| = K k_o}$. 
As $u$ is increased, the stability changes from DS to ES. See illustration in \Fig{fBogoFiveSites}a for an ${L_s=5}$ ring.
In the GPE regime, we write again the Landau criterion  
\beq
|v_o| \ \ < \ \ c_s,
\eeq
where $v_o$ is the flow velocity. 
For a chain, the analogous definition would be 
${v_o = K k_o = K(\pi/L)m_o }$. 
Recalling that ${c_s=\sqrt{u_L}K/L}$, it follows that the stability border that is implied by the Landau criterion 
coincides with the GPE border of \Eq{eGPE}. 

{\em We therefore conclude that the Landau stability border coincides with the GPE border.} For a ring, the negative frequencies change sign, and we get ES. For a chain, they do not change sign but rather accumulate to zero.

The GPE regime is furthermore bordered by DNLSE instabilities. For an $L_s=5$ chain, we saw in \Fig{fBogoFiveSites}b that as $u$ is increased, there is a transition at $u_c$ from DS to instability. Looking further, we see that for much larger $u$, we get DS again. More generally, for the first excited orbital for chains of length ${L_s=3,5,7}$, we get the results that are displayed in \Fig{fBogoComplex} featuring several DS regions. 

For long chains, the DNLSE border is pushed to much larger values of $u$, hence exposing a fully developed GPE-like regime. Looking in \Fig{fBogoComplex}b, one observes that for a chain with an even number of sites, the critical value is smaller. This shows that for an even number of sites, the SP wavefunction ${ \psi }$ does not have a node and is therefore somewhat more uniform. Irrespective of that, we see that for large $L_s$ we get ${u_{c}{=}\infty}$, hence the critical value of ${u_{L}{=}u_cL}$ goes to infinity {\em faster than $L$}, meaning that all the condensates are DS in the GPE limit.

\begin{figure*}
\begin{overpic}[width=4cm]{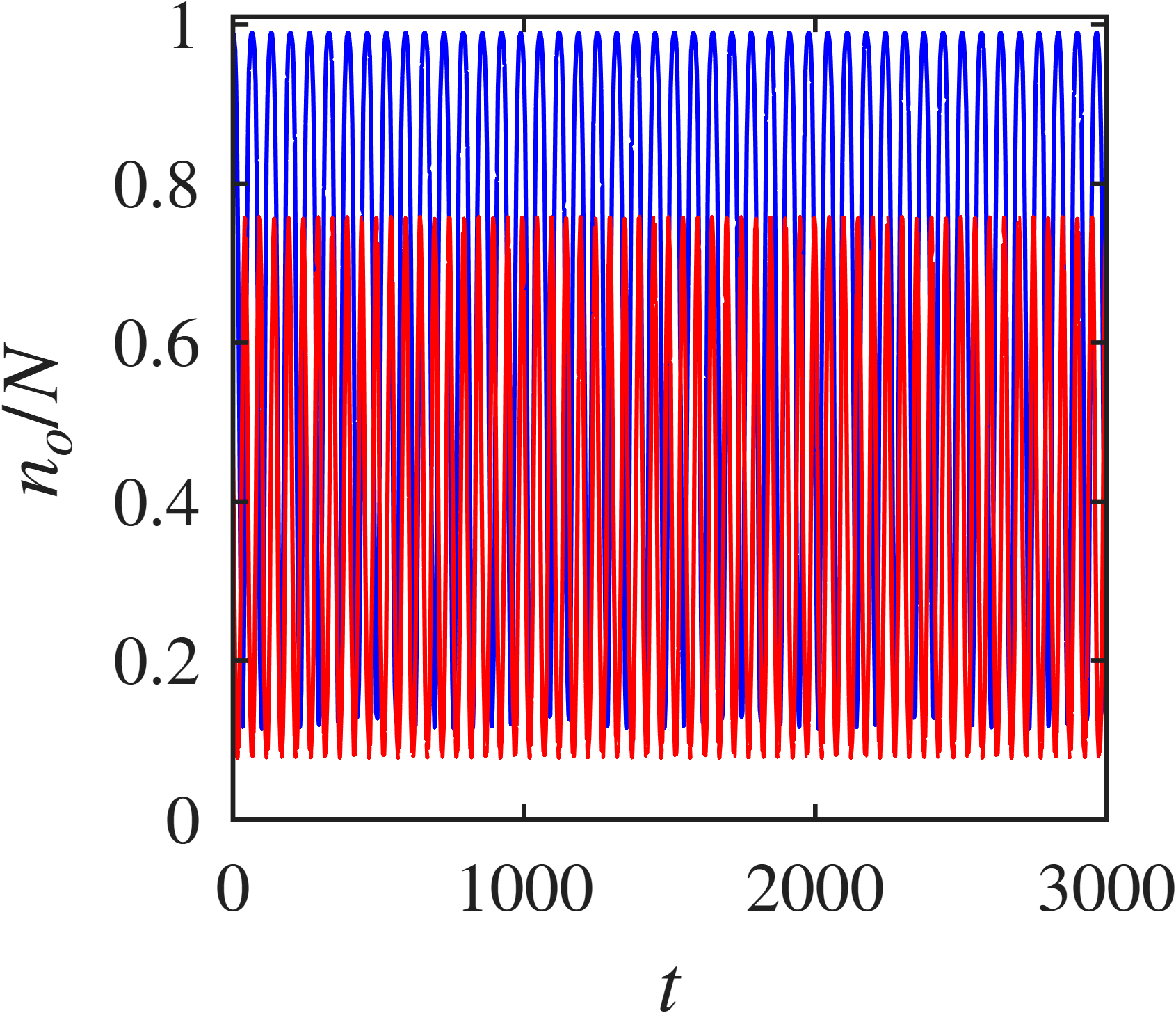}
\put(-3.0,80){(a)}
\end{overpic}
\ \ \
\begin{overpic}[width=4cm]{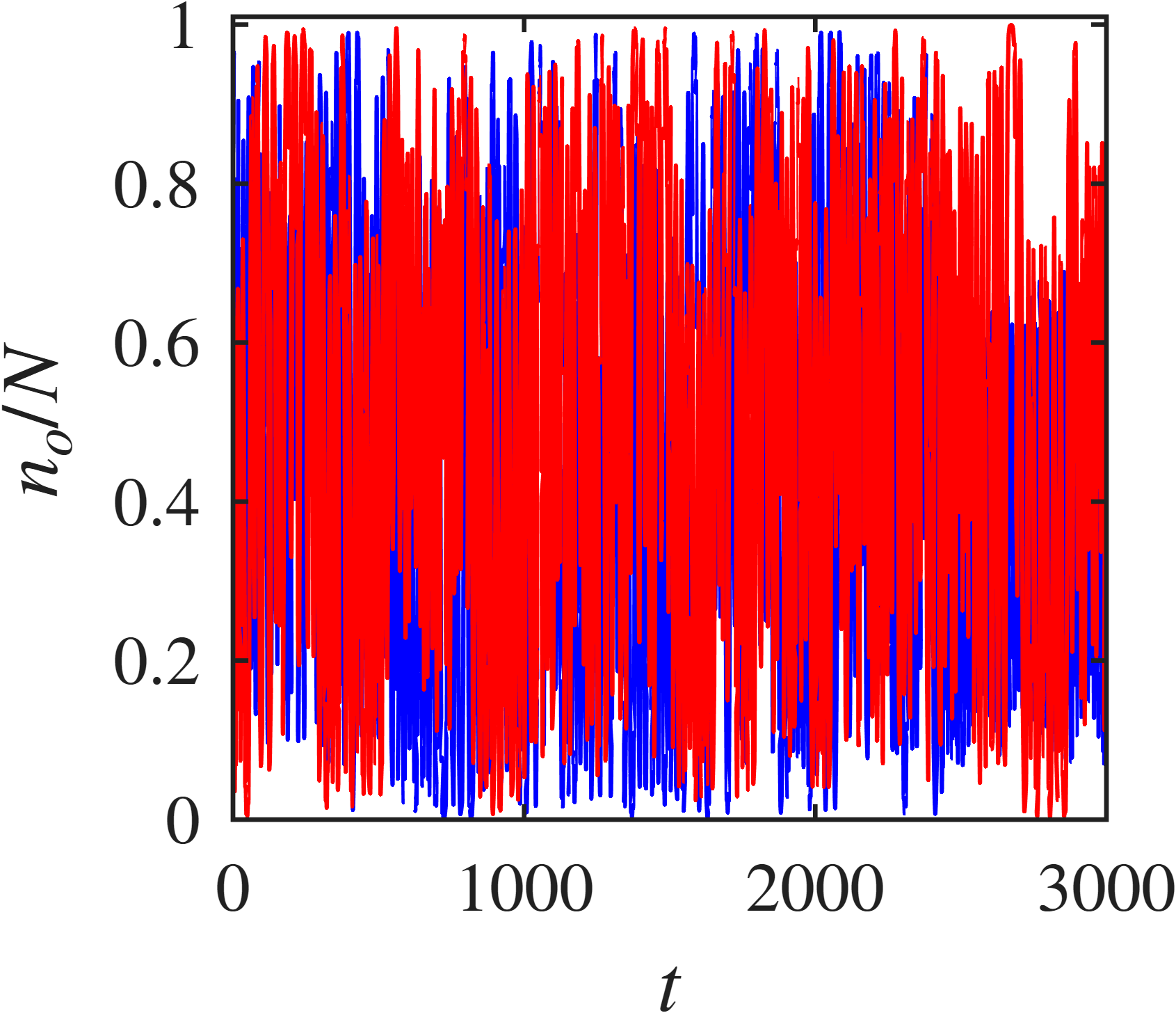}
\put(-3.0,80){(b)}
\end{overpic}
\ \ \
\begin{overpic}[width=4cm]{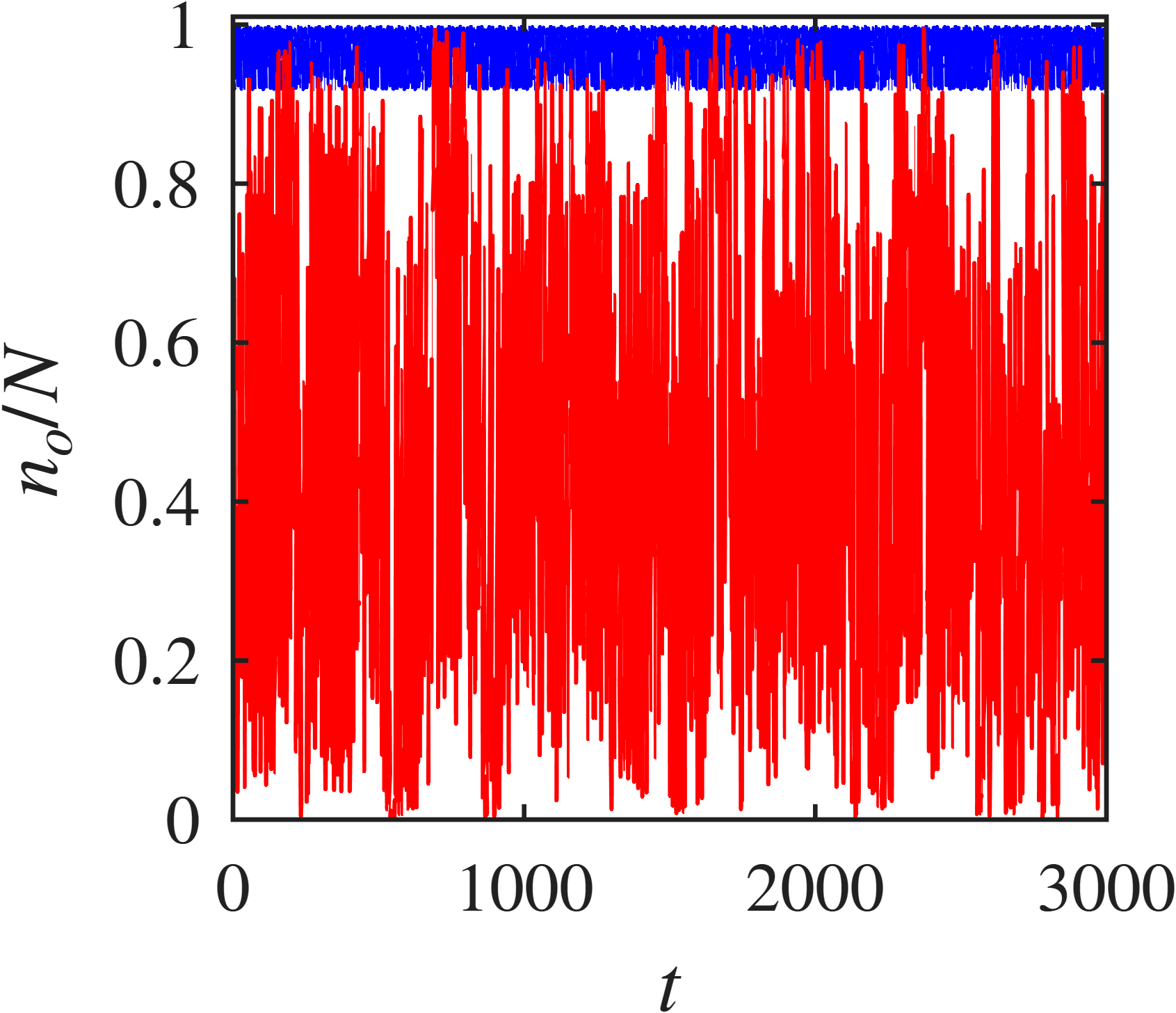}
\put(-3.0,80){(c)}
\end{overpic}
\caption{
{\bf Representative classical trajectories for 3~site chain.}
The trajectories are launched at $n_o/N{=}1$ (blue) and at $n_o/N{=}0.5$ (red). 
Plot of $n_o/N$ versus time, from left to right: 
{\bf (a)}~ 
$u{=}0.5$~(quasi-regular phase-space that contains an unstable SP);
{\bf (b)}~
$u{=}1.5$~(chaotic phase-space that contains an unstable SP);
{\bf (c)}~
$u{=}3.5$~(separated chaotic sea and stable SP island).
}
\label{fTraj3}
%
\ \\
%
\begin{overpic}[width=4cm]{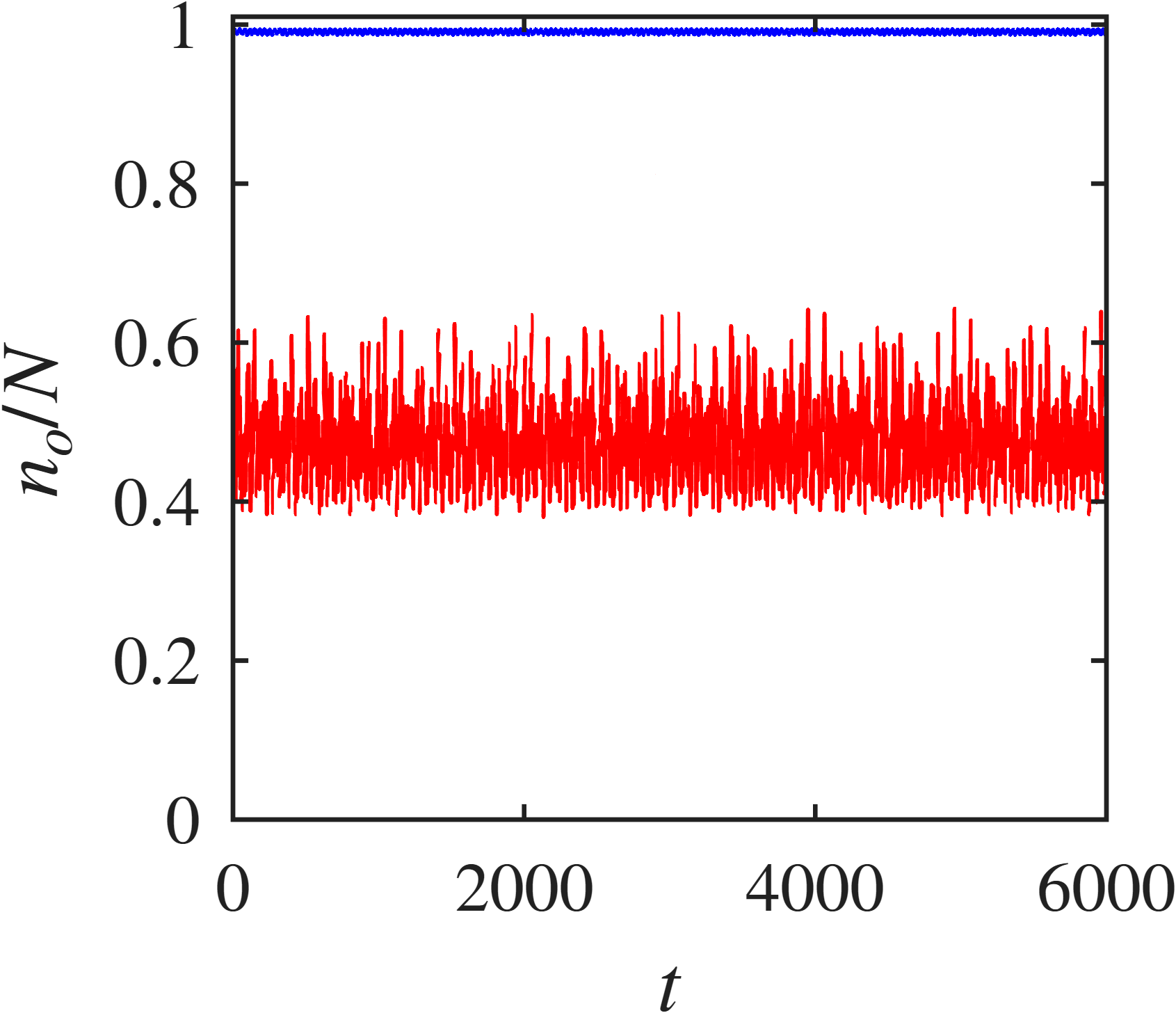}
\put(-3.0,80){(a)}
\end{overpic}
\ \ \
\begin{overpic}[width=4cm]{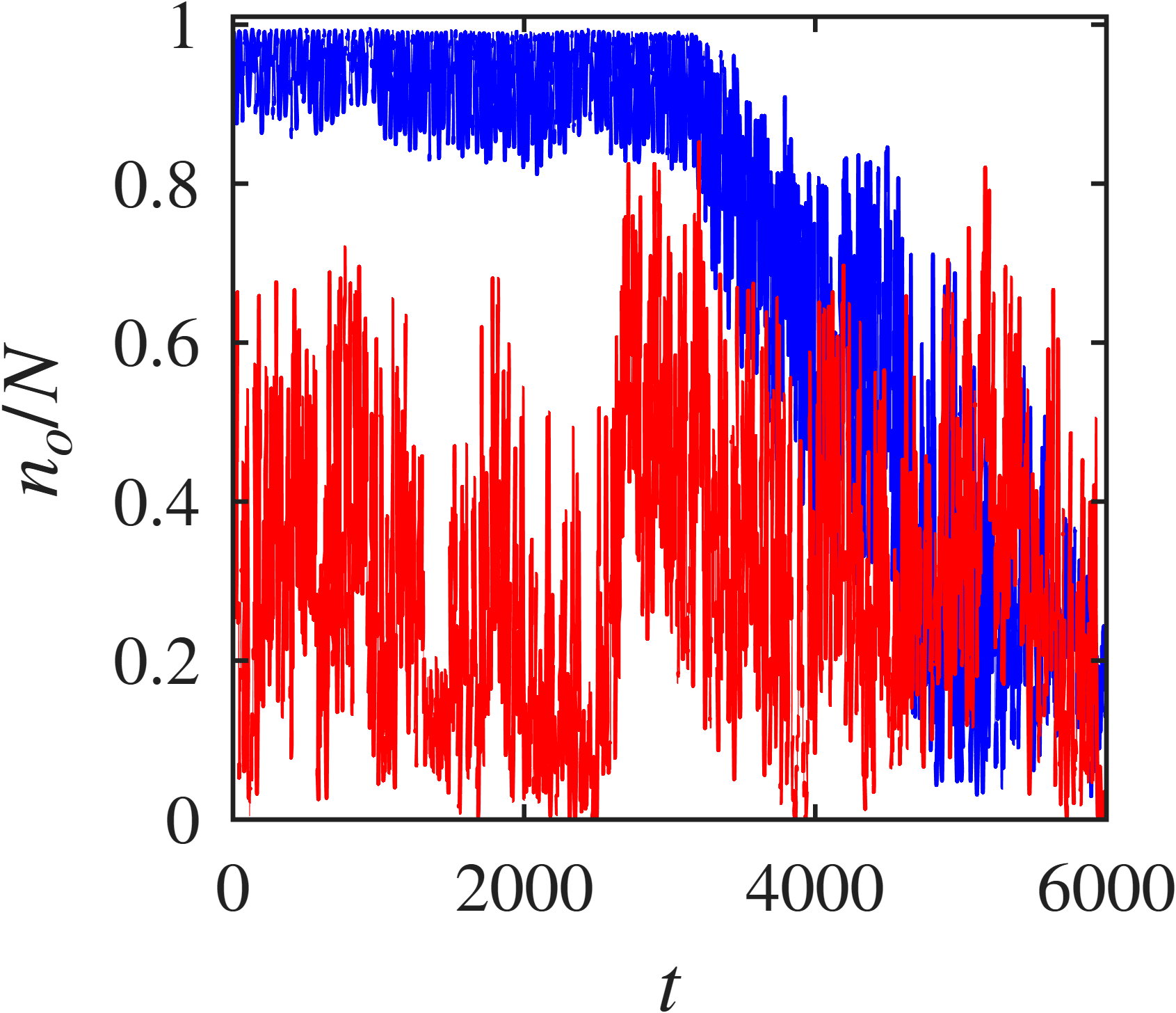}
\put(-3.0,80){(b)}
\end{overpic}
\ \ \
\begin{overpic}[width=4cm]{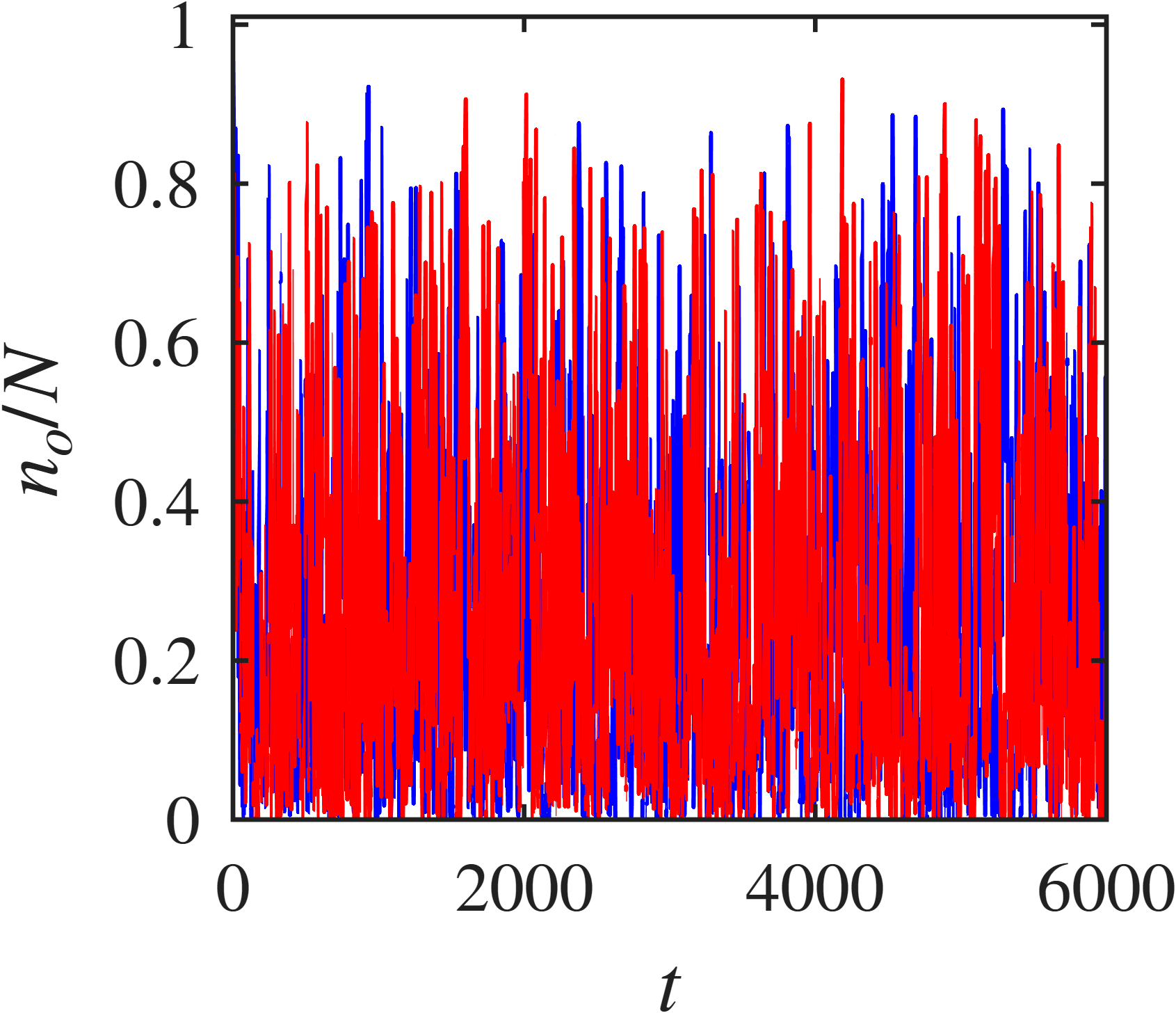}
\put(-3.0,80){(c)}
\end{overpic}
\caption{
{\bf Representative classical trajectories for 5~site chain.}
The trajectories are launched at $n_o/N{=}1$ (blue) and at $n_o/N{=}0.5$ (red).
Plot of $n_o/N$ versus time, from left to right: 
{\bf (a)}~
$u{=}0.5$ (quasi-regular phase-space that contains stable SP);
{\bf (b)}~
$u{=}1.5$ (just after the stability threshold ${u_c \approx 1.3}$);
{\bf (c)}~
$u{=}3.5$ (chaotic phase-space that contains unstable SP). 
}
\label{fTraj5}
%
%
\ \\  
%
%
\centering
\begin{overpic}[width=4cm]{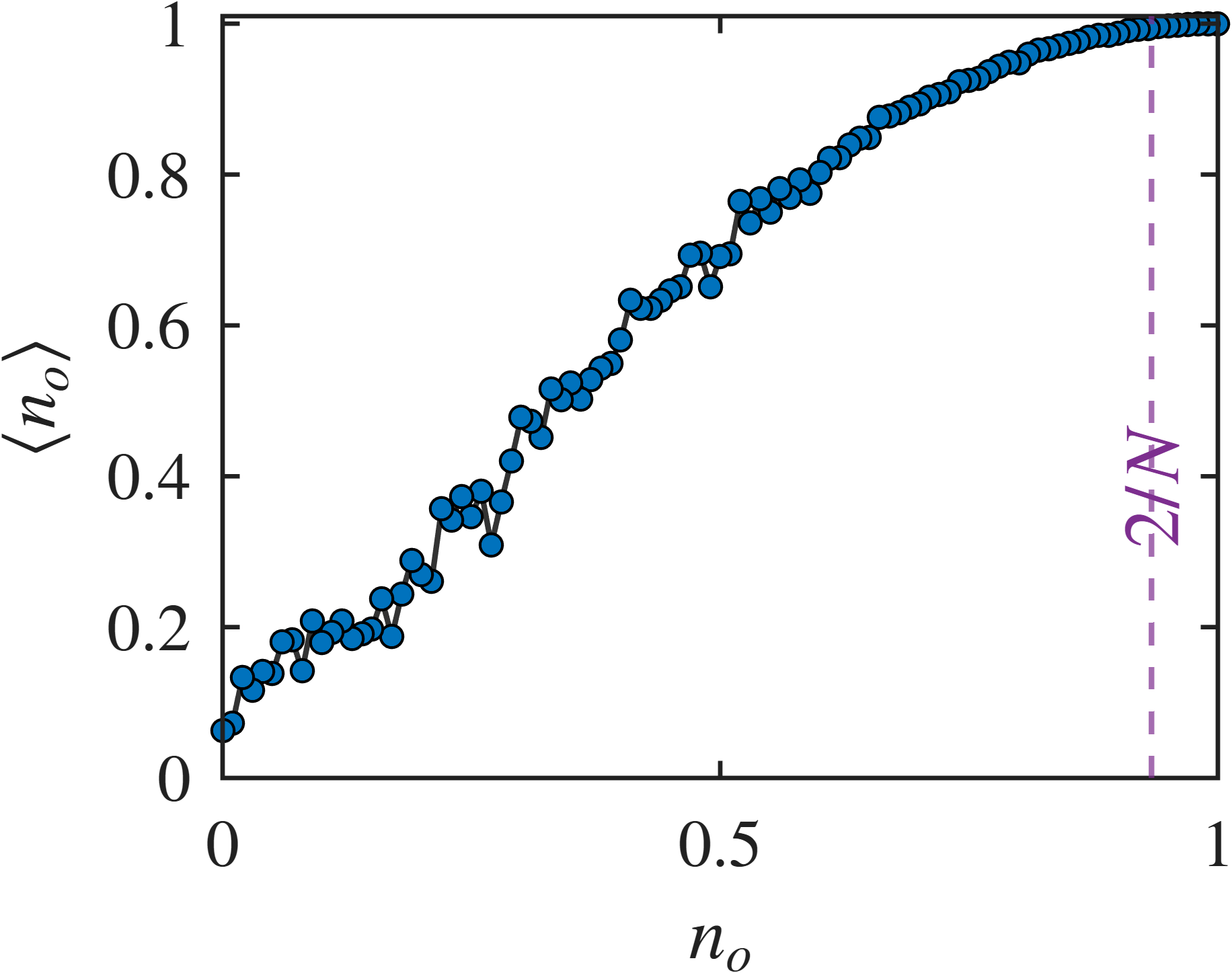} 
\put(-3.5,74){(a)}
\end{overpic}
\ \ \
\begin{overpic}[width=4cm]{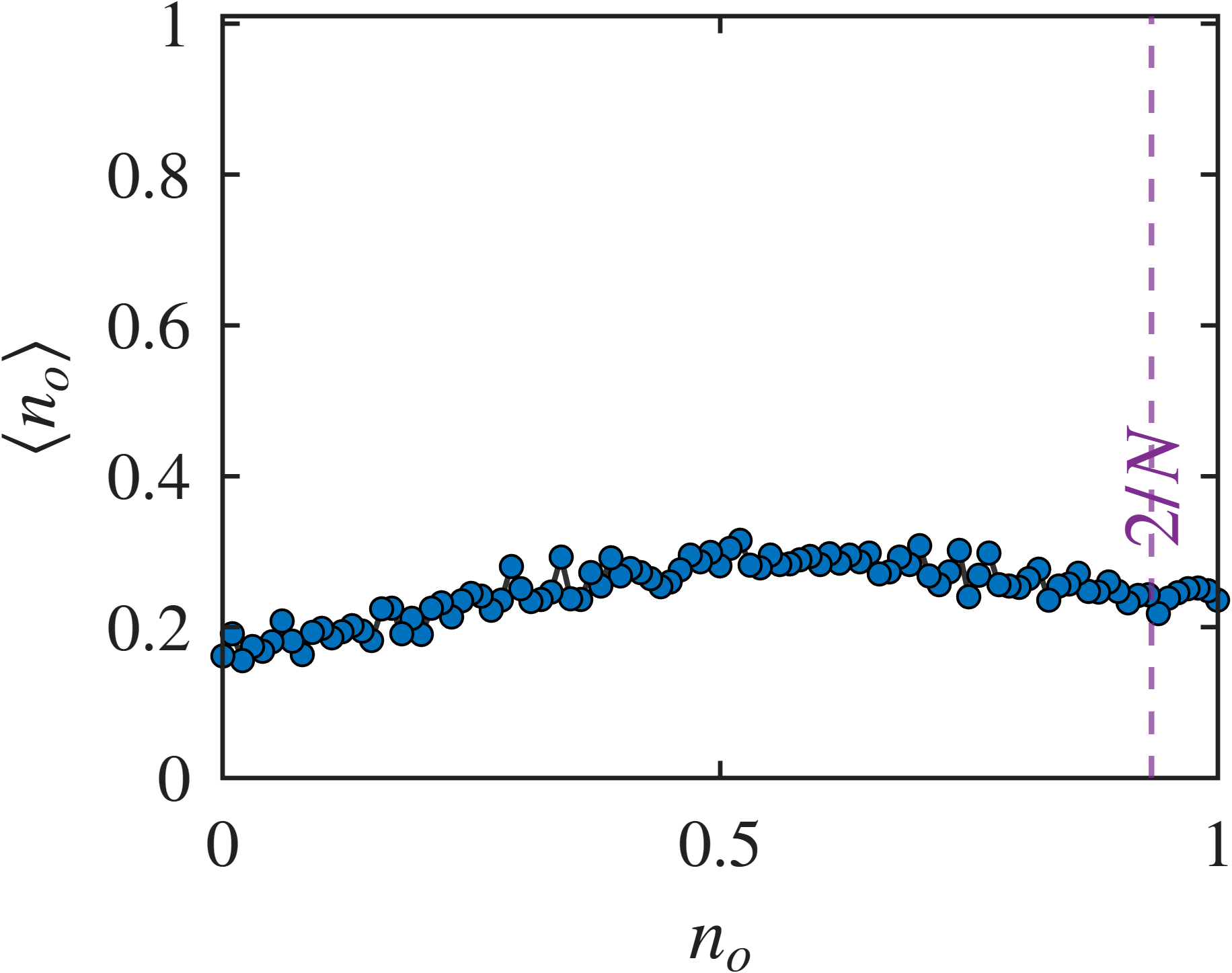}
\put(-3.5,74){(b)}
\end{overpic}
\ \ \ 
\begin{overpic}[width=4cm]{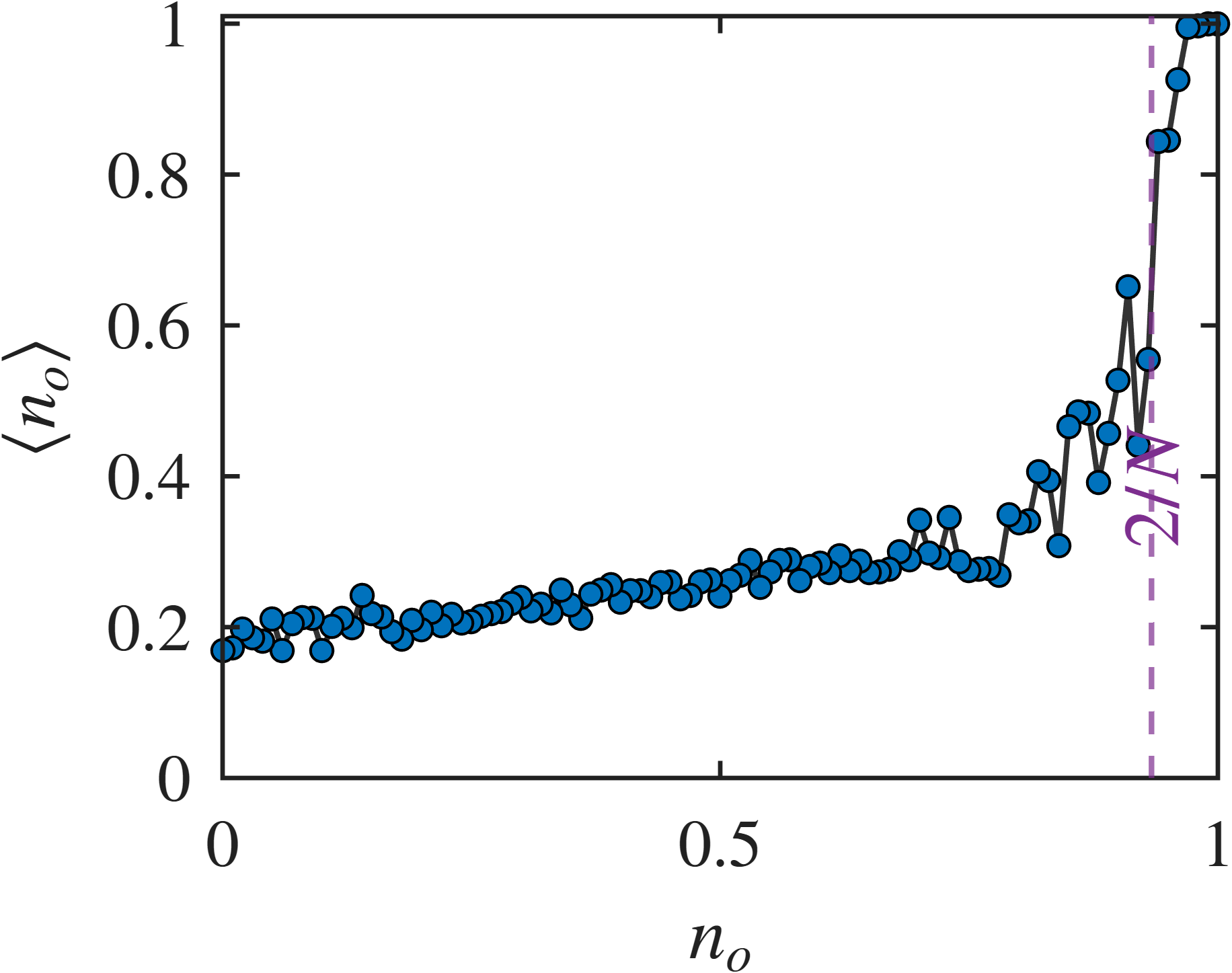}
\put(-3.5,74){(c)}
\end{overpic}
\caption{{\bf The stability island for 5~site chain.}
The size of the stability island is determined via classical simulations. The trajectory is launched at ${n_o}$, and after a long run, the average of $n_o$ is calculated. We consider $L_s{=}5$ chain with: 
{\bf (a)}~$u{=}0.5$ (quasi-regular phase-space that contains stable SP); 
{\bf (b)}~$u{=}3.5$ (unstable SP that is contained in a chaotic sea);
{\bf (c)}~$u{=}7.5$ (stable SP separated from the chaotic sea).
The quantum uncertainty border $2/N$ for ${N{=}30}$ is indicated by the vertical dashed line, implying that the stability island in the second stability range is barely resolved by the quantum eigenstates.}
\label{fTrajErg}
%
\ \\
%
\centering
\includegraphics[width=12cm]{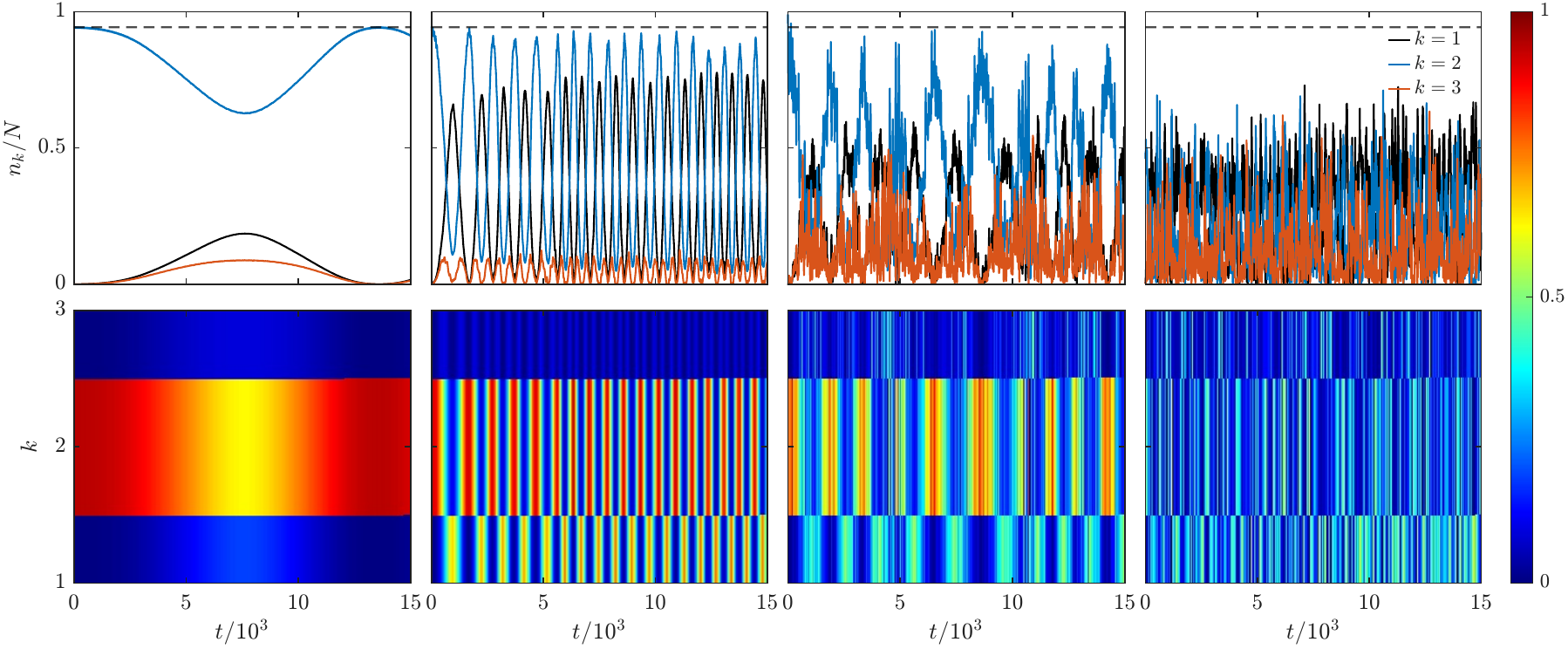}
\caption{{\bf Representative classical trajectories for $51$~site chain.} 
The non-vanishing coordinates $n_k$ are plotted as a function of time. The lower panels show the same plots in image format. 
The interaction is $u=5$.
In the legend, ``$k{=}1,2,3$" means ${k{=}(\pi/L)m}$ with ${m{=}1,2,3}$.  
The dashed horizontal line $n_{\text{SP}}/N{=}0.94$ indicates the location of the $m{=}2$ SP. This value reflects the depletion of the condensate due to the solitonic deformation of the SP wavefunction. The trajectories are launched in the vicinity of this SP with a displaced value for the $n_o$ coordinate. 
From left to right: $n_o/N{=}0.94$  
and $n_o/N{=}0.93$, 
and $n_o/N{=}1.0$ 
and $n_o/N{=}0.45$. 
Note that the other coordinates are rescaled such that $\sum n_k{=}N$.}
\label{fTrajL}
\end{figure*}

\section{Classical trajectories}
\label{sec:traj}

Having diagnosed whether an SP is stable or not, we now zoom out and want to know whether it is located within a stable island or embedded in a chaotic sea. For this purpose, we inspect long trajectories that are launched either in the vicinity of the SP or somewhere else, away from the SP, but at the same energy. Representative trajectories are plotted in \Fig{fTraj3} and \Fig{fTraj5} and \Fig{fTrajL} for ${L_s=3,5,51}$ chains. We discuss the figures in the following paragraphs.  

The $L_s=3$ chain is somewhat non-generic because it features SP instability for arbitrarily small~$u$. Nevertheless, we see that for weak interaction ($u=0.5$) the motion is quasi-regular, and only for stronger interaction ($u=1.5$) it becomes chaotic. This is clear by inspection of \Fig{fTraj3}ab, but also can be quantified via Fourier analysis (power spectrum of the motion) or via Lyapunov analysis (see later section). In the chaotic regime, we observe ergodicity, meaning the loss of ``memory" of the initial conditions. Strangely enough, as we further increase the interaction ($u=3.5$), see \Fig{fTraj3}c, a stability island is born, well separated from the chaotic sea. This is inverse to the DS-to-instability transition.   

The $L_s=5$ chain is rather generic. For small interaction ($u=0.5$), the motion is quasi-regular, and the SP is dynamically stable, see \Fig{fTraj5}a. For strong interaction ($u=3.5$), the motion is chaotic, the SP is dynamically unstable, and one observes ergodicity, see \Fig{fTraj5}c. Nevertheless, this system has more than 2 dof. Therefore, the topology of phase-space does not allow the existence of a strictly isolated island. In principle, one expects Arnold diffusion. However, such a type of escape is very slow (exponentially large with respect to the perturbation) and therefore cannot be observed in practice. What we can do is to lower the interaction to a value that is still larger but very close to the critical value ($u_c=1.3$). Indeed, for $u=1.5$, we see in \Fig{fTraj5}b that the trajectory launched at the SP eventually escapes, indicating ergodicity.      

In order to gain a more quantitative perspective on the route towards ergodicity, we plot in \Fig{fTrajErg} the long-time expectation value $\braket{n_o}$ versus the initial $n_o(t{=}0)$. We see that the lack of ergodicity is reflected in the quasi-regular regime (panel~a). In the ergodic regime (panel~b), the expectation value $\braket{n_o}$ depends only weakly on the initial $n_o$. In contrast, we see how the existence of a stability island is reflected in the DS regime (panel~c). The crossover (as a function of $u$) is affected by the practical limitation on the time of the simulation. As already pointed out, the time for Arnold-type ergodication is exponentially large, and therefore would not be observed even if Nature were classical. For a quantum system, such long time scales are totally irrelevant because they exceed the Heisenberg time.     

Representative trajectories for a long chain ($L_s=51$) are presented in \Fig{fTrajL}. Here, the SP is displaced. If the trajectory is launched very close to the SP, the motion is bounded and involves essentially a single mode. If the launching point is further away from the SP, we observe a crossover from quasi-integrability to chaos. An additional observation is that even in the latter case, chaos in a large chain, the motion actually involves a rather small number of modes.

\begin{figure*}
\centering
\begin{overpic}[width=5cm]{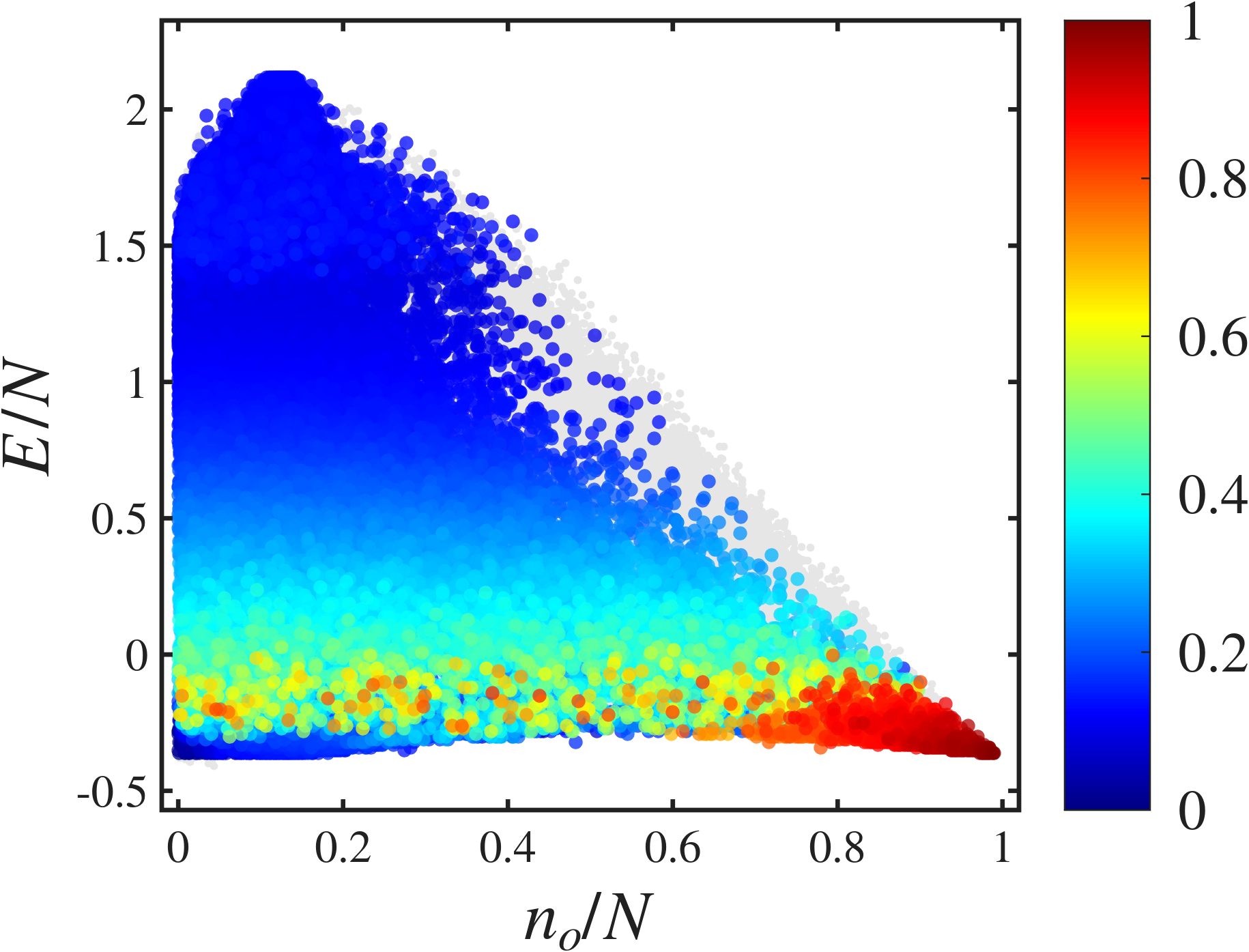} 
\put(-2.5,70){(a)}
\put(83,80){$\braket{n_{o}}$}
\end{overpic}
\ 
\begin{overpic}[width=5cm]{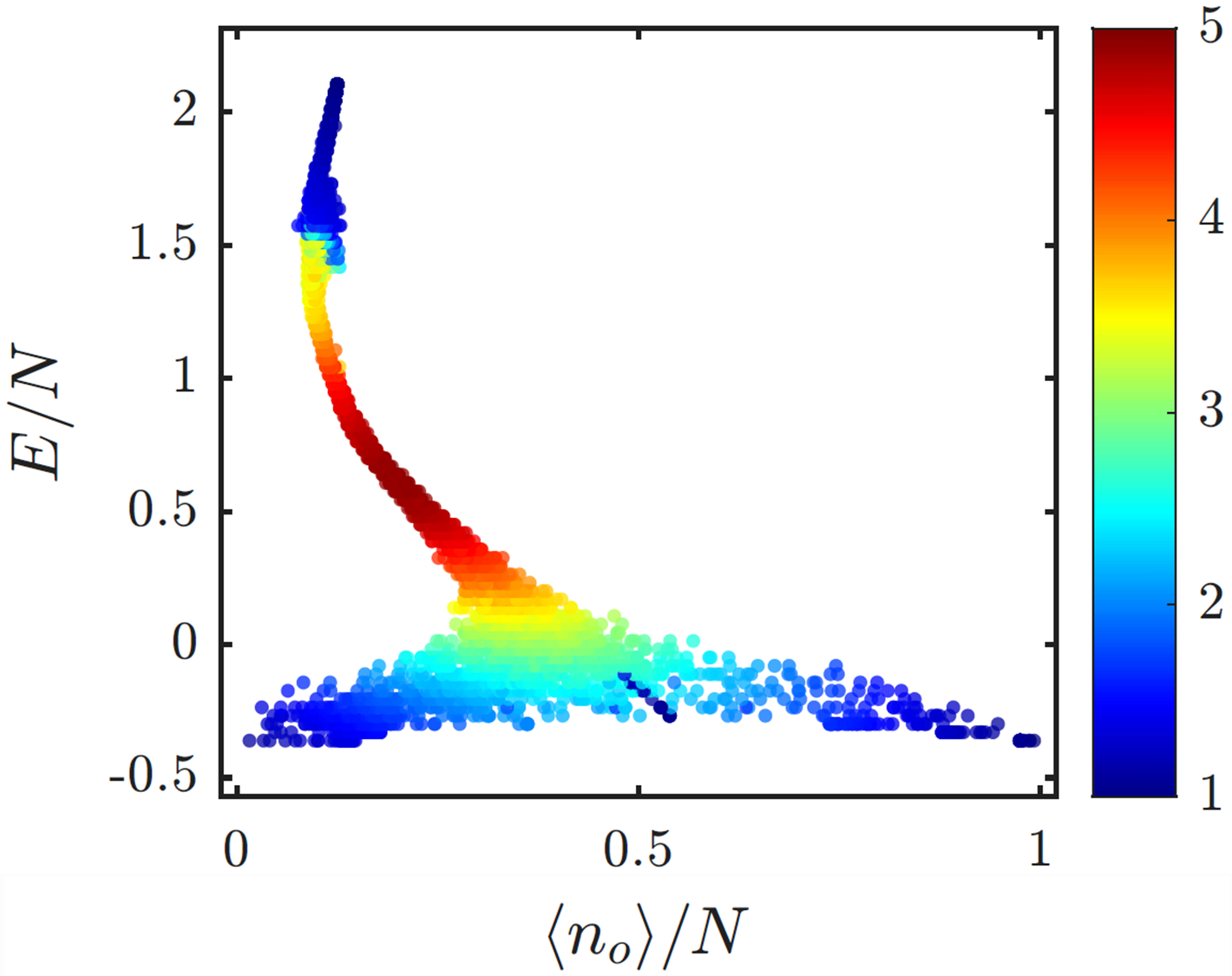}
\put(-2.5,70){(b)}
\put(86,80){$1/\mathcal{S}$}
\end{overpic}
\ 
\begin{overpic}[width=5cm]{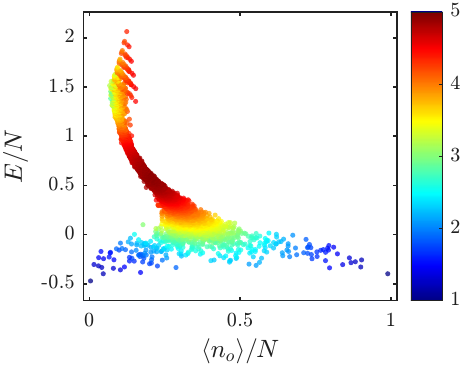}
\put(-2.5,70){(c)}
\put(86,80){$1/\mathcal{S}$}
\end{overpic}
\includegraphics[width=5cm]{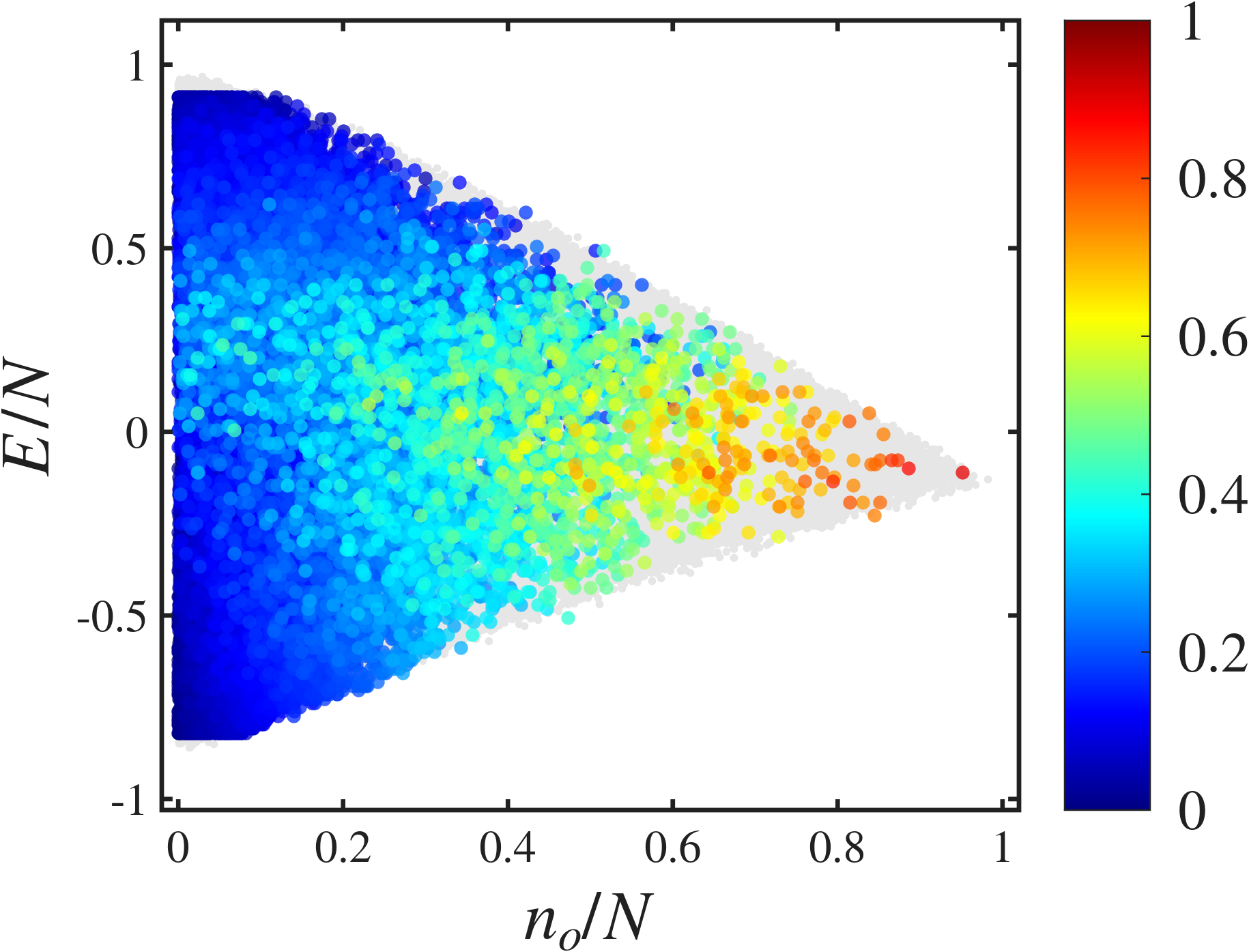}
\
\includegraphics[width=5cm]{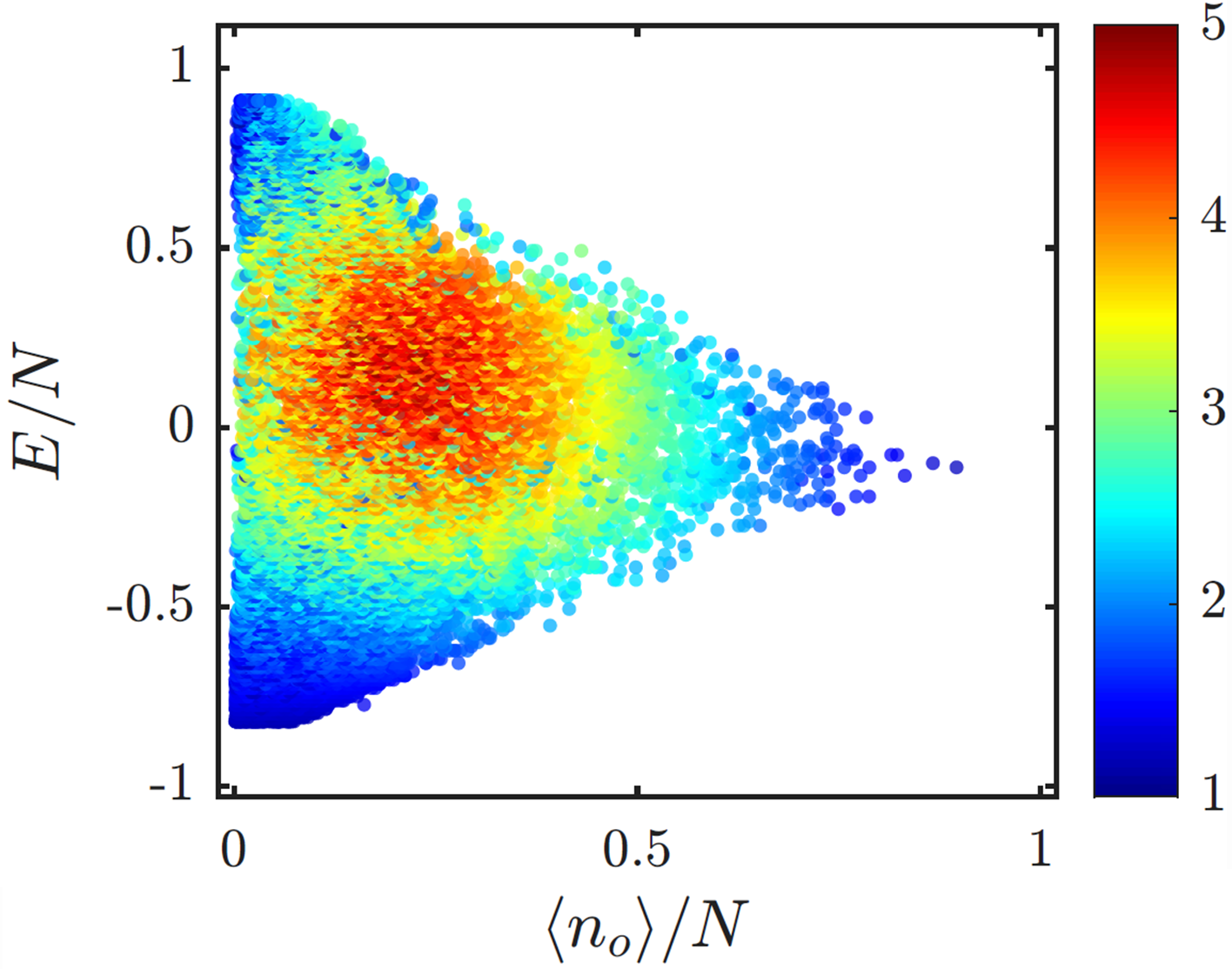}
\
\includegraphics[width=5cm]{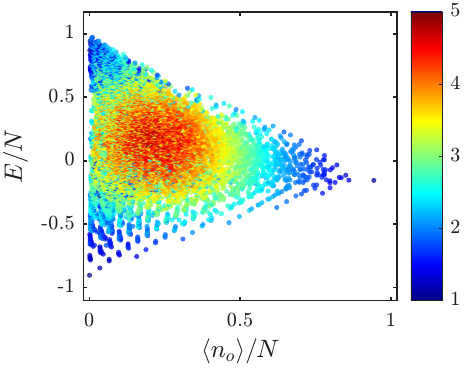}

\includegraphics[width=5cm]{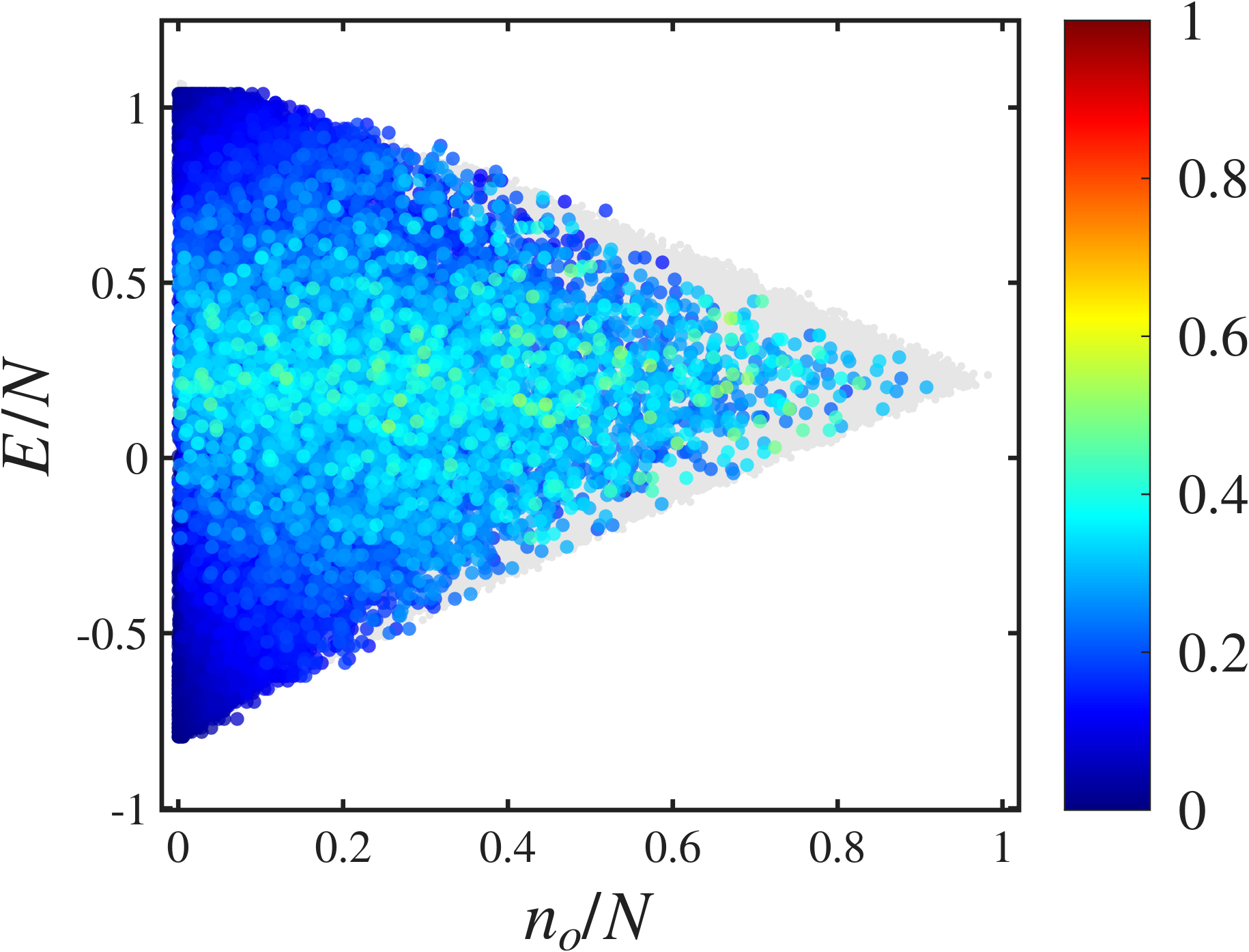} 
\
\includegraphics[width=5cm]{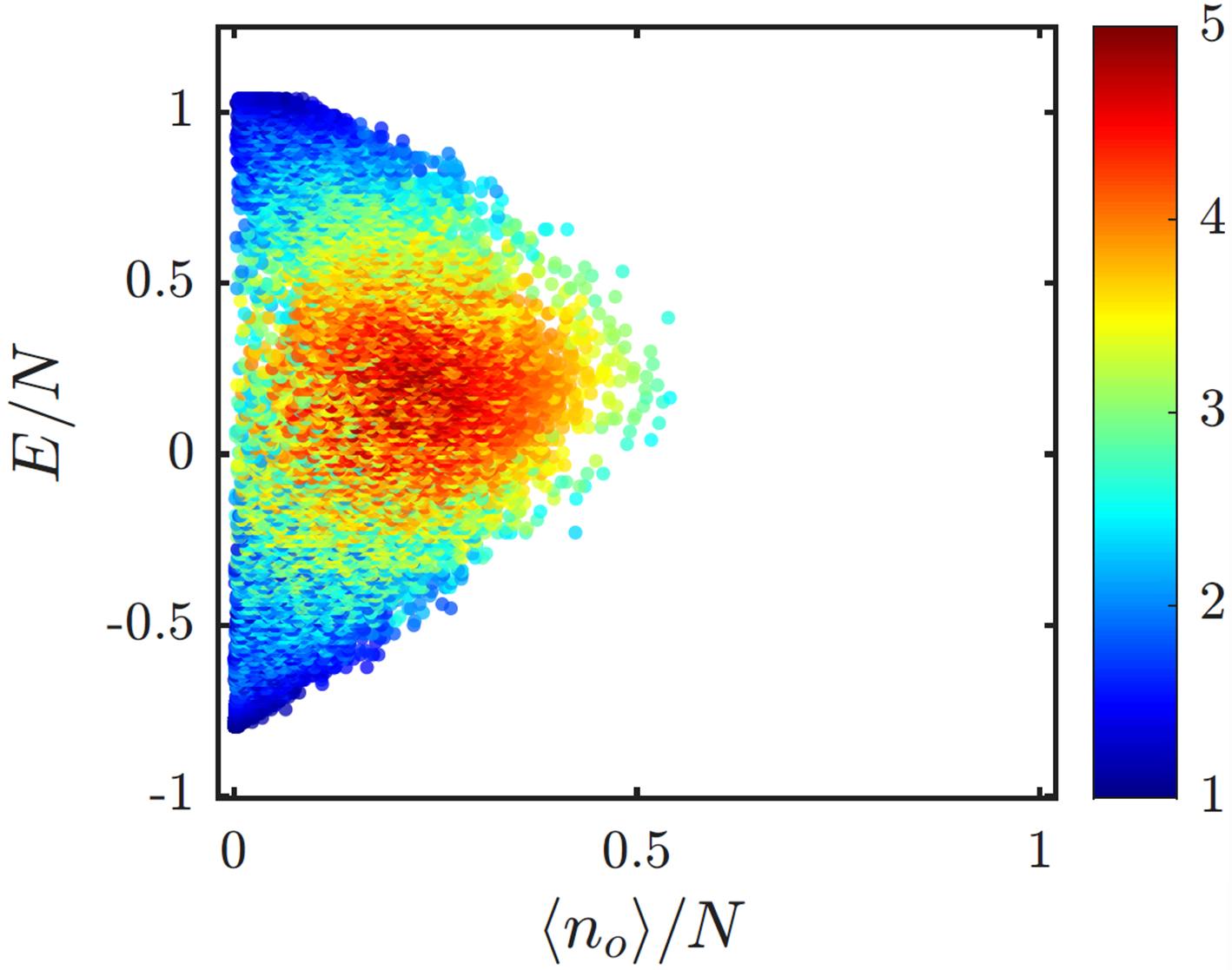} 
\
\includegraphics[width=5cm]{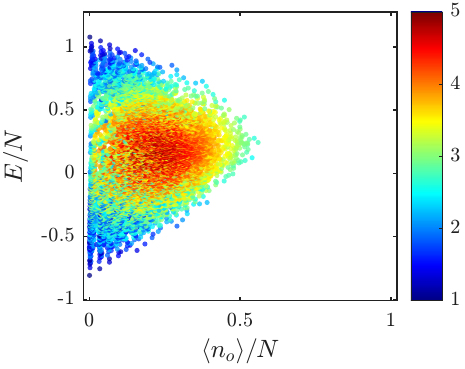} 

\includegraphics[width=5cm]{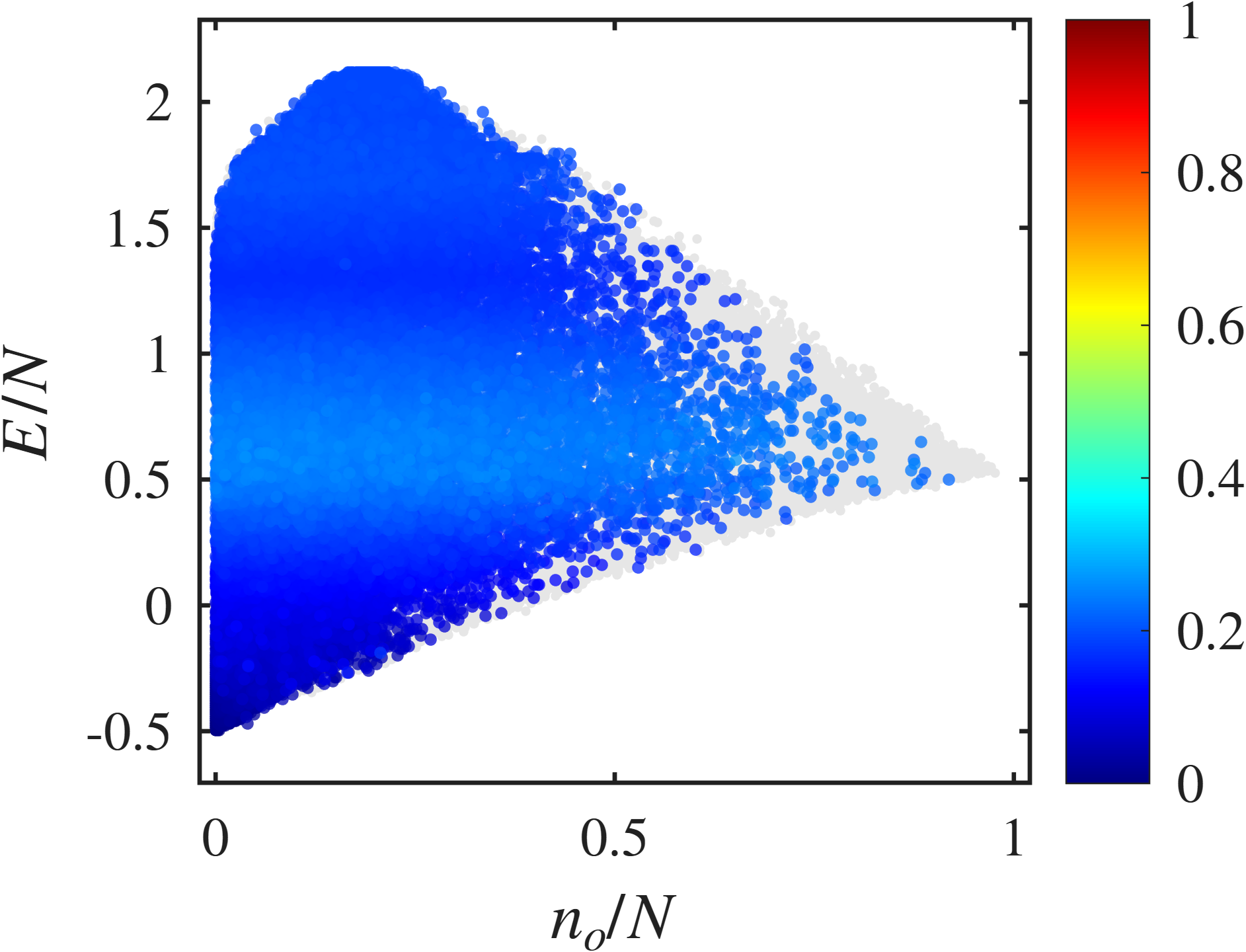} 
\
\includegraphics[width=5cm]{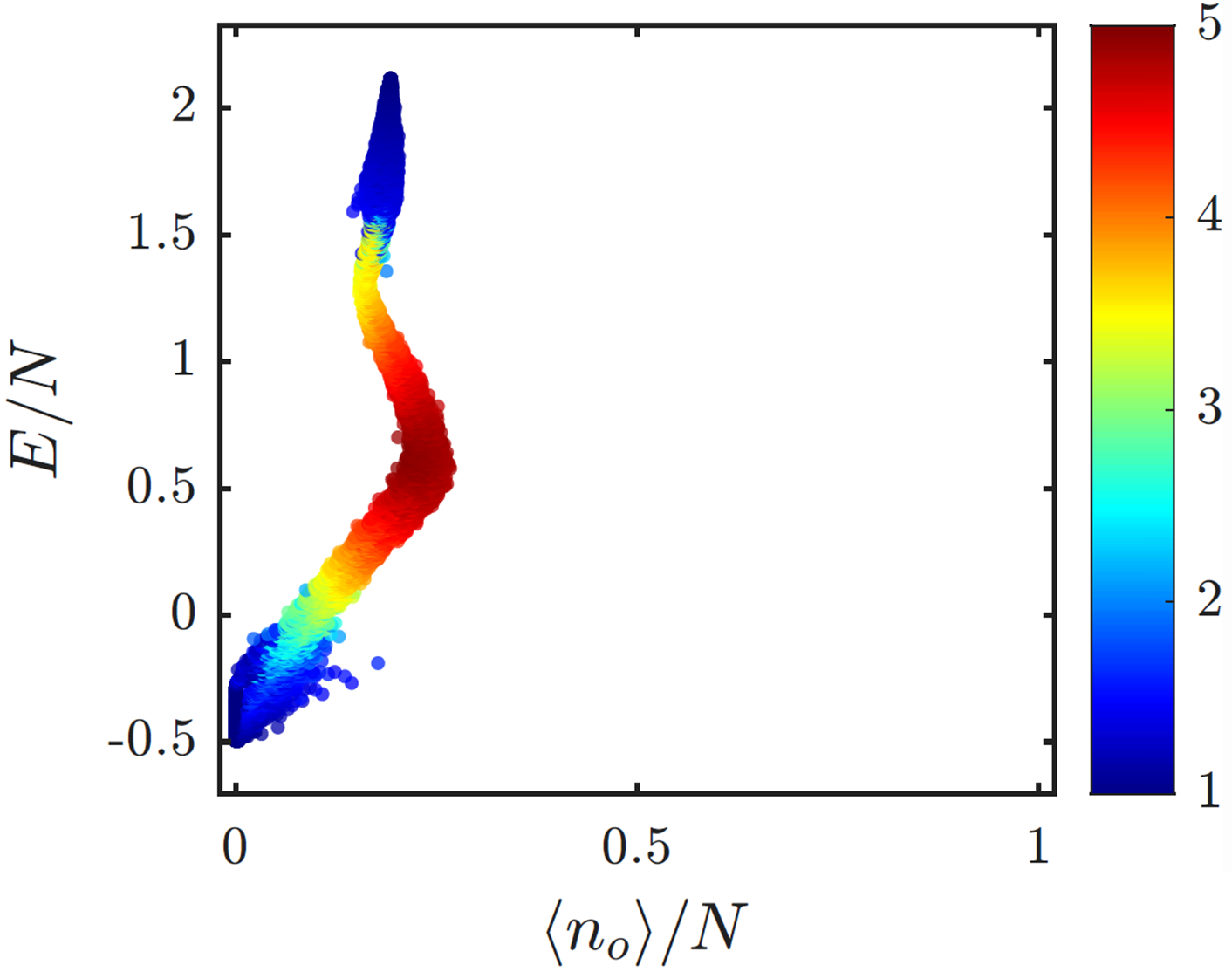}
\
\includegraphics[width=5cm]{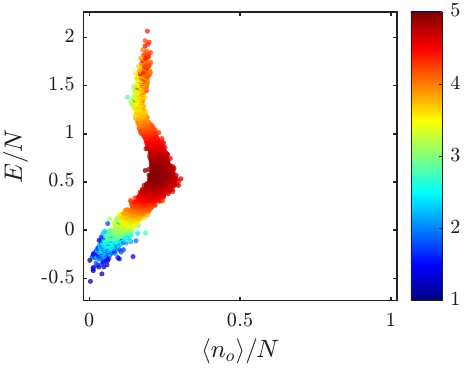} 

\caption{
{\bf Tomographic spectra for a 5~site Ring.} 
The columns are for: 
(a) Classical ${(n_{o},E)}$ energy-landscape. 
The plots display a uniform distribution of phase-space points $(E, n_{o})$. Some of the points are used as initial conditions for a long trajectory of duration $t{=}2500$, and color-coded according to the temporal average $\braket{n_{o}}$.  
(b) Classical $(\braket{n_{o}},E)$ spectrum, 
color-coded by inverse classical purity $1/\mathcal{S}$. 
(c) Quantum $(\braket{n_{o}},E)$ spectrum for $N{=}30$ particles,   
color coded by inverse purity $1/\mathcal{S}$. 
Only $P{=}0$ eigenstates are included. 
The rows from top to bottom are for: 
${(u{=}4, \phi{=}1.1 \pi)}$~demonstrating ES of the SP;
${(u{=}1, \phi{=}2.1 \pi)}$~demonstrating DS of the SP;
${(u{=}1, \phi{=}2.7 \pi)}$~where the SP becomes unstable;
${(u{=}4, \phi{=}2.7 \pi)}$~where, due to the increased chaos, we get ergodicity, similar to as seen in the upper part of the first-row spectrum.}

\vspace*{4cm}

\label{fTomoR}
\end{figure*}

\begin{figure*}
\centering
\begin{overpic}[width=5cm]{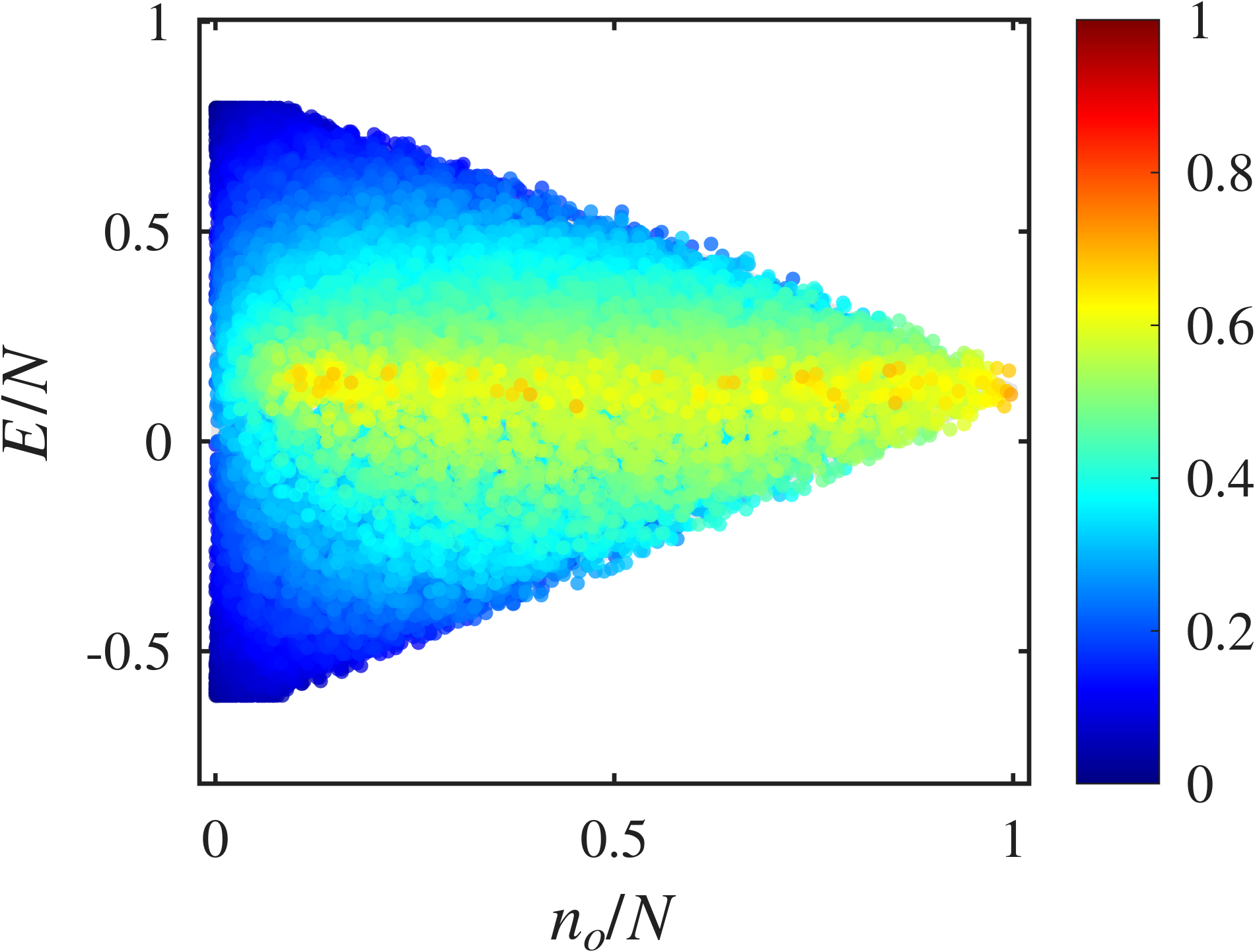} 
\put(-2.5,70){(a)}
\put(82,78){$\braket{n_{o}}$}
\end{overpic}
\
\begin{overpic}[width=5cm]{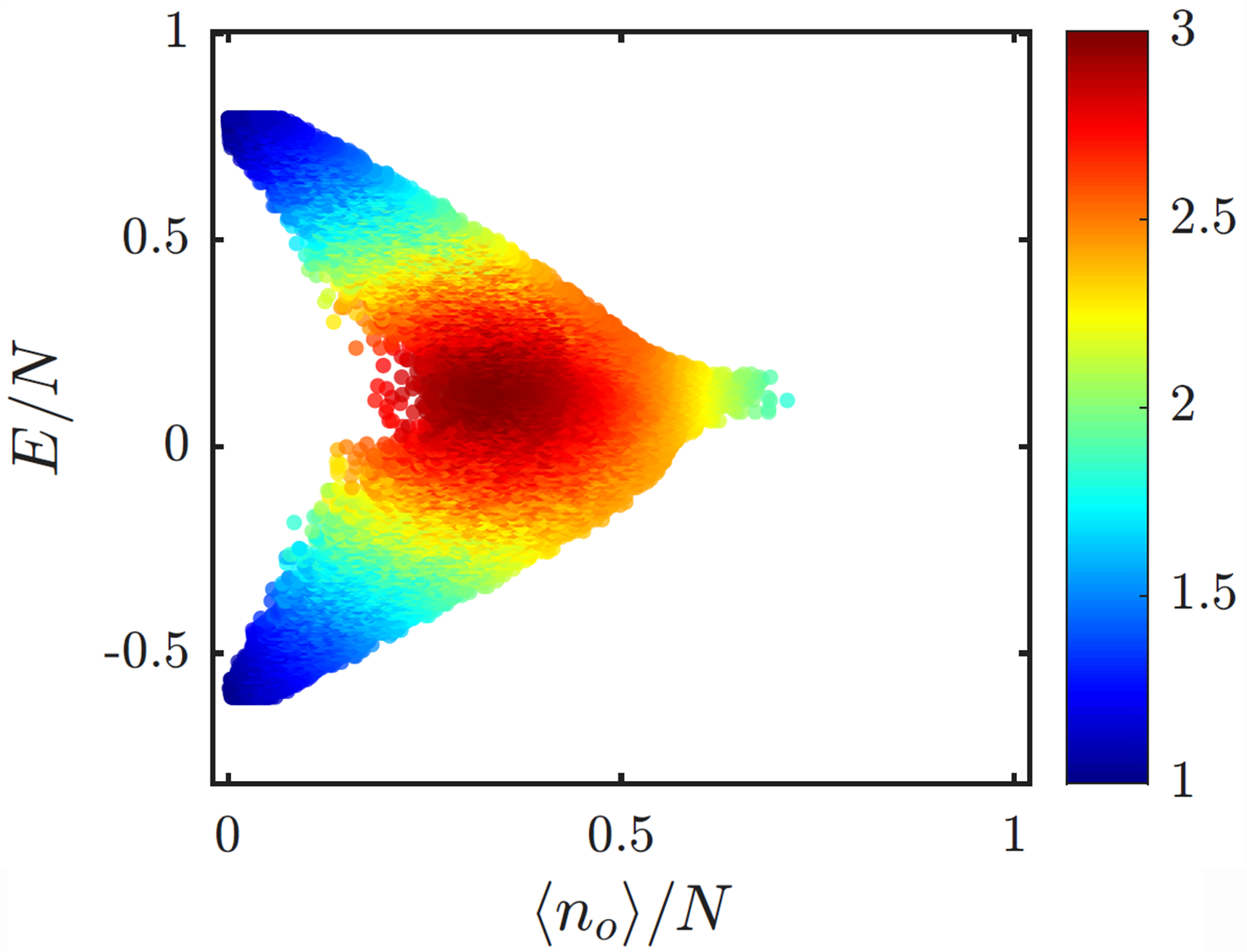}
\put(-2.5,70){(b)}
\put(83,77){$1/\mathcal{S}$}
\end{overpic}
\
\begin{overpic}[width=5cm]{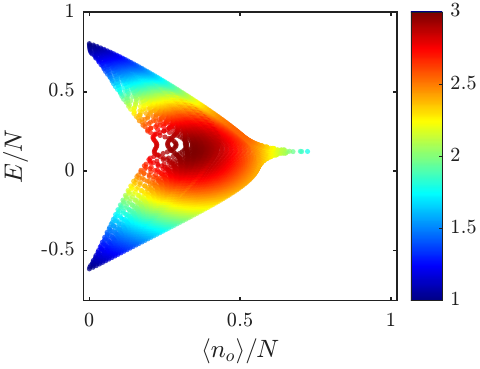}
\put(-2.5,70){(c)}
\put(83,77){$1/\mathcal{S}$}
\end{overpic}

\includegraphics[width=5cm]{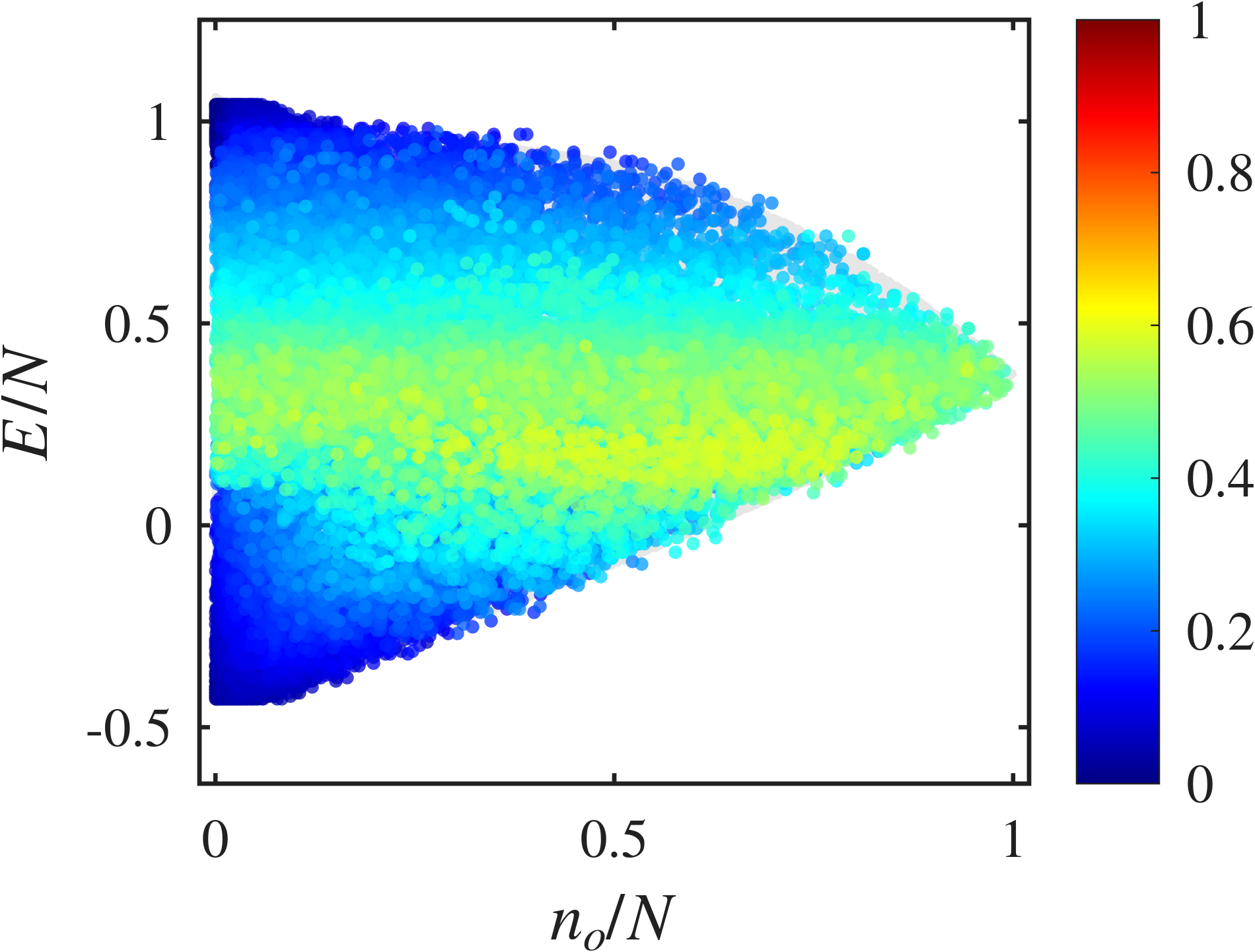}
\
\includegraphics[width=5cm]{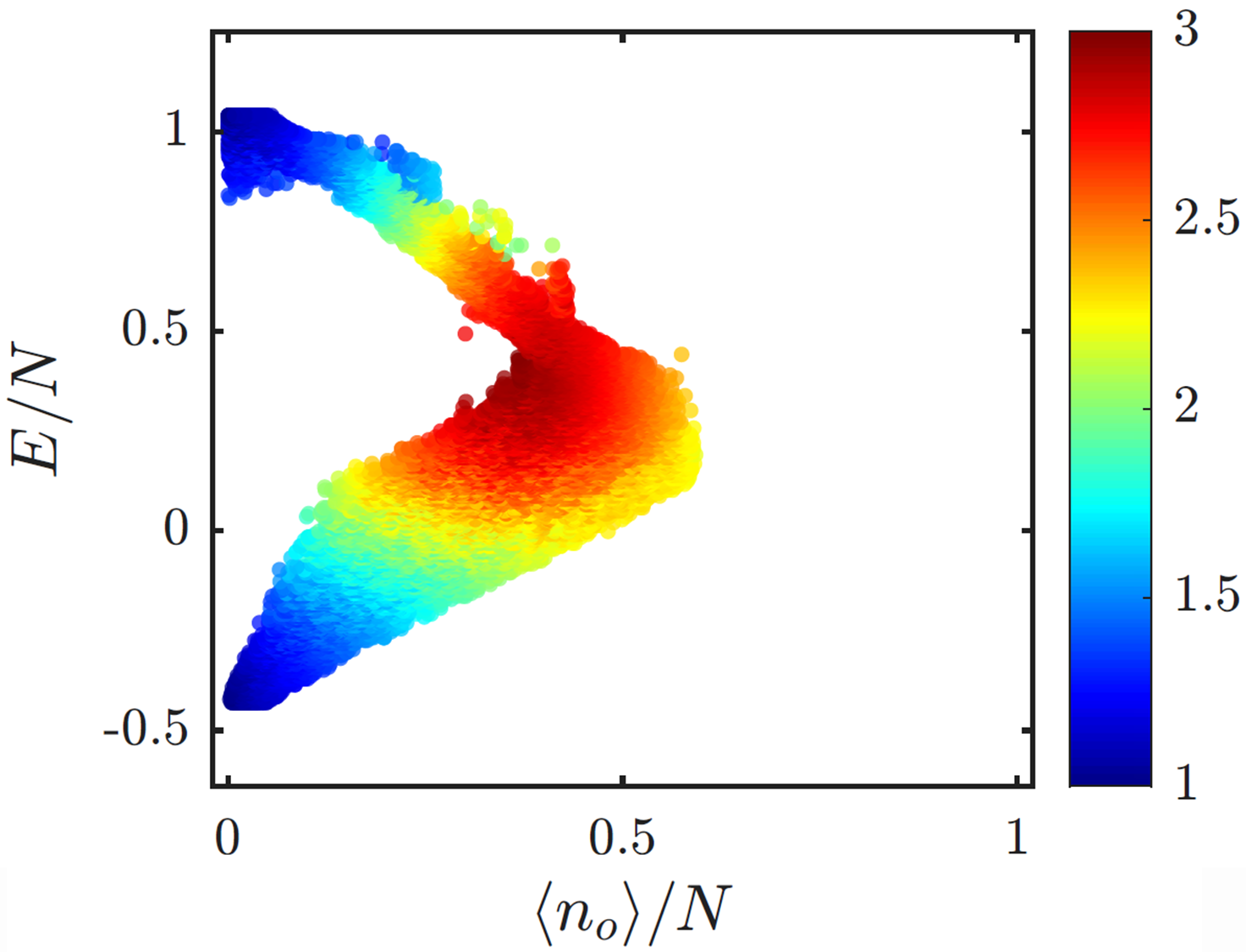} 
\
\includegraphics[width=5cm]{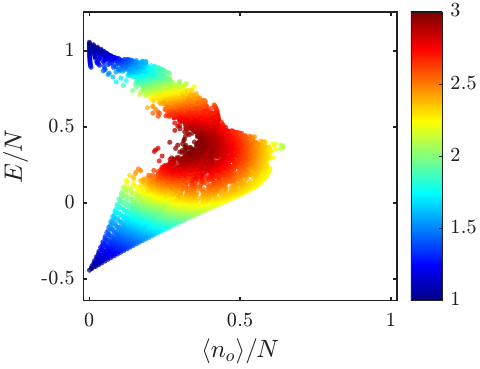} 

\includegraphics[width=5cm]{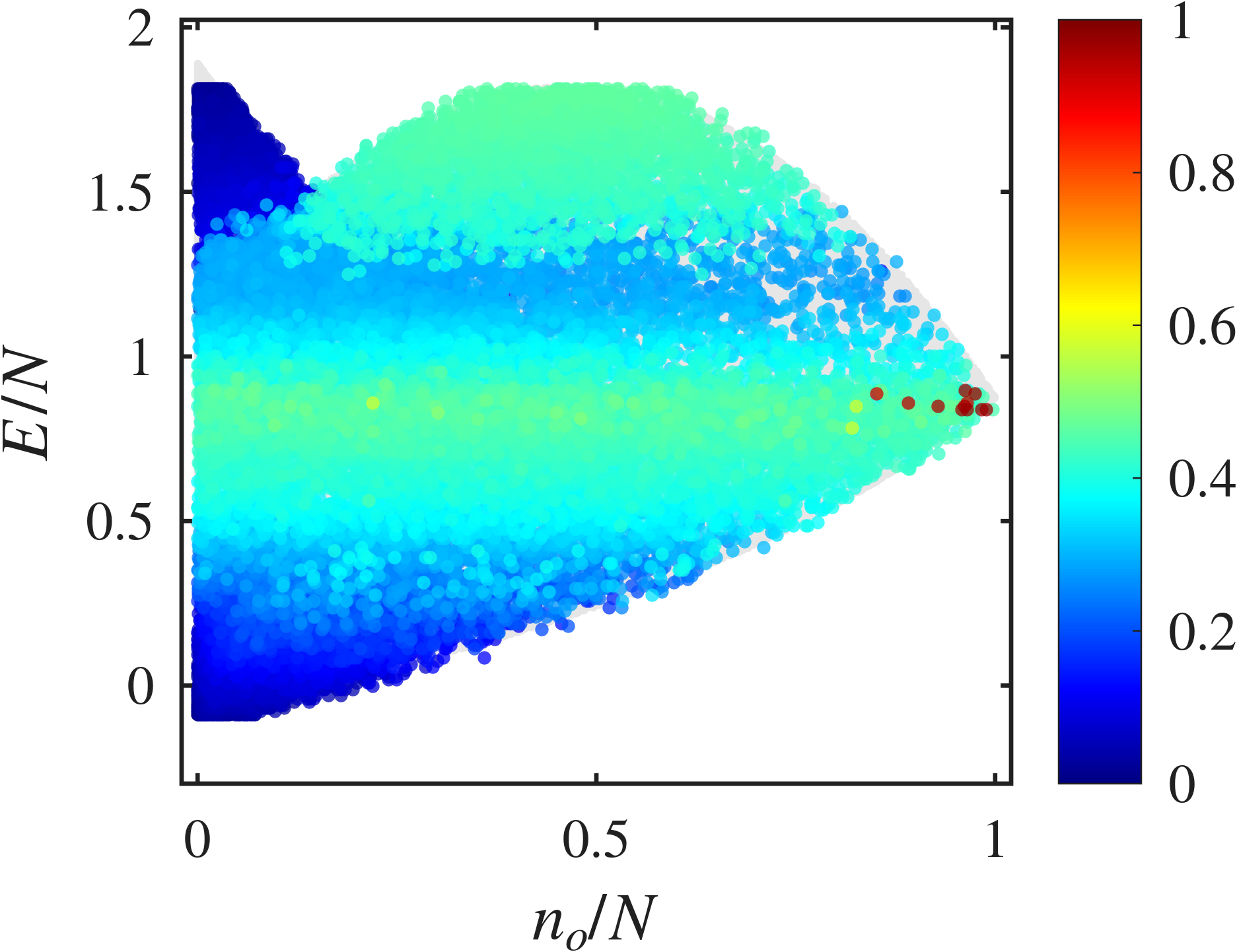}
\
\includegraphics[width=5cm]{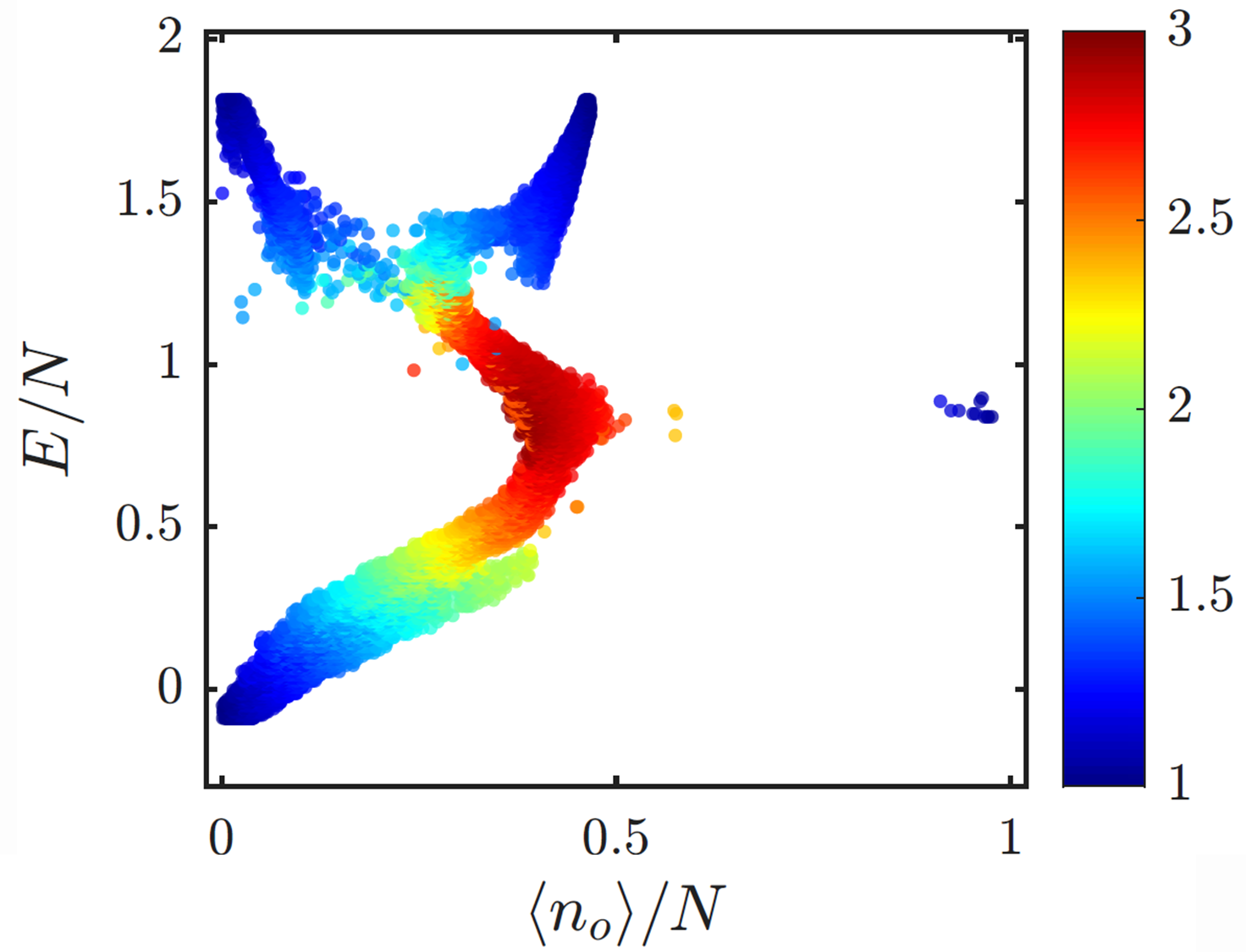} 
\
\includegraphics[width=5cm]{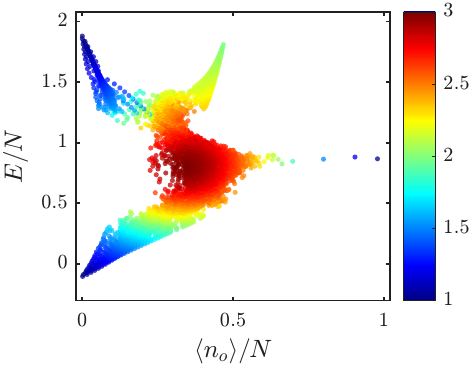} 

\caption{{\bf Tomographic spectra for a 3~site Chain.} 
The columns are arranged as in \Fig{fTomoR}: (a) Energy landscape, (b) Classical spectrum, and (c) Quantum
spectrum.
In column~(c), the spectrum is for $N{=}150$ particles, and odd-parity states are excluded.
The rows from top to bottom are for:   
$u{=}0.5$~(quasi-regular phase-space that contains an unstable SP),
$u{=}1.5$~(chaotic phase-space that contains an unstable SP), 
$u{=}3.5$~(separated chaotic sea and stable SP island).
}

\label{fTomoC3}
\end{figure*}

\begin{figure*}
\centering

\begin{overpic}[width=5cm]{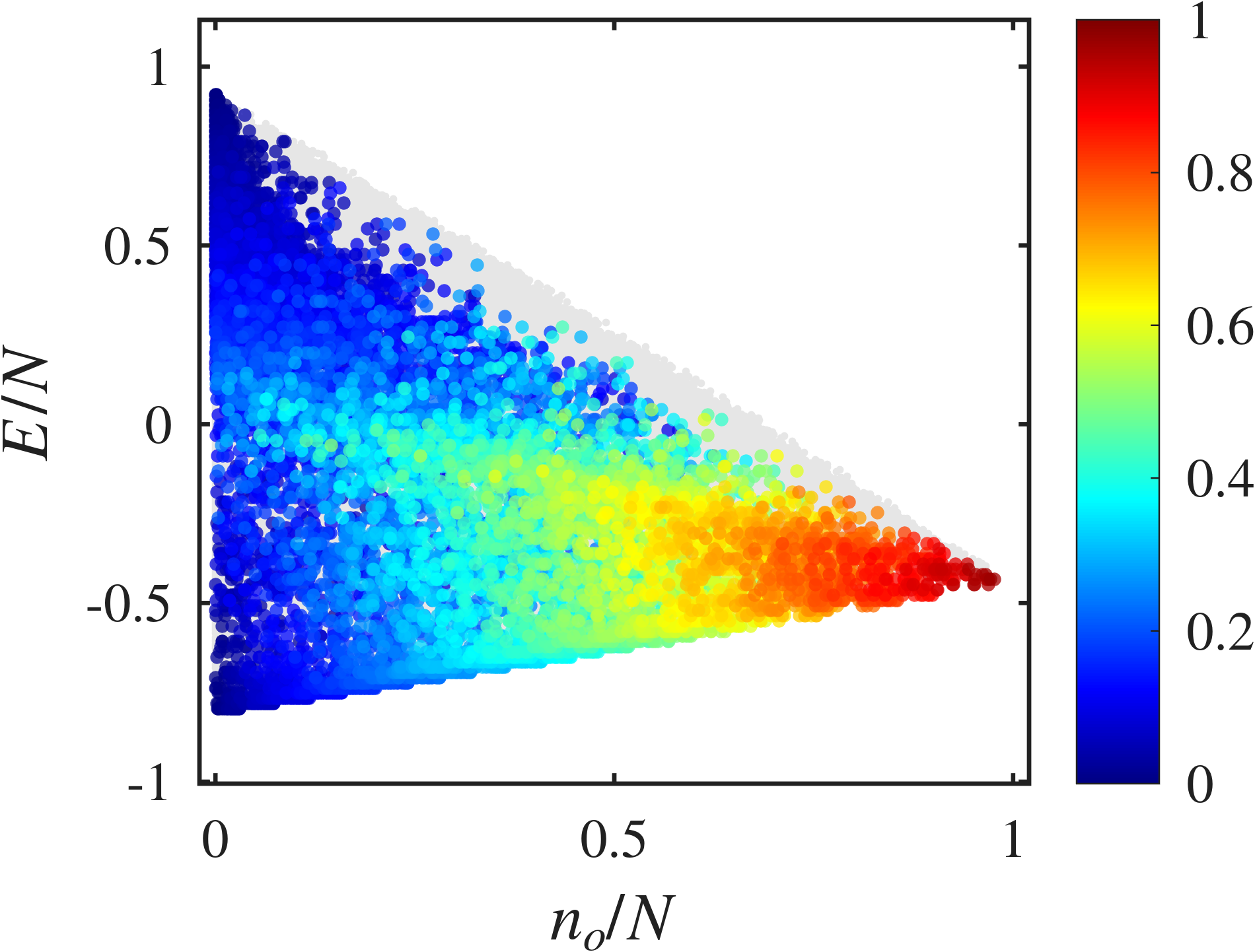}
\put(-2.5,70){(a)}
\put(83,78){$\braket{n_{o}}$}
\end{overpic}
\
\begin{overpic}[width=5cm]{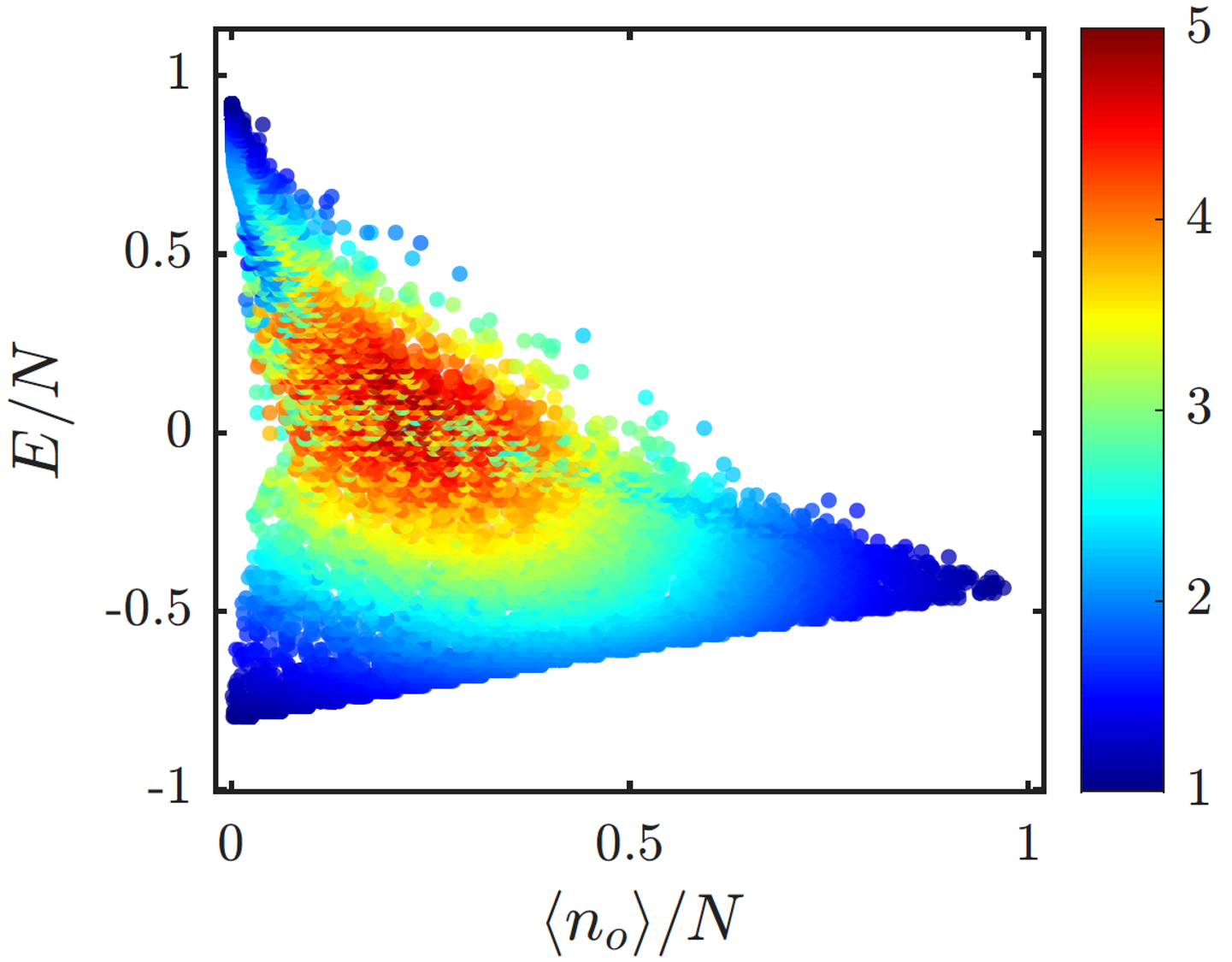} 
\put(-2.5,70){(b)}
\put(86,80){$1/\mathcal{S}$}
\end{overpic}
\
\begin{overpic}[width=5cm]{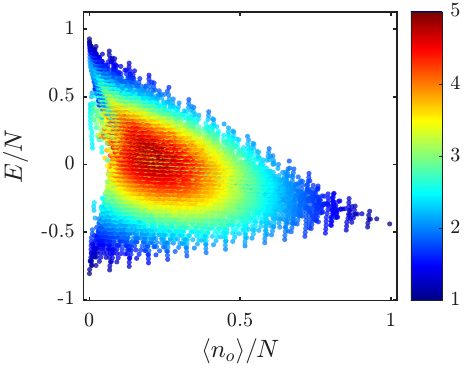}
\put(-2.5,70){(c)}
\put(86,80){$1/\mathcal{S}$}
\end{overpic}

\includegraphics[width=5cm]{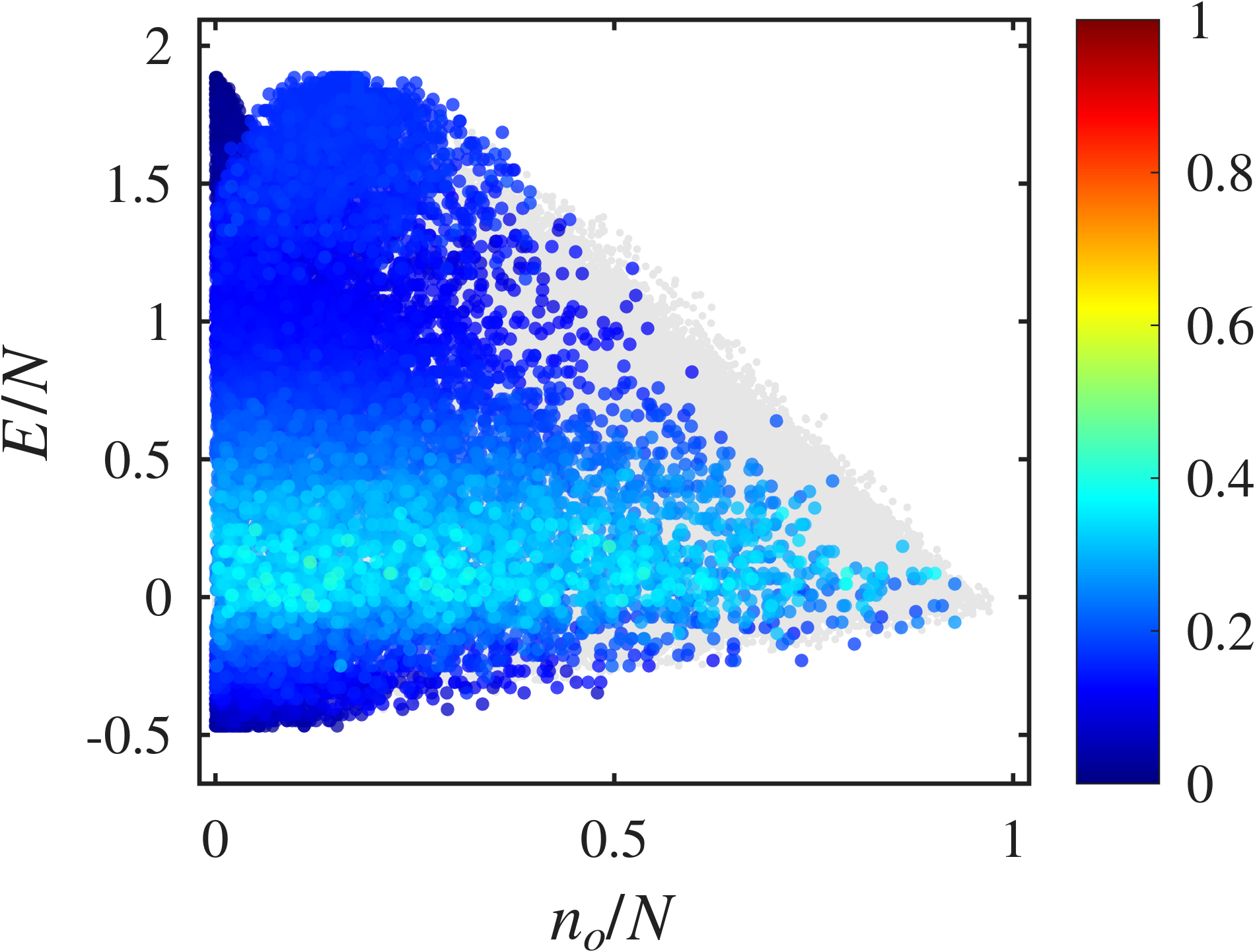}
\
\includegraphics[width=5cm]{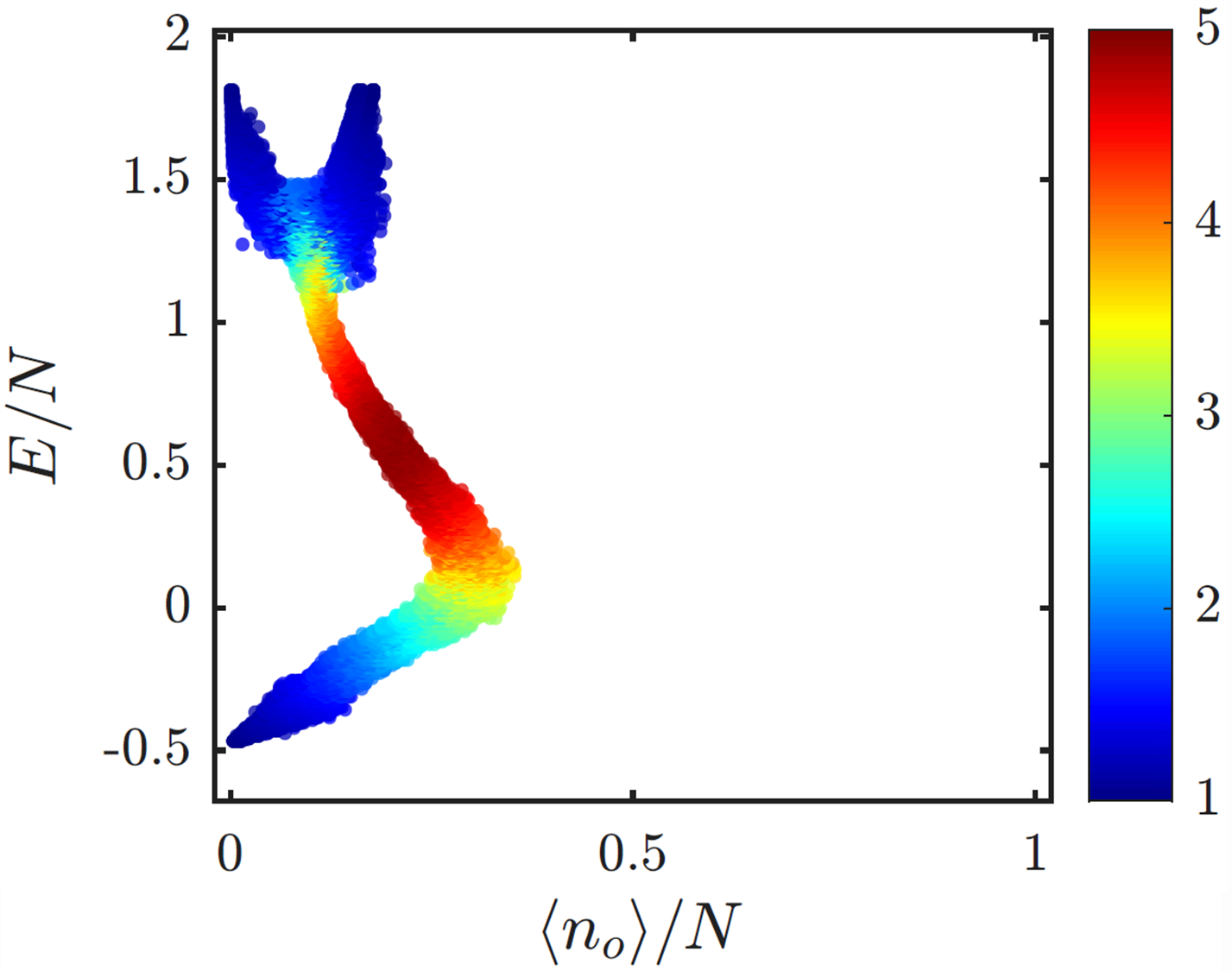} 
\
\includegraphics[width=5cm]{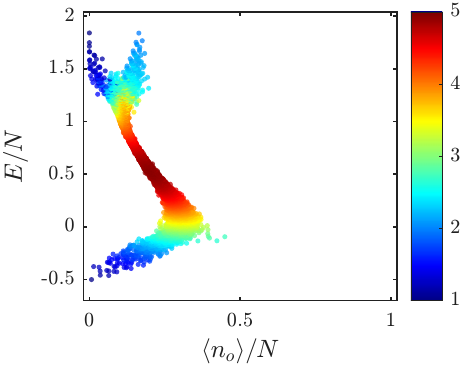} 

\includegraphics[width=5cm]{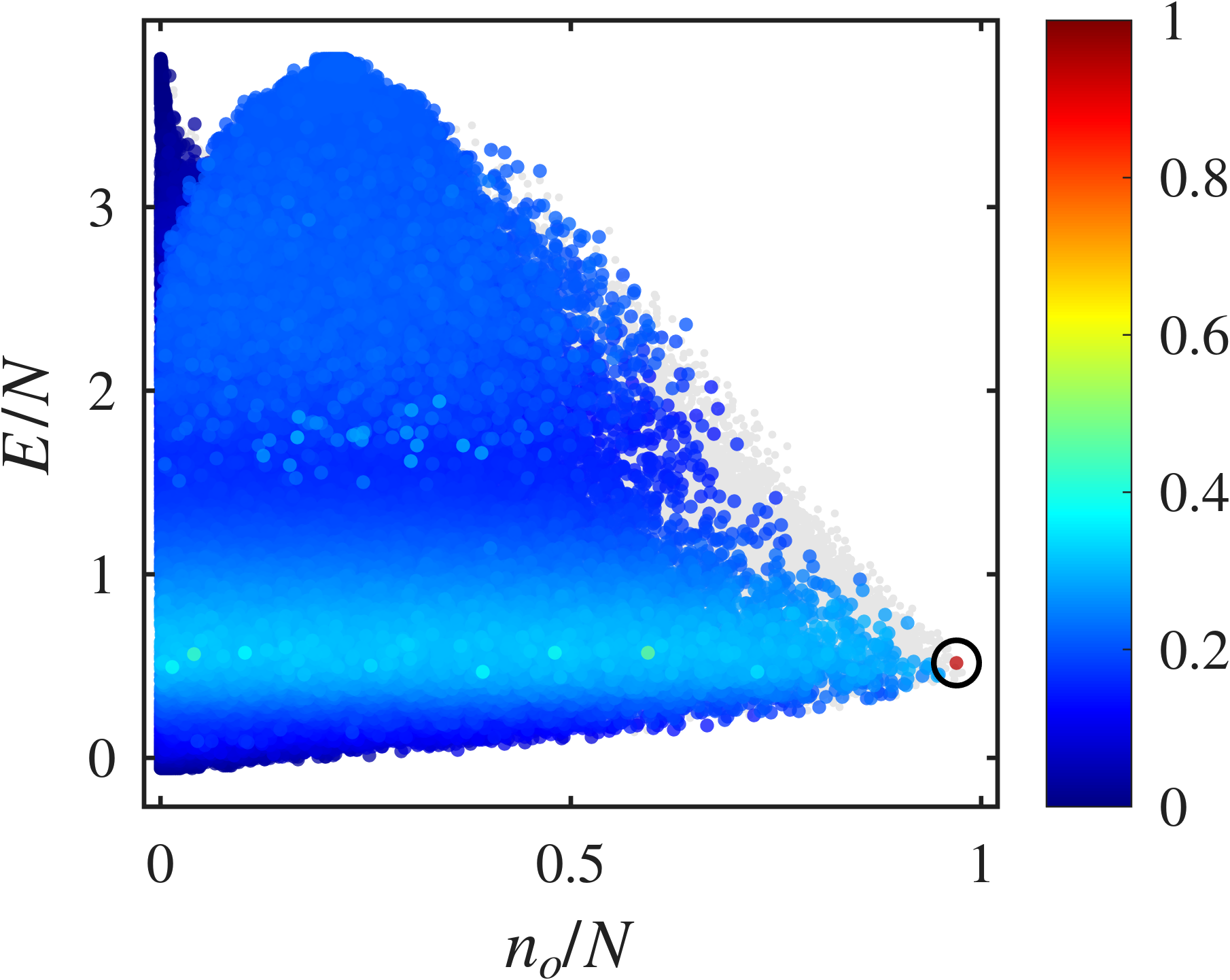} 
\
\includegraphics[width=5cm]{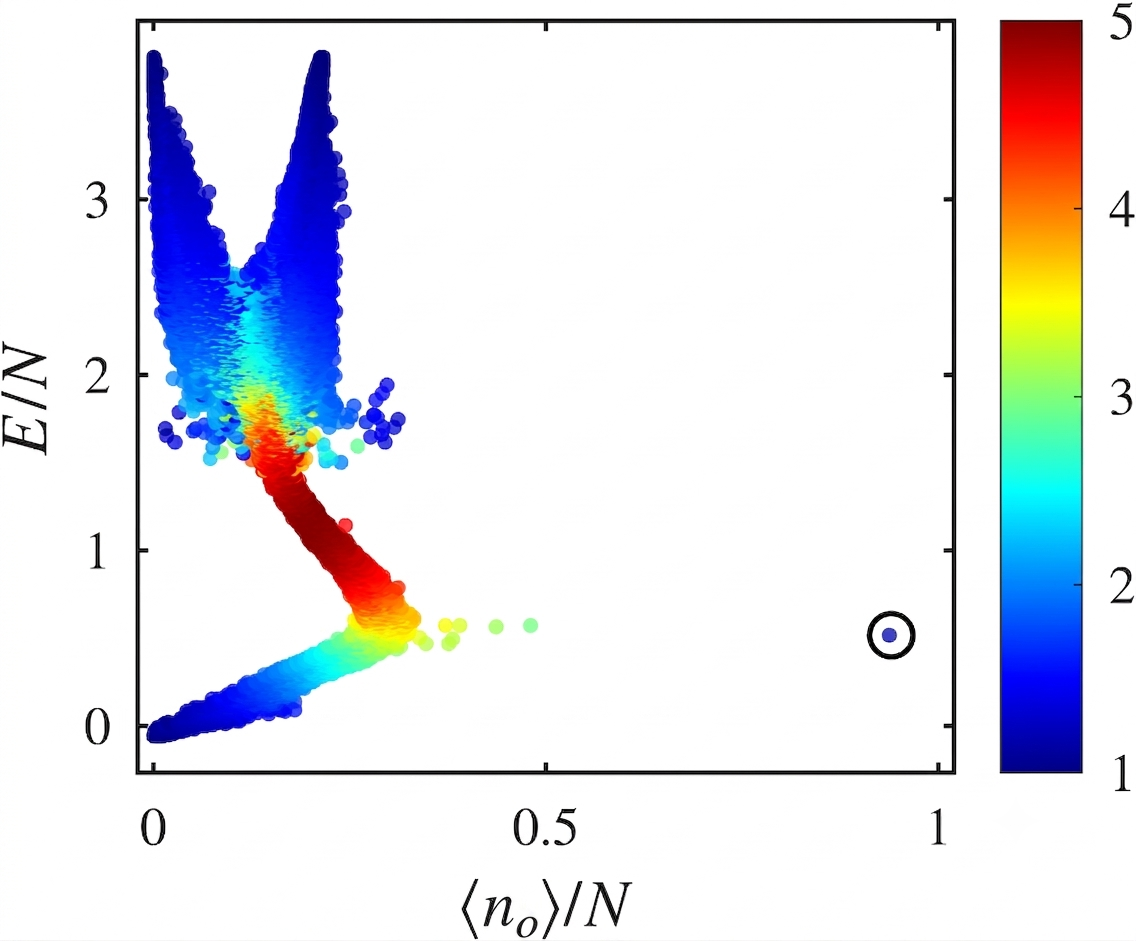} 
\
\includegraphics[width=5cm]{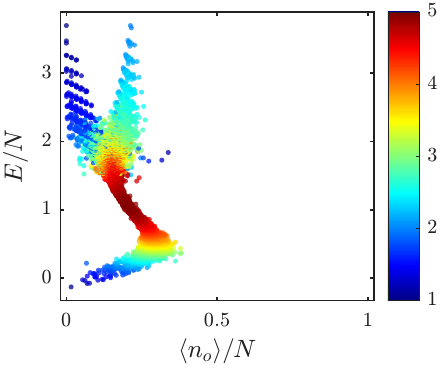}

\caption{{\bf Tomographic spectra for a 5~site Chain.} 
The columns are arranged as in \Fig{fTomoR}: In column~(c), the spectrum is for $N{=}30$ particles, and odd-parity states are excluded.
The rows from top to bottom are for:   
${u=0.5}$~(quasi-regular phase-space that contains stable SP), 
${u=3.5}$~(chaotic phase-space that contains unstable SP), 
and ${u=7.5}$~(separated chaotic sea and stable SP island). 
For weak interaction, there is no indication of chaos, and we identify a dynamically stable condensate. For stronger interaction (2nd and 3rd rows), the ergodization is an indication of chaos, and in the upper part of the spectrum, one identifies self-trapping. In the 3rd row, the quantum spectrum fails to resolve the stability island (encircled in the left panels, with no corresponding eigenstate in the right panel). 
}

\label{fTomoC5}
\end{figure*}

\section{Spectrum tomography}
\label{sec:tomo}

A tomographic image of the spectrum provides insight into the underlying phase-space structure that supports the quantum eigenstates. The following figures display quantum tomographic images of the spectrum (panels of the 3rd column), which are compared with classical tomography (panels of the 1st and 2nd columns).     
\Fig{fTomoR} is for an $L_s{=}5$ ring;  
\Fig{fTomoC3} is for an $L_s{=}3$ chain; 
and \Fig{fTomoC5} is for an $L_s{=}5$ chain.

Each point in the Quantum tomographic image represents an eigenstate, whose vertical position is the energy $E_{\nu}$. 
The extra dimensions (horizontal axis and color-code) are exploited to reveal properties of the eigenstates. 
Here, specifically, the horizontal axis and the color-code correspond, respectively, to the occupation of an inspected orbital labeled as~``o", and to the purity of the eigenstate. Namely, each eigenstate is represented by a point ${(E_{\nu},\braket{n_o}_{\nu})}$ that is color-coded by $\mathcal{S}_{\nu}$. 
For the ring we inspect, without loss of generality, 
the $m{=}0$ orbital, that becomes an excited orbital once ${\phi>\pi}$. For the chains, we inspect the first excited orbital ($m{=}2$).

The quantum spectrum is obtained via diagonalization of the Hamiltonian matrix, which is a rather cheap procedure from a numerical perspective. We compare it to the classically generated spectrum. The latter is numerically expensive. It is obtained as follows: We select a large random set of points in phase-space, then use a representative subset of the points as initial conditions for long trajectories. For each trajectory, we calculate the temporal average of $n_o(t)$, which we regard as the classical analog of $\braket{n_o}_{\nu}$. Then we construct a classical tomographic image of the spectrum by plotting a point ${(E,\braket{n_{o}})}$ for each trajectory.

\subsection{Ring}

The first row of \Fig{fTomoR} illustrates ES of the inspected condensate. For  ${\phi=1.1\pi}$, the metastable eigenstate is located in a local minimum. The minimum is shallow, and therefore, for presentation purpose, we chose a rather strong interaction (${u=4}$) that allows reasonable resolution. For this value of $u$, the upper part of the spectrum is chaotic, and due to ergodicity $\braket{n_{o}}$ is roughly the same for all initial conditions at the same energy. The dispersion merely reflects that our `long' trajectories have a finite time duration. The uppermost eigenstates are self-trapped states, hence featuring ${\braket{n_{o}} = N/5}$.   

The second row of \Fig{fTomoR} (${\phi=2.1\pi}$) illustrates DS of the inspected condensate, while in the 3rd row (${\Phi=2.7\pi}$) the SP is unstable, and therefore a metastable condensate does not exist. However, $u{=}1$ is a rather low value, and therefore ergodicity is not achieved. The spread of $\braket{n_o}_{\nu}$ reflects the underlying quasi-regular structure of phase-space. In order to demonstrate ergodicity due to chaos for the same $\phi$, we display in the 4th row the spectrum for ${u=4}$.

\subsection{Chain}

\Fig{fTomoC3} displays tomography for an ${L_s=3}$ chain with ${N=150}$  particles. Recall that such a chain is non-generic. For small interaction ($u=0.5$), the SP is unstable, but the dynamics is quasi-regular. For stronger interaction ($u=1.5$), chaos is still rather weak, and therefore ergodization is poor. For strong interaction (${u=3.5}$), the chaos is strong enough to induce ergodization in the chaotic sea, but on the other hand, the SP gains DS, hence we have an island of stability. The quantum spectrum features {\em hybrid} eigenstates and therefore the sea/island classification is blurred. In the upper part of the spectrum, self-trapping in the middle site is conspicuously distinguished from self-trapping in the two opposing sites.   

\Fig{fTomoC5} displays tomography for an ${L_s=5}$ chain with ${N=30}$  particles. This is our {\em generic} example for a chain. It exhibits a GPE-like DS regime for weak interactions, which is demonstrated with $u=0.5$ in the upper row. Irrespective of that, the dynamics is quasi-regular. For stronger interaction ($u=3.5$), the SP becomes unstable, and due to chaos, we get ergodicity. The upper part of the spectrum reflects self-trapping. For larger interaction ($u=7.5$), the SP becomes again DS, which is the same scenario as discussed for the $L_s{=}3$ ring. 
But here the quantum spectrum does not follow the classical spectrum. The stability island is too small to be resolved, and a DS condensate does not appear. 
The reason for this is the small value of~$N$. In \Fig{fTrajErg} we show that the uncertainty width is comparable with the size of the stability island, and apparently the latter is too small to accommodate an eigenstate. 

{The above tomography illuminates generic features of {\em high-dimensional chaos}. The ergodicity within the chaotic sea becomes much better, which is reflected in the very narrow dispersion of $\braket{n_o}$. Consequently, the stability island is better resolved: it is a well-separated tiny region with ${\braket{n_o}\sim N}$. At the same time, its borders become ill-defined topologically, as discussed in connection with \Fig{fTrajErg} in \Sec{sec:traj}.}

\section{Spectral fingerprints of metastability}
\label{sec:spectral}

We would like to inspect quantitatively whether or how the stability of the SP is reflected in the tomography of the quantum spectrum. For demonstration, we use the following procedure. 
Given a tomographic $(\braket{n_o},E)$ spectrum, 
the subscript ``max" is used to indicate the many-body eigenstate $\ket{E_{\nu}}$ for which $\braket{n_o}_{\nu}$ is maximal. 
The values of $\braket{n_o}_{\nu}$ and $\mathcal{S}_{\nu}$ 
for this state are denoted as  
$n_{\max}$ and $\mathcal{S}_{\max}$.
If $u=0$ we get ${n_{\max} = N}$ and ${\mathcal{S}_{\max}=1}$.
We discuss below what happens for reasonably large~$u$.

\subsection{Ring}

Let us focus on the 5~site Ring.  
As the control parameter $\phi$ is varied, the condensate experiences ES$\leadsto$DS$\leadsto$instability transitions. We label the inspected orbital by ``o". Here, ``o" refers to the $m=0$ orbital.
The dependence of $n_{\max}$ and $\mathcal{S}_{\max}$ on $\phi$ is presented in \Fig{fig:fgpRing} for ${u{=}1}$ and for ${u{=}4}$.
Up to ${\phi{=}\pi}$ the inspected condensate is trivially stable,
simply because it is the ground state. For larger ${\phi}$, we still have ES up to $\phi{=}1.75\pi$. Then we observe a moderate drop in $n_{\max}$. Once $\phi$ crosses the instability threshold at $\phi=2.5\pi$, there is a much sharper drop, indicating that the metastable condensate has been diminished.  Note that for the larger value of $u$, it is difficult to resolve the two transitions. 

Subsequently, for ${u=1}$, there is recovery of the stability as the $m=0$ orbital approaches the top of the energy landscape. This recovery becomes obvious once we notice that, for ${\Phi>4\pi}$, the top state is in fact the ground state of the ${U \mapsto -U}$ Hamiltonian. The latter features a self-trapping transition for large $u$, hence recovery is avoided. This self-trapping was also apparent in the tomographic spectrum.

\subsection{Chain}

\begin{figure*}[h!]
\centering

\begin{overpic}[width=7.5cm]{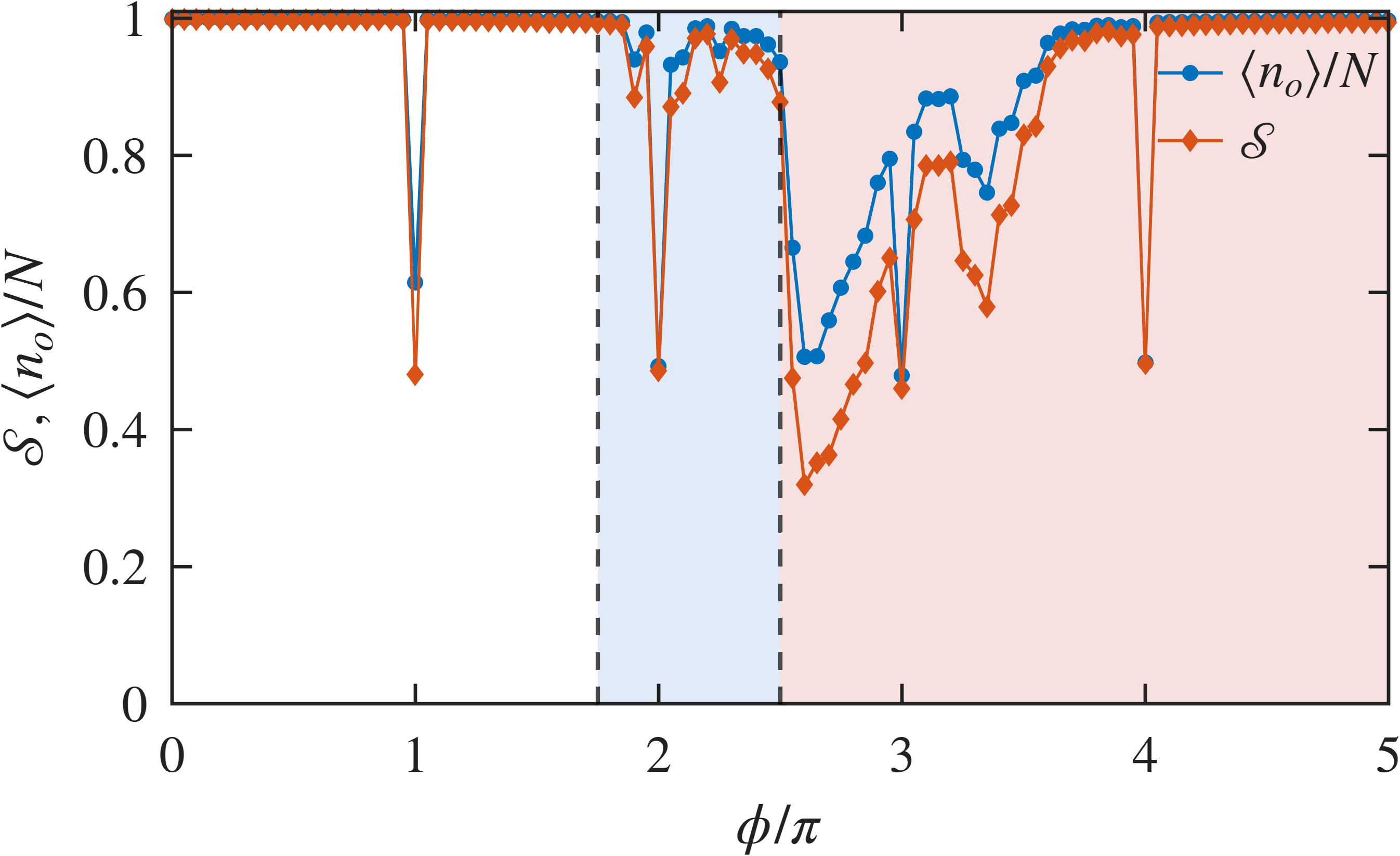}
\put(-1.0,58){(a)}
\end{overpic}
\ \ \ \ 
\begin{overpic}[width=7.5cm]{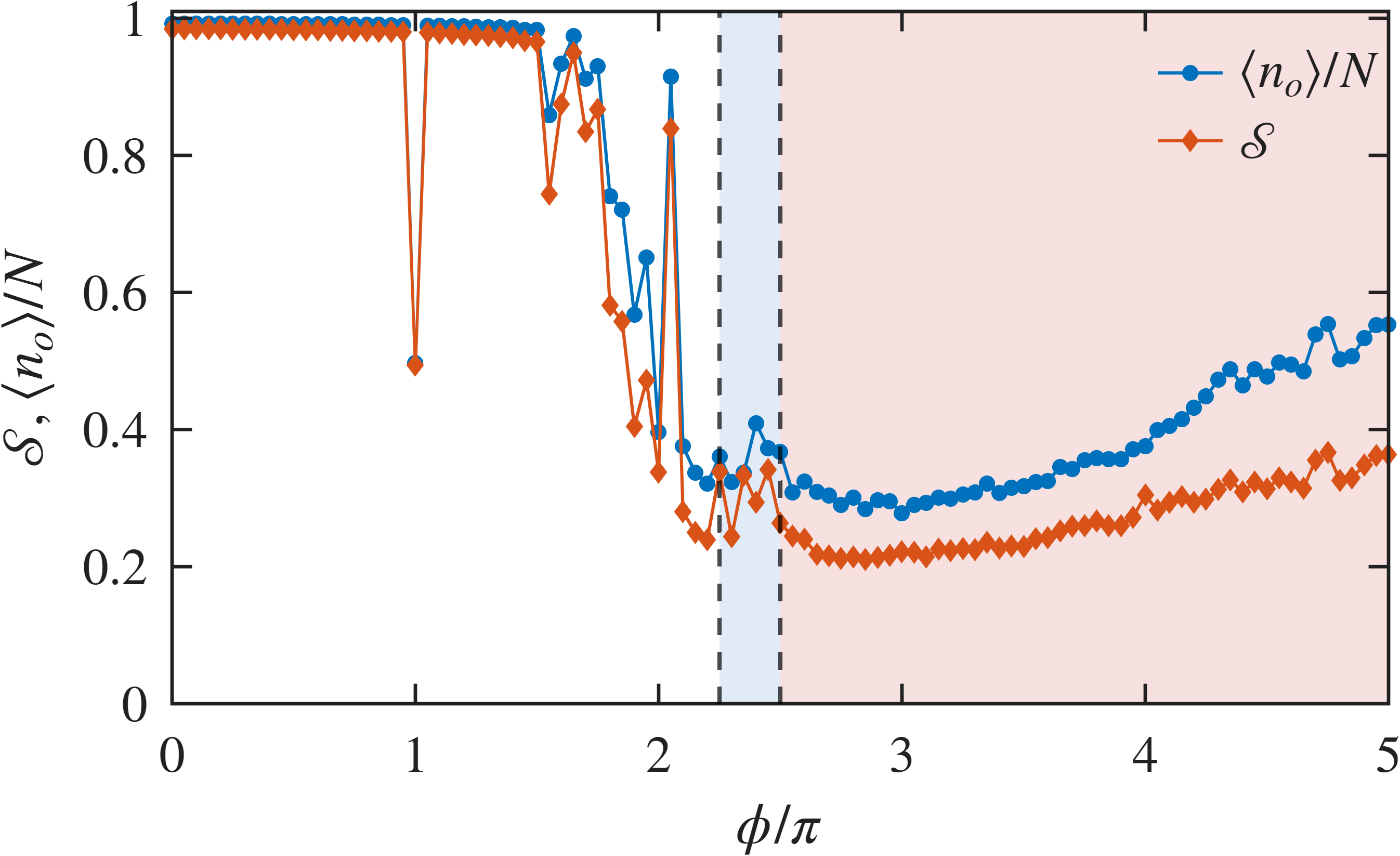}
\put(-1.0,58){(b)}
\end{overpic}
\caption{
{\bf Fingerprints of stability for a Ring.}
We consider a 5~site ring with $N=30$ particles. 
The interaction is $u=1$ (left) and $u=4$ (right). We keep only the $P{=}0$ eigenstates. The inspected orbital ``o" is $m=0$. 
The calculation of $\braket{n_o}_{\nu}$ and $\mathcal{S}$ 
is carried out for the ``max" many-body eigenstate (see text).  
The plot shows the dependence of $n_{\max}$ and $\mathcal{S}_{\max}$ on $\phi$, 
reflecting the level crossing at ${\phi=\pi}$ and the ES$\leadsto$DS$\leadsto$instability transitions 
that are indicated by dashed lines based on the Bogoliubov analysis of \Fig{fBogoRing}.
}
\label{fig:fgpRing}
%
%
\ \\ \ \\  
%
%
\centering

\begin{overpic}[width=7.5cm]{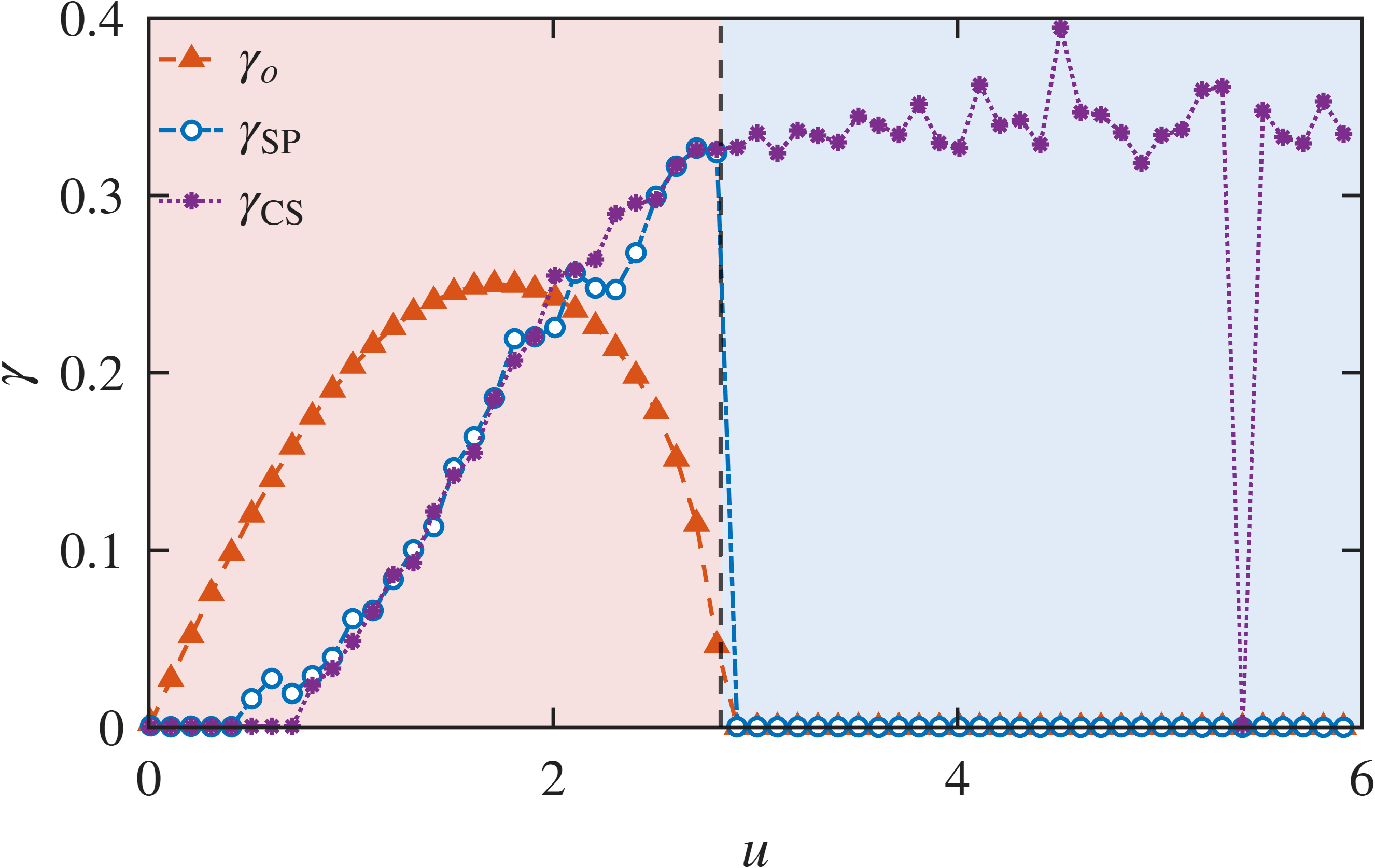}
\put(-1.75,60){(a)}
\end{overpic}
\ \ \ \
\begin{overpic}[width=7.5cm]{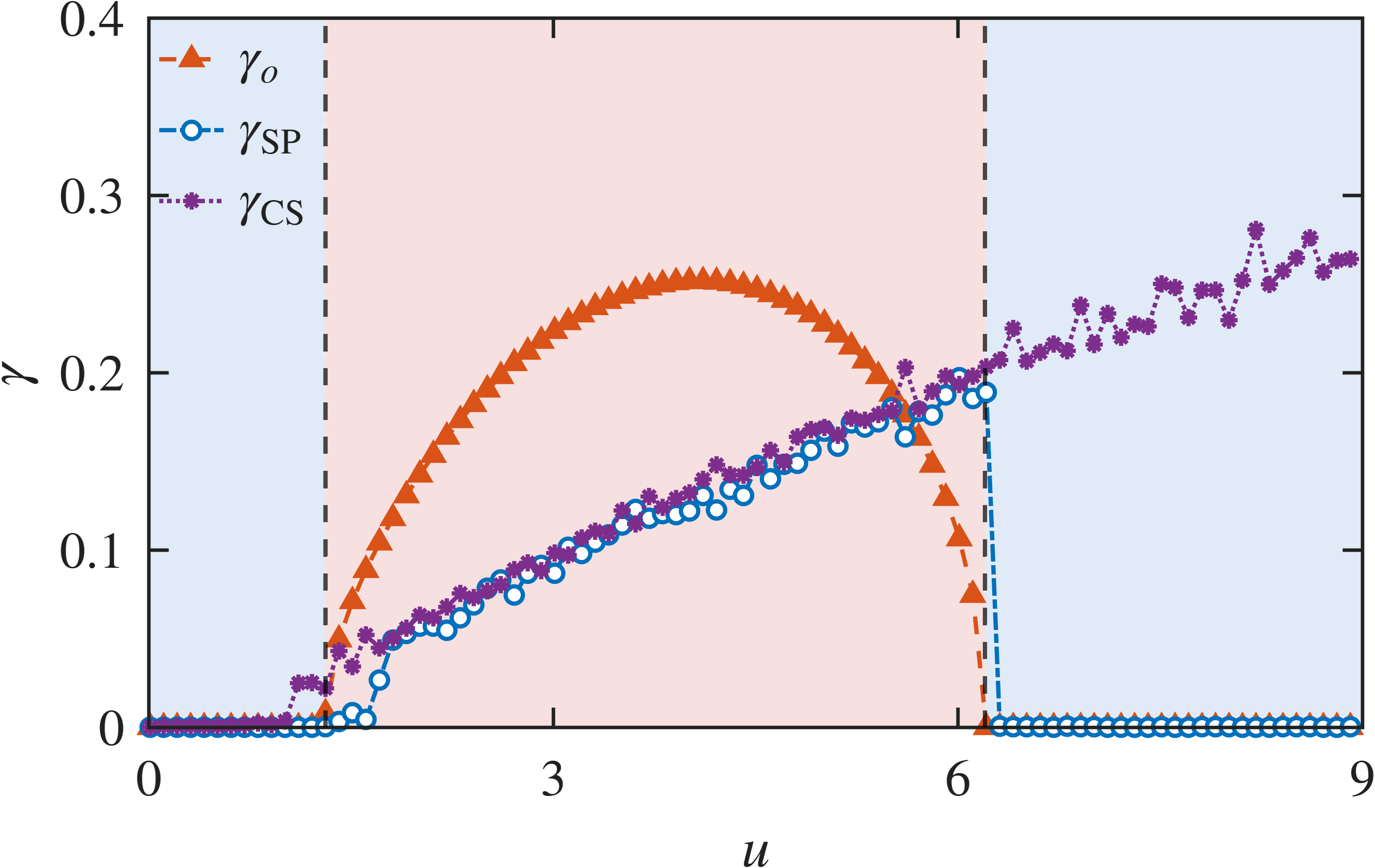}
\put(-1.75,60){(b)}
\end{overpic}

\begin{overpic}[width=8.0cm]{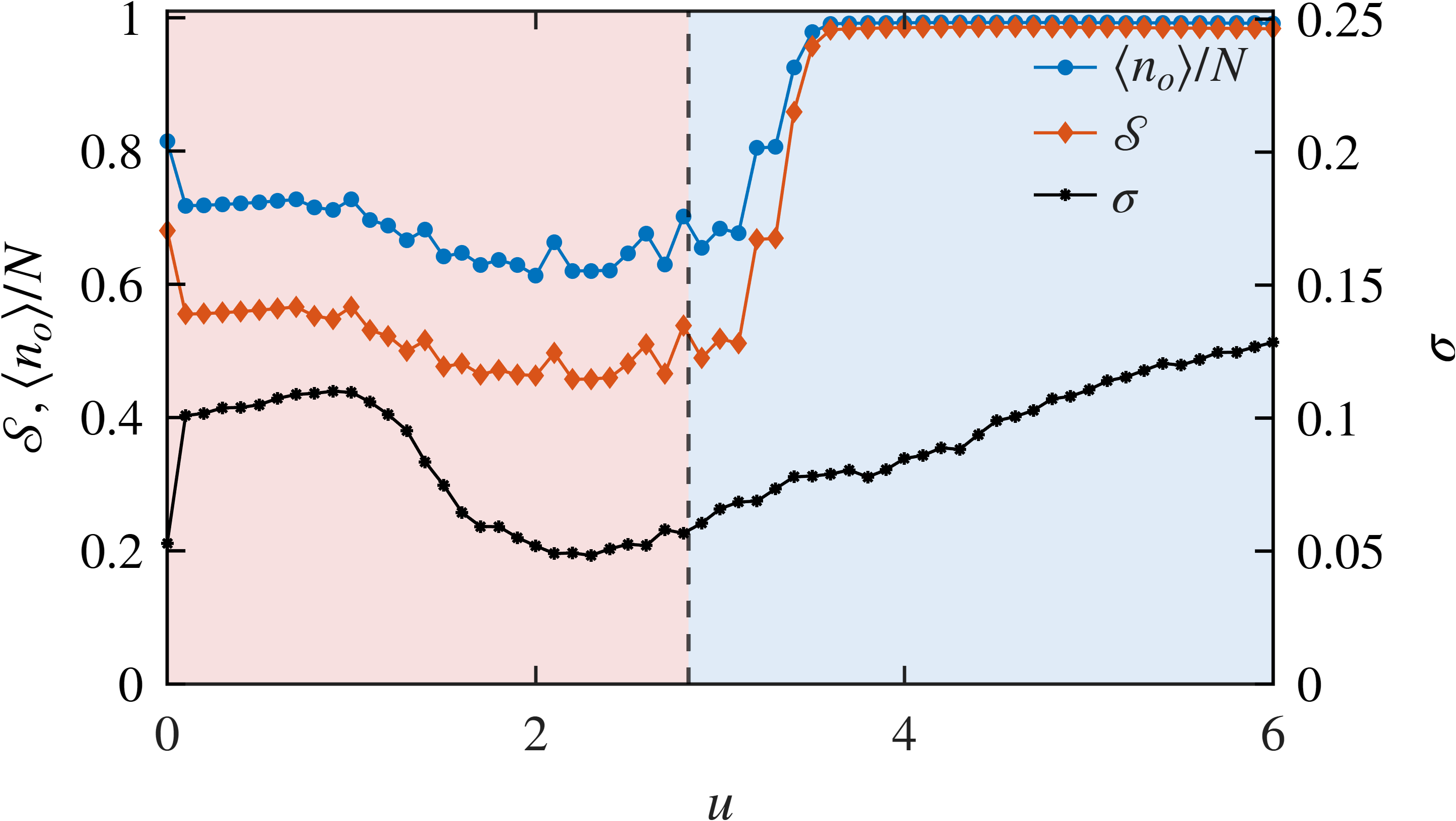}
\put(-1.0,53){(c)}
\end{overpic}
\ \ \ \
\begin{overpic}[width=8.0cm]{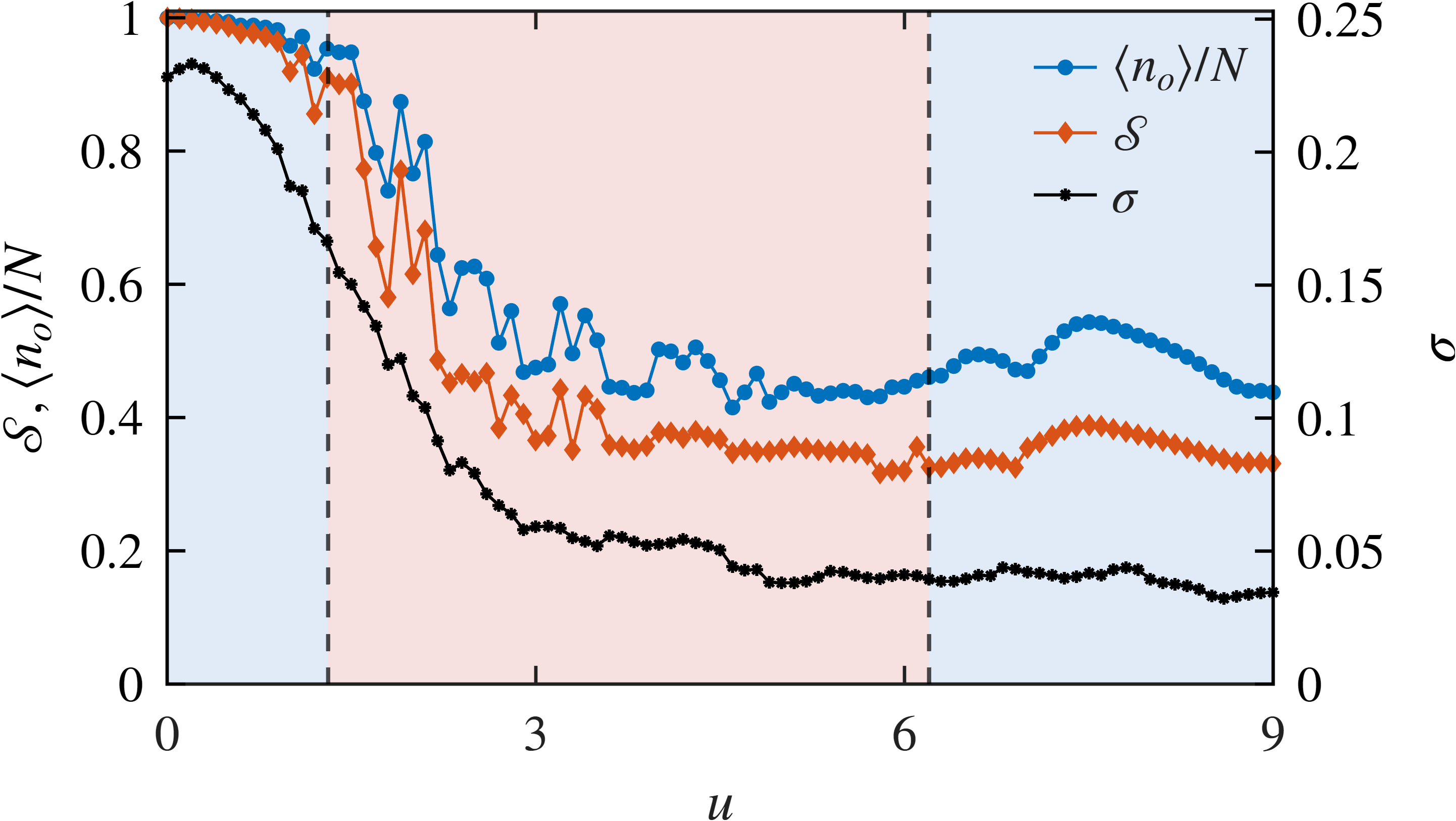} 
\put(-1.0,53){(d)}
\end{overpic}

\caption{
{\bf Fingerprints of stability for a Chain.}
We inspect the $m{=}2$ orbital and the associated metastable condensate.  
Left and Right panels are for 3~site and 5~site chains, respectively, 
featuring  instability$\leadsto$DS and DS$\leadsto$instability$\leadsto$DS  transitions. In the quantum analysis, the number of particles is ${N{=}150,20}$ respectively, and we keep only the symmetric eigenstates.
{\em First row:} Dependence of $\gamma_{o}, \gamma_{\text{SP}}, \gamma_{\text{CS}}$ on $u$. Note that ${\gamma_{\text{SP}}\sim 0}$ if the SP is embedded in a quasi-regular island, whereas ergodization leads to ${\gamma_{\text{SP}}\sim \gamma_{\text{CS}}}$. 
{\em Second row:} Dependence of $n_{\max}$ and $\mathcal{S}_{\max}$ and $\sigma$ on $u$, where $\sigma$ is the ergodicity measure defined in \Eq{eergo}. The secondary instability$\leadsto$DS transition for the 5~site chain is barely reflected in the quantum case because the effective ${\hbar\sim 1/N}$ with $N{=}20$ is rather large.
}
\label{fig:fgpChain}
\end{figure*}

The dependence of $n_{\max}$ and $\mathcal{S}_{\max}$ on $u$ for a 3~site chain and for a 5~site chain is presented in \Fig{fig:fgpChain}.
We inspect the $m{=}2$ orbital and the associated metastable condensate.  
The upper panels summarize the results of the classical stability analysis. Three classical measures are used to
characterize the stability of the SP as a function of the interaction $u$. The first measure is $\gamma_o$, the largest positive imaginary part of the eigenvalues of the Bogoliubov matrix $W$ \Eq{eW}. It tells us whether we have DS or instability. Note that the 3~site chain lacks a GPE-like stability regime, meaning that ${u_c=0}$. 

The two other measures, which are inspected in the upper panels of \Fig{fig:fgpChain}, are Lyapunov exponents. $\gamma_{\text{CS}}$ is extracted from a trajectory that is launched in the chaotic sea, while $\gamma_{\text{SP}}$ is the Lyapunov exponent of a trajectory that is launched in the vicinity of the SP. If the SP is unstable, and no stability island exists, one expects ${\gamma_{\text{SP}} \sim \gamma_{\text{CS}} }$, which is indeed what we see. It is important to note that the Lyapunov exponent is not quantitatively correlated with $\gamma_o$; notably, we might have weak or strong chaos irrespective of the stability of the unstable SP.

The second row of panels in \Fig{fig:fgpChain} is based on the quantum tomography of the spectrum. It shows the dependence on $u$ of $n_{\max}$ and  $\mathcal{S}_{\max}$ and of an ergodicity measure $\sigma$ that will be defined in the next section.  One observes that the instability$\leadsto$DS transition for the 3~site chain, and the  DS$\leadsto$instability transition for the 5~site chain are well reflected. In contrast, the secondary instability$\leadsto$DS transition for the 5~site chain is barely reflected in the quantum case. This is hardly surprising, because the effective ${\hbar\sim 1/N}$ with $N=20$ is rather large, and therefore the classical stability island is barely resolved as implied by \Fig{fTrajErg}.    

\begin{figure}
\centering
\begin{overpic}[width=6cm]{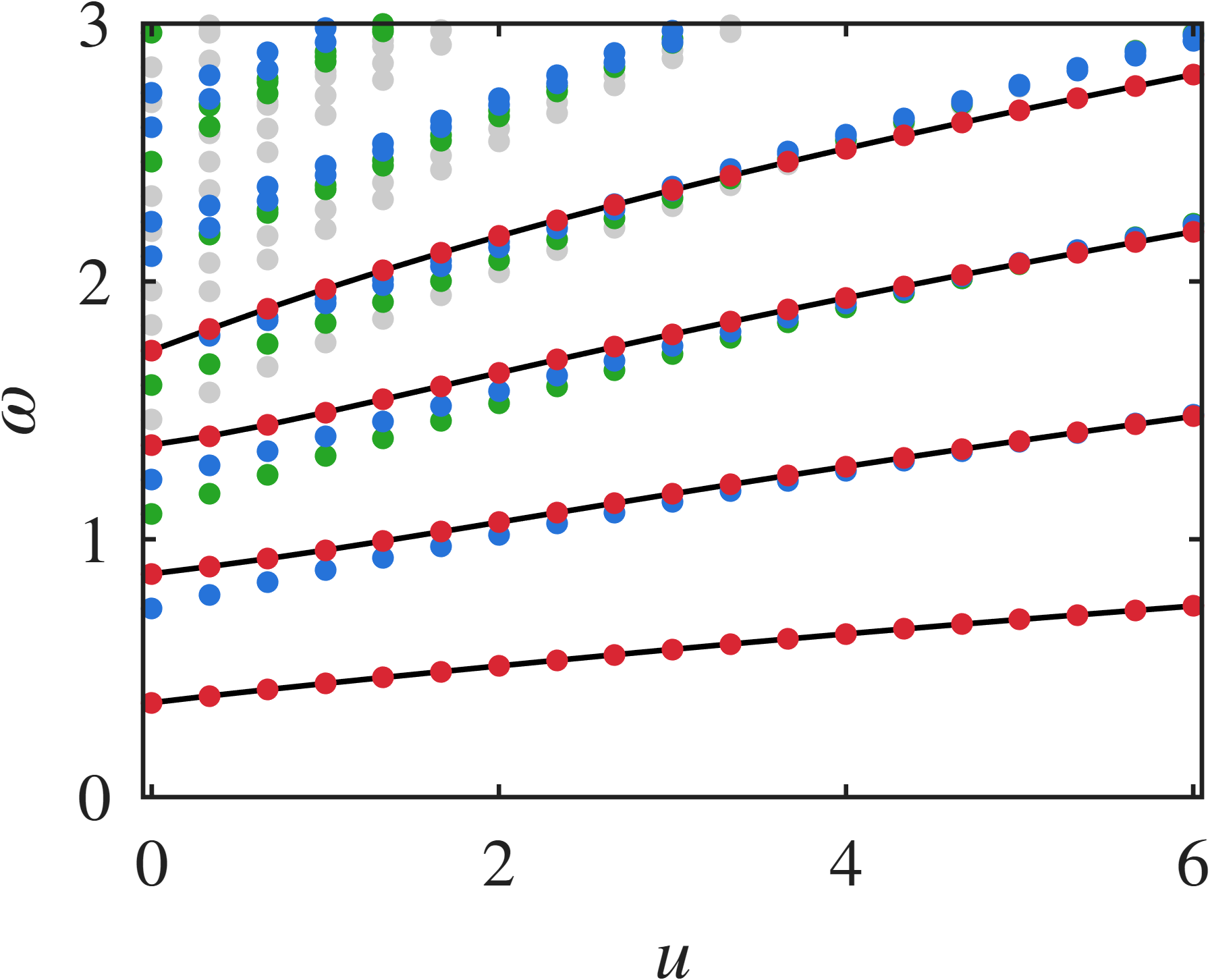}
\put(-3.0,76){(a)}
\end{overpic}

\begin{overpic}[width=6cm]{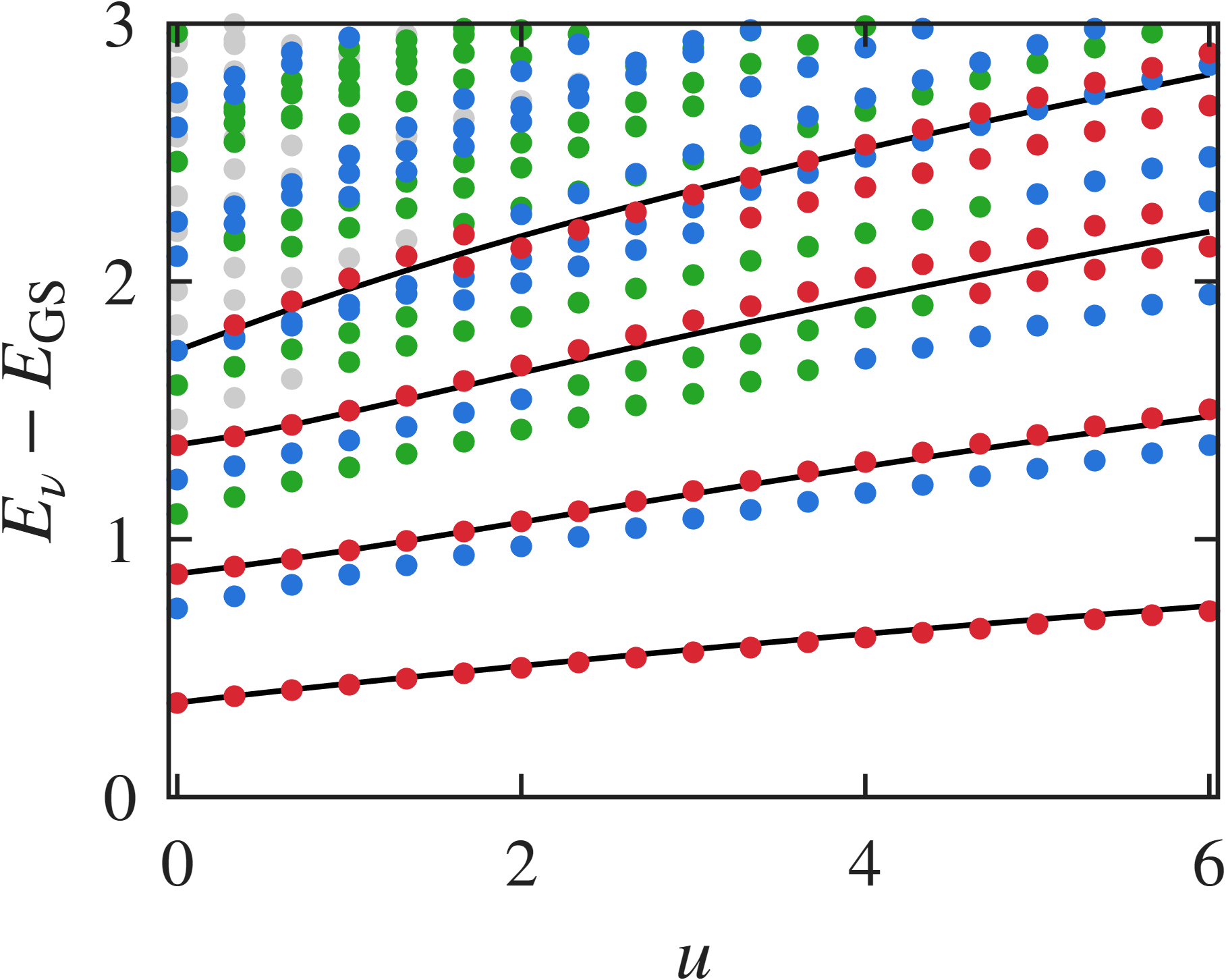}
\put(-2.5,74){(b)}
\end{overpic}

\begin{overpic}[width=6cm]{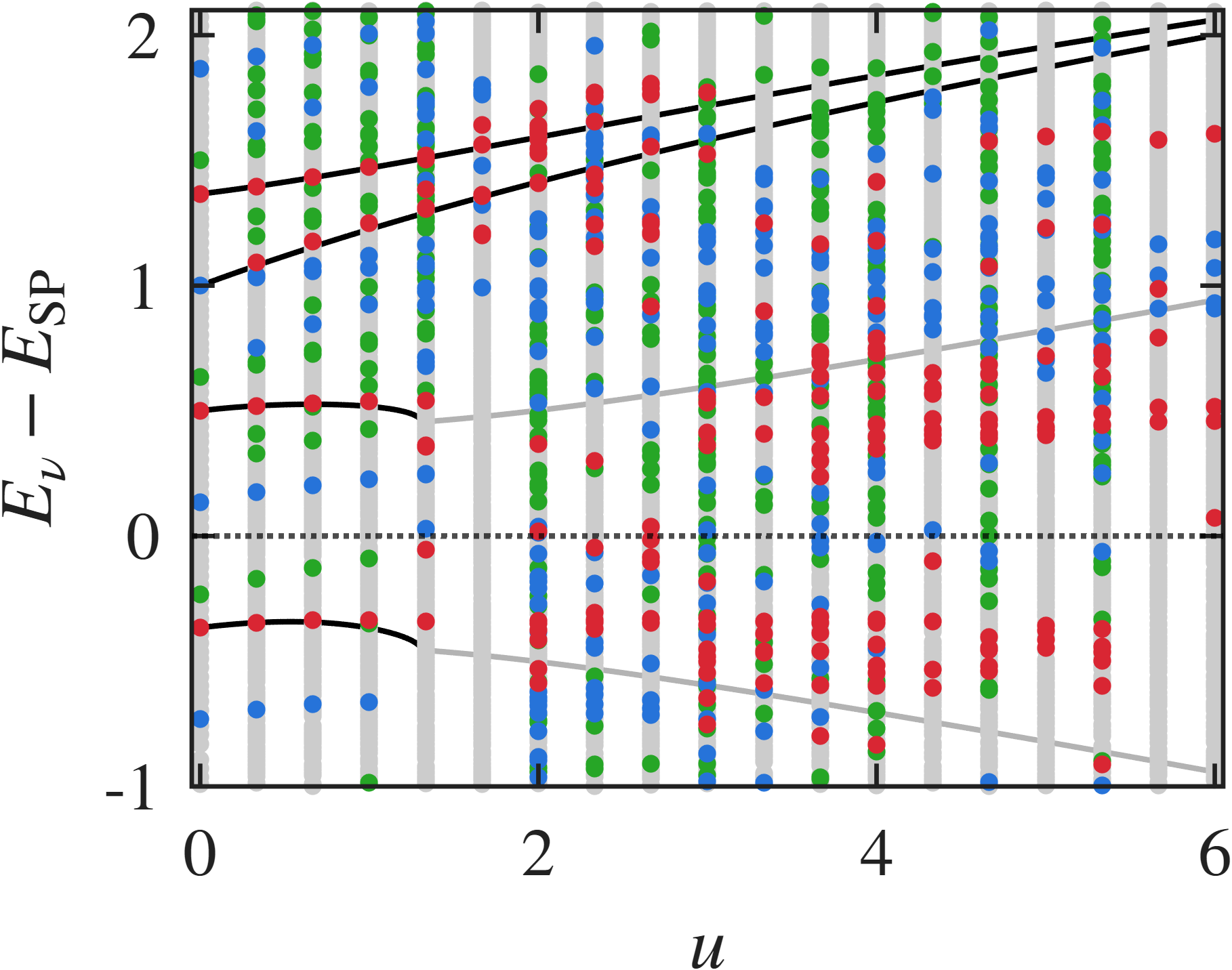}
\put(-2.5,73){(c)}
\end{overpic}

\caption{
{\bf Many-body excitation energies for a 5-site chain.} 
We consider the same chain of \Fig{fTomoC5} with $N{=}20$ particles. 
(a)~Expected excitation spectrum based on Bogoliubov frequencies. Single-phonon excitations are in red, $n_{\text{phonons}}=2,3$ are in blue and green, while higher phonon occupations are shown in gray.
(b)~The exact excitation energies ($E_{\nu} - E_{\text{GS}}$) obtained through exact diagonalization. The $n_{\text{phonons}}$  color-code is determined as explained in the main text.  
(c)~The same for the $m_o{=}2$ condensate. The solid black lines represent the real Bogoliubov frequencies, while the gray lines represent the real part once they become complex.}
\label{fig:BogoCompare}
\end{figure}

\subsection{Excitations}

Speaking about ``spectral signature", the natural common inclination is to expect that there is a relation between many-body level spacings and the Bogoliubov frequencies. We have not emphasized this perspective because it has a rather limited validity in practice, and does not look useful or illuminating. Nevertheless, for completeness, we demonstrate this perspective.      

On the basis of Bogoliubov's diagonalization procedure, one expects to have the following approximation for eigenstates that reside in phase space in the vicinity of the SP,  
\beq
E_{\nu} = E_{\text{SP}} + \sum_{\nu_q} \nu_q \omega_q. 
\eeq
Here ${\nu_q=0,1,2,\cdots}$ are approximate quantum numbers and ${ \nu = \{ \nu_q \} }$ is regarded as a composite index. The sum ${ n_{\text{phonons}} = \sum_q \nu_q }$ distinguishes between single-phonon excitations and multi-phonon excitations. In \Fig{fig:BogoCompare}a, we display the expected spectrum, as a function of $u$, in the vicinity of the ground state SP. The points are color-coded according to  ${ n_{\text{phonons}} }$.   
In \Fig{fig:BogoCompare}b, we plot the many-body excitation energies based on the exact diagonalization. Here, the points are color-coded as single-phonon or multi-phonon excitations based on their spectral weight.

The {\em spectral weight} of an excitation is calculated as follows. 
We define the number-conserving Bogoliubov quasi-particle operator \cite{NCG,NCC} as 
${ C^\dagger_q = u_q A_q - v_q A^{\dagger}_{-q} }$, 
where ${ A_q = \frac{1}{\sqrt N}\bm{a}^{\dagger}_{k_o+q} \bm{a}_{k_o} }$, and the real coefficients ${(u_q,v_q)}$ are determined via the diagonalization of $\bm{W}$. 
This operator is applied ${ n_{\text{phonons}} }$ times on the interacting ground state, hence generating synthetic multi-phonon excitations.  
Then, each exact eigenstate is projected onto those synthetic excitations and color-coded according to the dominant overlap. In particular, 1-phonon excitations are colored in red, 
while states with ${ n_{\text{phonons}} >3 }$ are color-coded in gray. 

Comparing panel~(a) and panel~(b) of \Fig{fig:BogoCompare}, we see that, as expected, the Bogoliubov approximation provides a rather satisfactory approximation for the spectrum. Encouraged, we turn to apply the same procedure for the excitations in the vicinity of the energy $E_{\text{SP}}$ of an excited condensate. The comparison shows that the usefulness of the Bogoliubov approximation is rather limited.

\begin{figure*}
\centering
\begin{overpic}[width=6cm]{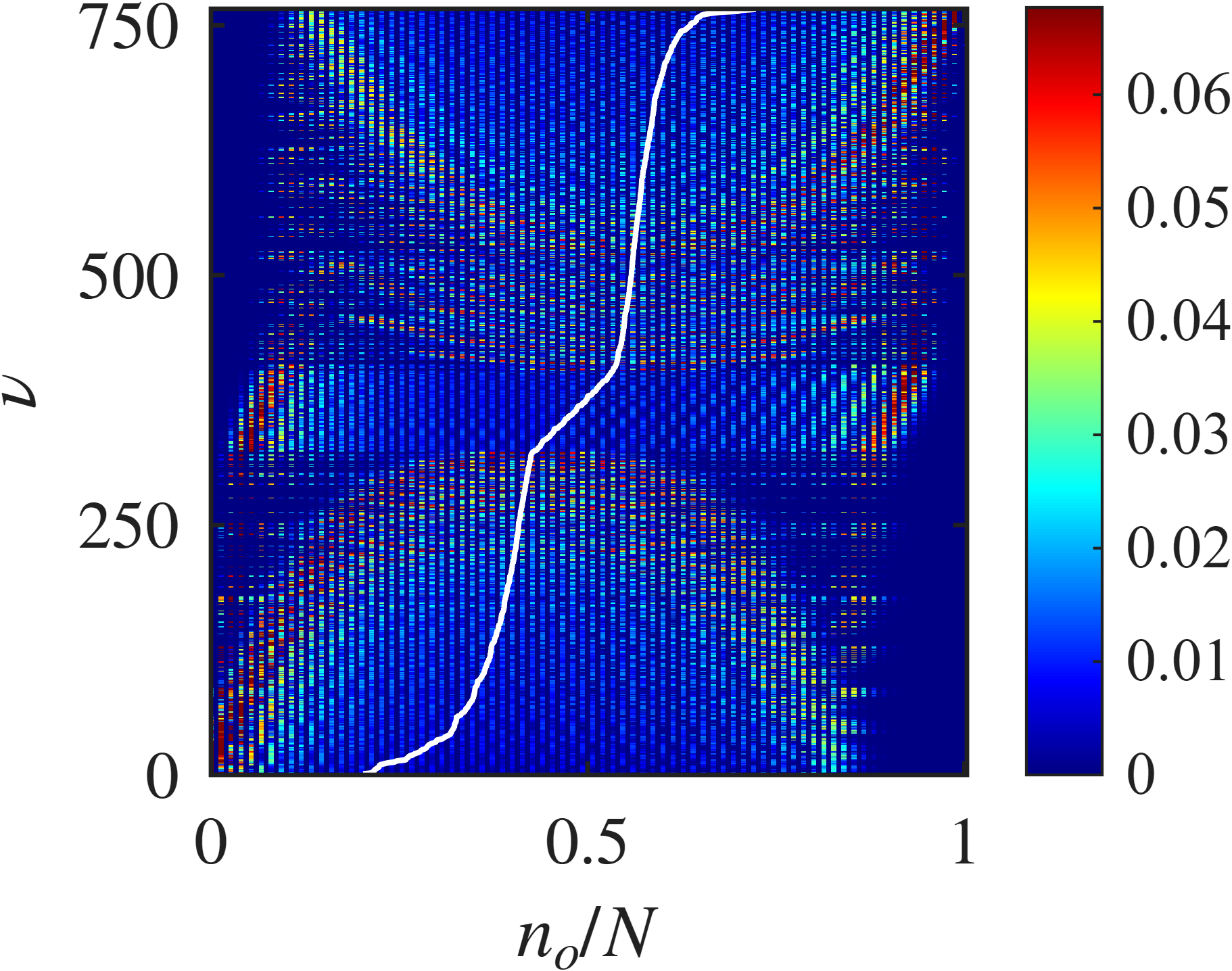} 
\put(-3.0,75){(a)}
\end{overpic}
\begin{overpic}[width=6cm]{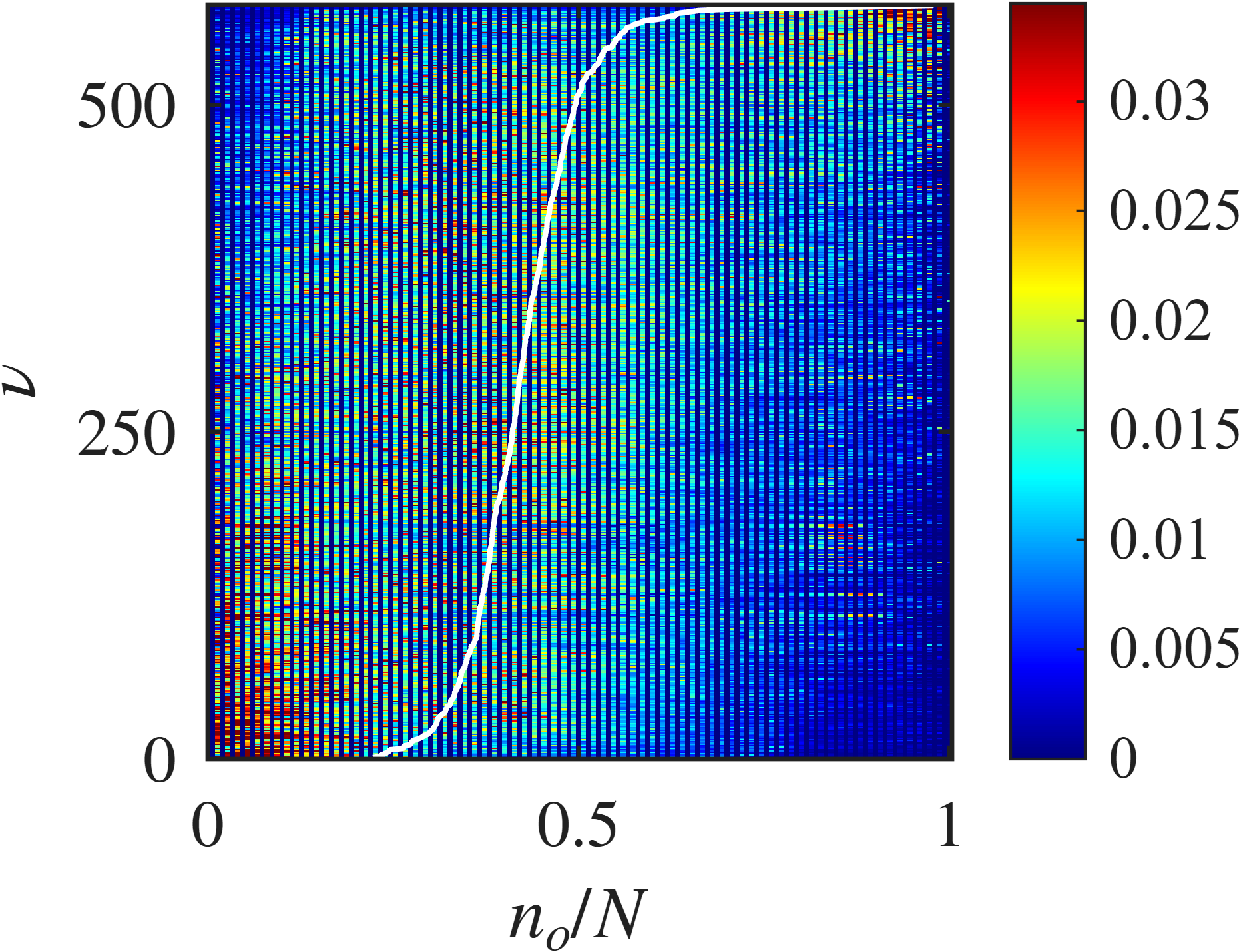}
\put(-3.0,75){(b)}
\end{overpic}
\ \ \ \ \ \ \ \ \ \ 
\begin{overpic}[width=6cm]{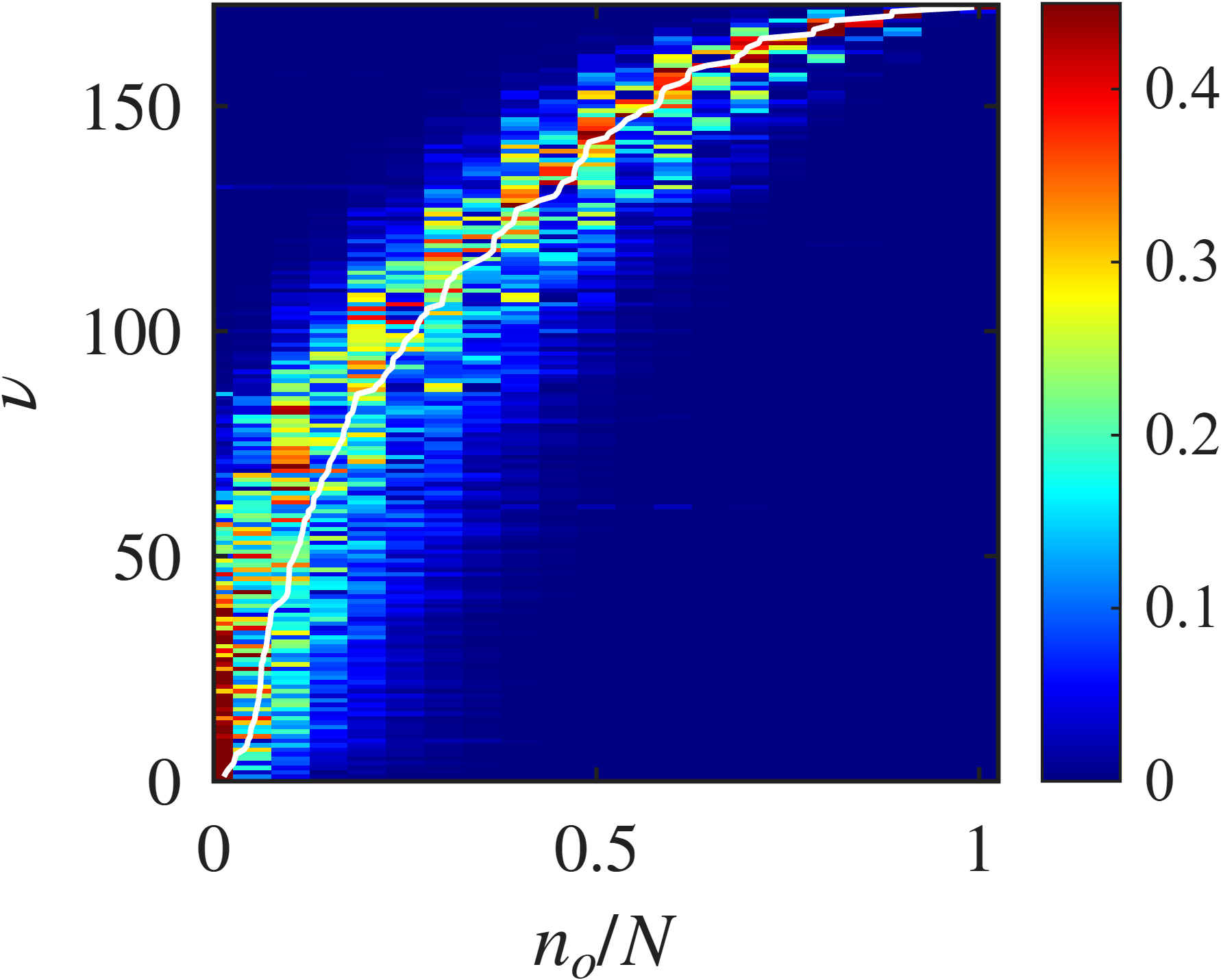} 
\put(-1.0,75){(c)}
\end{overpic}
\begin{overpic}[width=6cm]{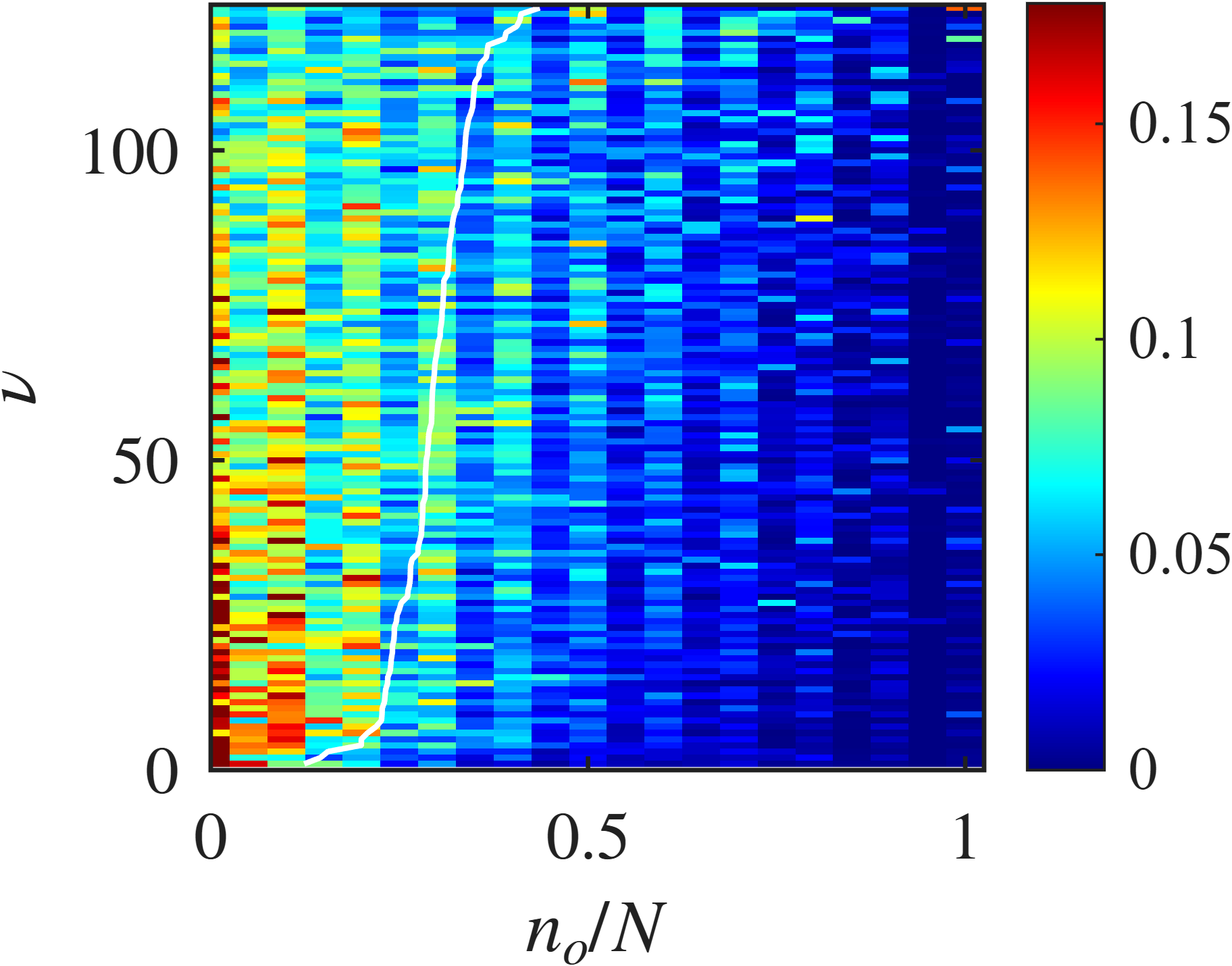} 
\put(-1.0,75){(d)}
\end{overpic}

\caption{{\bf Inspection of quantum ergodicity.}
Complementary to the tomography of \Fig{fTomoC3} and \Fig{fTomoC5}.  
Each row of an image is a $P^{\nu}(n_o)$ distribution. These rows are ordered by~$\braket{n_o}$. All the even-parity eigenstates within a narrow energy window around $E_{\text{SP}}$ are included.  
The upper panels are for the 3~site chain with ${N=150}$. 
The lower panels are for the 5~site chain with ${N=20}$. 
Left and right panels are for ${u=0.5}$ and ${u=3.5}$, respectively. 
}

\label{fQergImages}
%
\ \\ \ \\ 
%

\begin{overpic}[width=6cm]{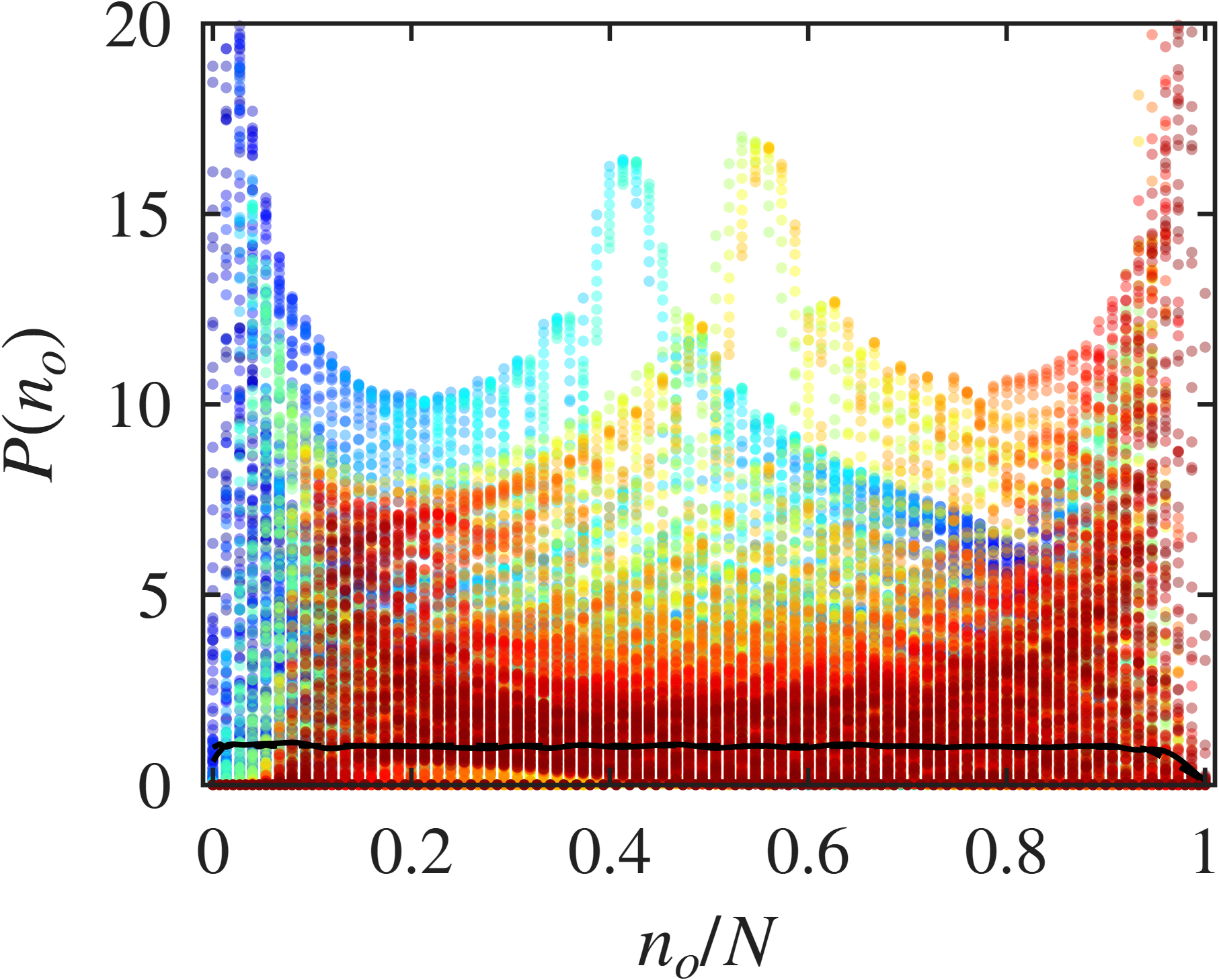} 
\put(-3.0,75){(a)}
\end{overpic}
\ \ \ \
\begin{overpic}[width=6cm]{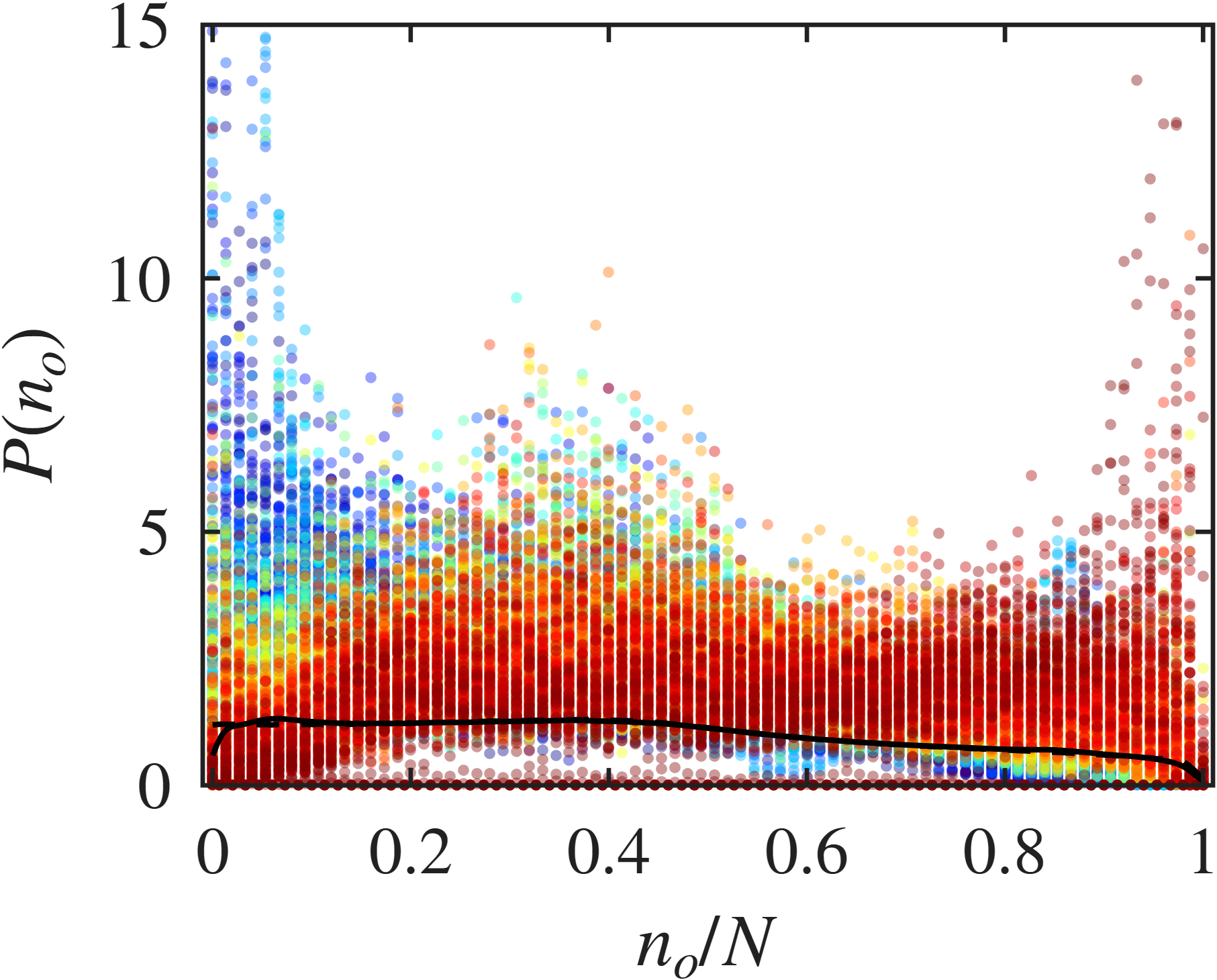}
\put(-3.0,75){(b)}
\end{overpic}

\begin{overpic}[width=6cm]{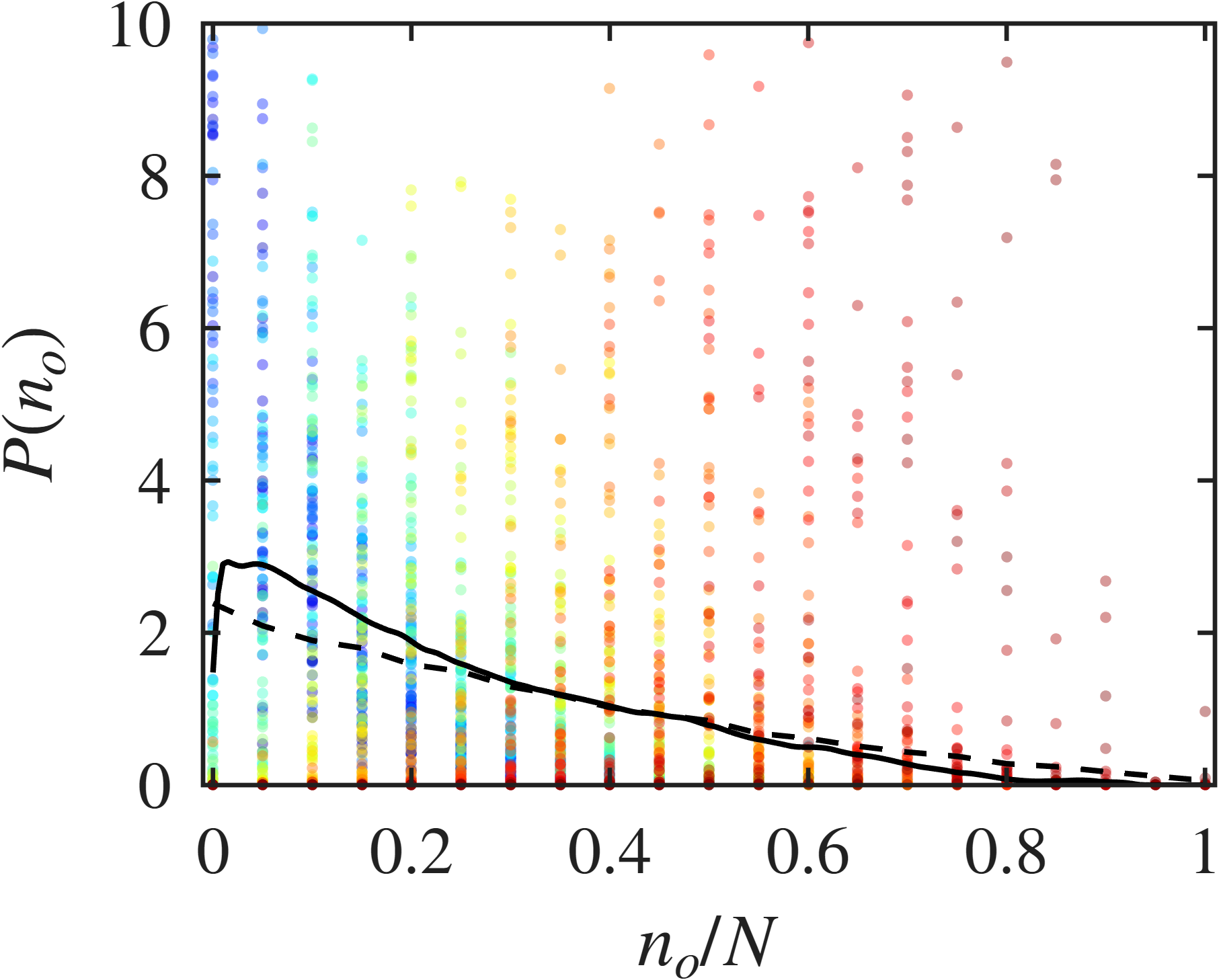}
\put(-3.0,75){(c)}
\end{overpic}
\ \ \ \
\begin{overpic}[width=6cm]{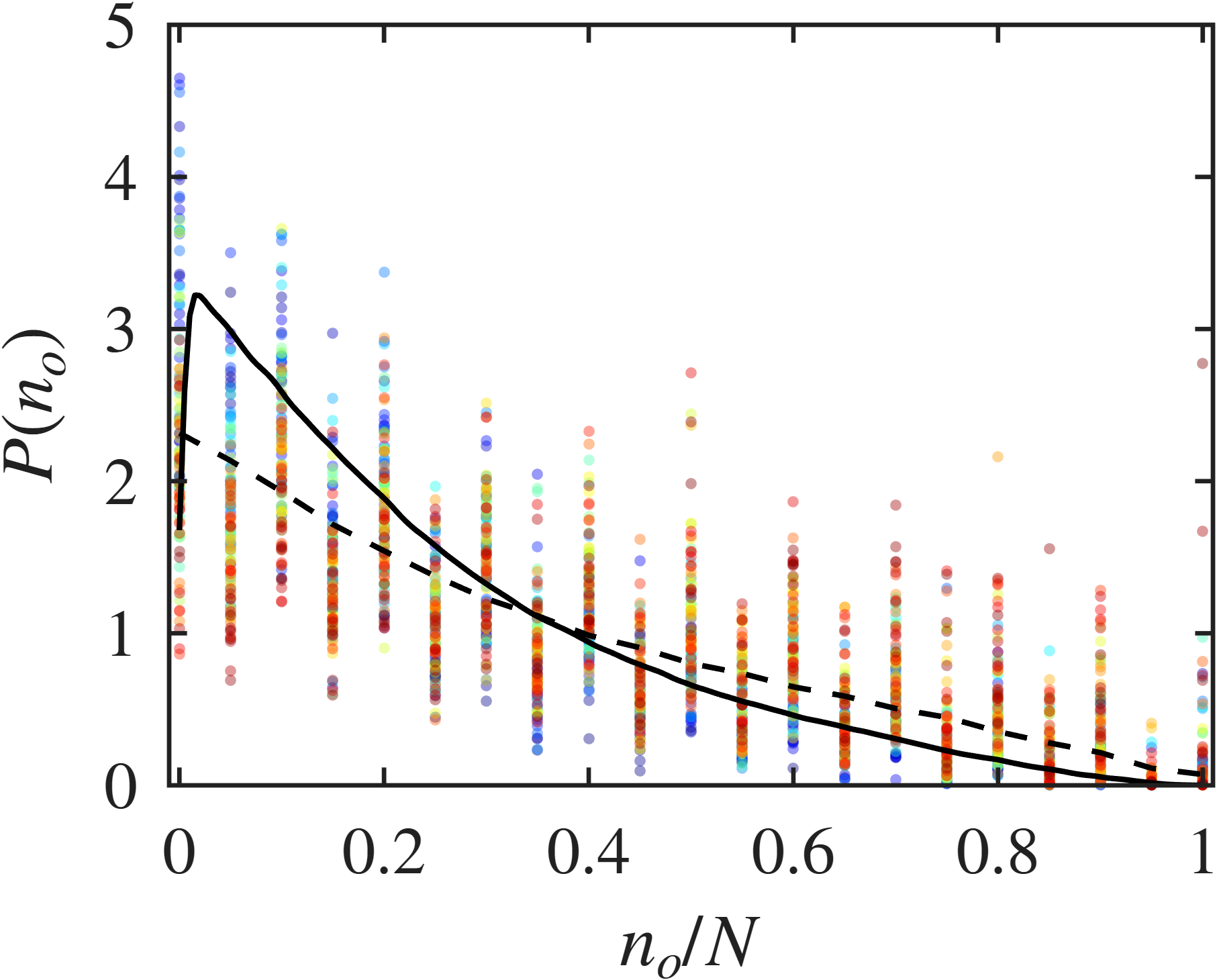} 
\put(-3.0,75){(d)}
\end{overpic}
\caption{
{\bf Optional view over quantum ergodicity.}
The $P^{\nu}(n_o)$ of representative eigenstates (colored lines) 
are compared with the classical (solid) and quantum (dashed)
microcanonical distributions 
(in the upper panels, the difference cannot be resolved). 
The data set is the same as that of \Fig{fQergImages},
and the panel arrangement is in one-to-one correspondence. The color code reflects~$\braket{n_o}$ -- it is used to distinguish the eigenstates.
}

\label{fQergPlots}
\end{figure*}

\section{Quantum ergodicity}
\label{sec:erg}

A superficial indication of ergodicity is narrow dispersion of the expectation value $\braket{n_o}$ within an energy window in the tomographic image of the spectrum. A more refined procedure is to look on the full distribution. \Fig{fQergImages} displays the $P^{\nu}(n_o)$ of eigenstates that belong to a narrow window around the energy~$E_{\text{SP}}$ of the inspected SP. The rows of each image are ordered by $\braket{n_o}$.  An optional view on the same data is provided by \Fig{fQergPlots}, where the $P^{\nu}(n_o)$ plots are distinguished by color. 

In \Fig{fQergPlots} we further compare the $P^{\nu}(n_o)$ of the individual eigenstates to the microcanonical distribution at the same energy, namely ${E \sim E_{\text{SP}} }$. The quantum microcanonical distribution is 
\beq
P^{E}(n_o) \ \ = \ \ \text{Average}_{\nu \in E} \Big[P^{\nu}(n_o)\Big], 
\eeq
where the average is taken over all the eigenstates within the energy window {${[E,E+dE]}$ that is narrow classically but contains a large number $N_E$ of states.   Namely, 
${\text{Average}_{\mu}[f_{\mu}] = (1/N_E)\sum_{\mu} f_{\mu}}$.} 
The corresponding classical distribution is obtained from long trajectories: specifically, we inspect the distribution of the $n_o$ coordinate over the simulation time.
We see that there is, in general, a rather good agreement between the classical and the quantum $P^{E}(n_o)$ irrespective of ergodicity. 
A slight systematic deviation that is observed in the lower panels of \Fig{fQergPlots} possibly reflects the difference between the Bose-Einstein and the Rayleigh-Jeans distribution \cite{KottosVardi}.

In order to characterize quantum ergodicity within the energy window of interest, we calculate the dispersion   
\beq \label{eergo}
\sigma \ \ = \ \ \sqrt{\text{Average}_{\nu, n_o}\Big[ \big(P^{\nu}(n_o)-P^{E}(n_o)\big)^2 \Big] }.
\eeq
{Here the average is also over the initial value of the  $n_o$ coordinate.}  
The results are displayed in \Fig{fig:fgpChain} as a function of~$u$. 
Small $\sigma$ indicates that ${P^{\nu}(n_o) \sim P^E(n_o)}$ for most eigenstates. This is what we expect if the underlying phase space is fully chaotic. Large $\sigma$ is possibly a signature for a structured phase space that features remnants of quasi-integrable regions.  Specifically, if we have stability island(s), we expect the eigenstates to be {\em localized} or possibly {\em hybrid}. The latter term implies that a quantum eigenstate can be a superposition of two localized pieces that have different phase-space locations. This is in the spirit of the chaos-assisted tunneling paradigm.

We turn to discuss details that are revealed by inspection of \Fig{fQergImages}.
For the $L_s=5$ chain, the interpretation is rather simple. For small $u$, the phase space is quasi-integrable, and this is reflected in the ordering of the localized eigenstates. For larger $u$, chaos dominates, and all the eigenstates look similar (the residual systematic difference between them is an artifact of the ordering by $\braket{n_o}$).

We now turn to look on the $L_s=3$ images of \Fig{fQergImages}. For small $u$, we observe that the eigenstates are in general {\em hybrids}. If we hypothetically could turn off the hybridization, then we would find just two types of eigenstates: those that are localized in $n_o$ and those that belong to the chaotic sea. Because of the hybridization, as an artifact of the row ordering, one might get the wrong impression that there are $\sim4$ types of eigenstates. 
Looking in \Fig{fig:fgpChain}, we see that for small $u$, the fluctuations ($\sigma$) increase as a function of $u$. This reflects that hybridization is reduced. Namely, for small $u$, there are near-degeneracies in the spectrum that encourage hybridization, but as $u$ is increased, these ``resonances" are diminished due to level repulsion, and therefore the wavefunctions, {\em counter-intuitively}, become less ``ergodic".

Inspecting further the dependence of $\sigma$ on $u$ in the ${L_s{=}3}$ case, as we enter the chaotic regime, the trend is reversed due to the induced ergodicity, and the fluctuations become smaller. Nevertheless, even when the chaotic sea dominates, there are still rather strong fluctuations, indicating remnants of quasi-regular regions. For much larger values of $u$, the trend reverses again: fluctuations become stronger as $u$ increases. This reversed trend reflects the crossover to the DS regime, where the SP becomes stable.

\section{Summary}
\label{sec:end}

Systems that have two degrees of freedom, such as the Bose-Hubbard trimer, are in some sense special. We can say that they feature low-dimensional chaos: they have phase-space with distinct quasi-regular and chaotic regions. This is because the energy surface has ${2d{-}1{=}3}$ dimensions, while the KAM barriers have $d{=}2$ dimensions. 
For $d{>}2$, the KAM barriers cannot divide the energy surface into distinct regions. Even if the non-linearity is small, there is always slow leakage, aka Arnold diffusion, via dynamical barriers. 

{Thus, in generic non-integrable high-dimensional Hamiltonian systems, phase-space is no longer partitioned by invariant barriers. Disregarding fragmentation due to energetic barriers, this facilitates transport across the whole energy surface (connectivity). However, in practice, ergodicity is still limited by the possibly slow rate of exploration, by the persistence of sticky quasi-regular structures, and by the associated wide distribution of dwell times.}

Consequently, one may speculate that trimers are distinct compared to rings and chains with more sites. But as far as eigenstate tomography is concerned, the distinction between low- and high-dimensional chaos is blurred for complementary reasons: (a)~In low-dimensional phase-space, quantum dynamics allows leakage via dynamical barriers; (b)~In high-dimensional phase-space, quantum dynamics does not allow Arnold diffusion via small openings; (c)~Quantum eigenstates might be hybrid. 

The present work has considered {\em generic} Bose-Hubbard systems that have more than 3 sites. We have inspected the various transitions between regimes where the SP that supports condensation is either ES, DS, or unstable. These transitions are quantified, as in \Fig{fig:fgpChain}, by both local Bogoliubov measures (notably $\gamma_o$) and by global Lyapunov measures (notably $\gamma_{\text{SP}}$ and $\gamma_{\text{CS}}$). The inspected {\em quantum} measures are respectively those that detect metastable condensates ($n_{\max}$ and $\mathcal{S}_{\max}$), and the quantum ergodicity measure ($\sigma$). 

For a ring, thanks to the interaction, the Bogoliubov frequencies of the supporting SP become positive, leading to ES. Chains, as opposed to rings, cannot support ES condensates, but nevertheless, they are non-chaotic in the GPE limit, and can support DS condensates in the GPE-like interaction regime. Their Bogoliubov frequencies accumulate to zero as the GPE border is crossed. However, because the number of sites is finite, there is an additional DNLSE threshold $u_c$ for the appearance of instability that can diminish the condensate. If $L_s$ is too small, one might encounter ${u_c{=}0}$ as opposed to the ${u_c{=}\infty}$ that characterizes the GPE limit.  

The ES, DS and instability regimes can be probed in state of the art cold-atom experiments. In a ring geometry \cite{atomtronics, exprRingRev, exprRingNIST}, the transitions between these regimes are controlled by the Sagnac phase $\Phi$, i.e. by the rotation frequency which is experimentally tunable. The signature of DS, as opposed to the ES, is the lack of thermodynamical stability, meaning that it is diminished by environmental decoherence. The survival of the superflow can be probed via time-of-flight imaging of the momentum distribution \cite{atomtronics} (see e.g. Fig. 15 there). For chains and for ``box" confined condensates \cite{Box1,Box2,Box3,Box4,Box5,Box6}, the relevant control parameter is the interaction strength $u{=}NU/K$, which is tunable via a Feshbach resonance or the lattice depth. Recent experiments on far-from-equilibrium dynamics \cite{turbPRA,turbPRL} provide motivation for further study of the relaxation dynamics.

In terms of methodology, our work has utilized an insightful semiclassical perspective (``tomography") for dealing with a condensed matter many-body problem. Just by glancing at the tomographic image, e.g. \Fig{fTomoC5}, we can identify the underlying structure of phase-space, to distinguish chaotic from quasi-regular regions, and to detect metastable, hybrid, and self-trapped states.  Our focus was on metastability versus ergodicity and how they depend on the number of sites. 
We emphasize the following: (a)~Our approach is both numerically efficient and informative, as opposed to methods that are based on level statistics; (b)~Our approach can be used for large rings, {whereas the common practice in ``quantum chaos" uses Poincaré sections whose visualization becomes challenging for more than two-degree-of-freedom systems. An exception that proves the rule is the visualization of classical structures and quantum states of four-dimensional maps as discussed by \cite{Richter}.}

In the arena of high-dimensional chaos, one has to acknowledge the famous numerical experiment of Fermi-Pasta-Ulam and Tsingou (FPUT). The large $L_s$ limit of a chaotic system can lead either to total microscopic chaos, as in the Boltzmann paradigm, or to integrability with diminished chaos, as in FPUT, which is also the case here in the GPE limit. Therefore, the clarification of the role of chaos for large rings is non-trivial.  
In the vast GPE/DNLSE literature, the role of chaos for finite $L_s$ has not been sufficiently explored. Our emphasis was on the many-body quantum problem, highlighting the mixed nature of the dynamics. {Accordingly, it is complementary to a recent work of the Freiburg group \cite{BHHchainChaos1,BHHchainChaos2}, that has explored the optimal route towards quantum chaos for the same Bose–Hubbard model. Their finding was that this is controlled by~$u$. As opposed to that, our purpose was to detect signatures of mixed phase-space, focusing on the manifestation of quasi-regular islands that can support metastable condensates, whose stability is controlled by the dimensionless parameters of \Eq{eGPE} and \Eq{eDNLSE}. We observe that with more degrees of freedom, such islands become better distinct from the ergodic sea, while their borders become ill-defined topologically. This stands in opposition to the very structured phase-space of two-degree-of-freedom systems, as reflected in the tomographic quantum spectrum.}

In some sense, the grand motivation for the present work is to provide tools and to achieve a better understanding regarding the validity or the failure of the semiclassical eigenfunction hypothesis of Percival. To the best of our understanding, while this hypothesis is widely accepted, a universal rigorous proof for all Hamiltonian systems does not exist. Rather, there are specific inquiries, mainly numerical. But alas, these inquiries usually refer to systems that have 2~dof, which we regard as non-generic in this context, notably the well-studied Mushroom Billiard \cite{Mushroom} and the Kicked Top \cite{KickedTop}. 
{An exception is Ref.\cite{Richter} where the semiclassical eigenfunction hypothesis was studied and visualized for a 4D map.} 
The failure of this hypothesis, even for relatively small systems, is likely to be of significance for studies of quantum thermalization and many-body localization, where the role of the underlying mixed phase-space is ill understood.  This holds in particular for strongly disordered systems, where thermalization can be modeled as the ergodization of small nearby chaotic subsystems that exchange particles and energy, sometimes referred to as hot spots~\cite{Basko}.


\ \\ 
{\bf Acknowledgments} --  
The research has been supported by the Israel Science Foundation, grant No.518/22.

\appendix

\section{Bogoliubov analysis for long chains}
\label{sec:BogoApp}

\begin{figure}
\centering

\begin{overpic}[width=7cm]{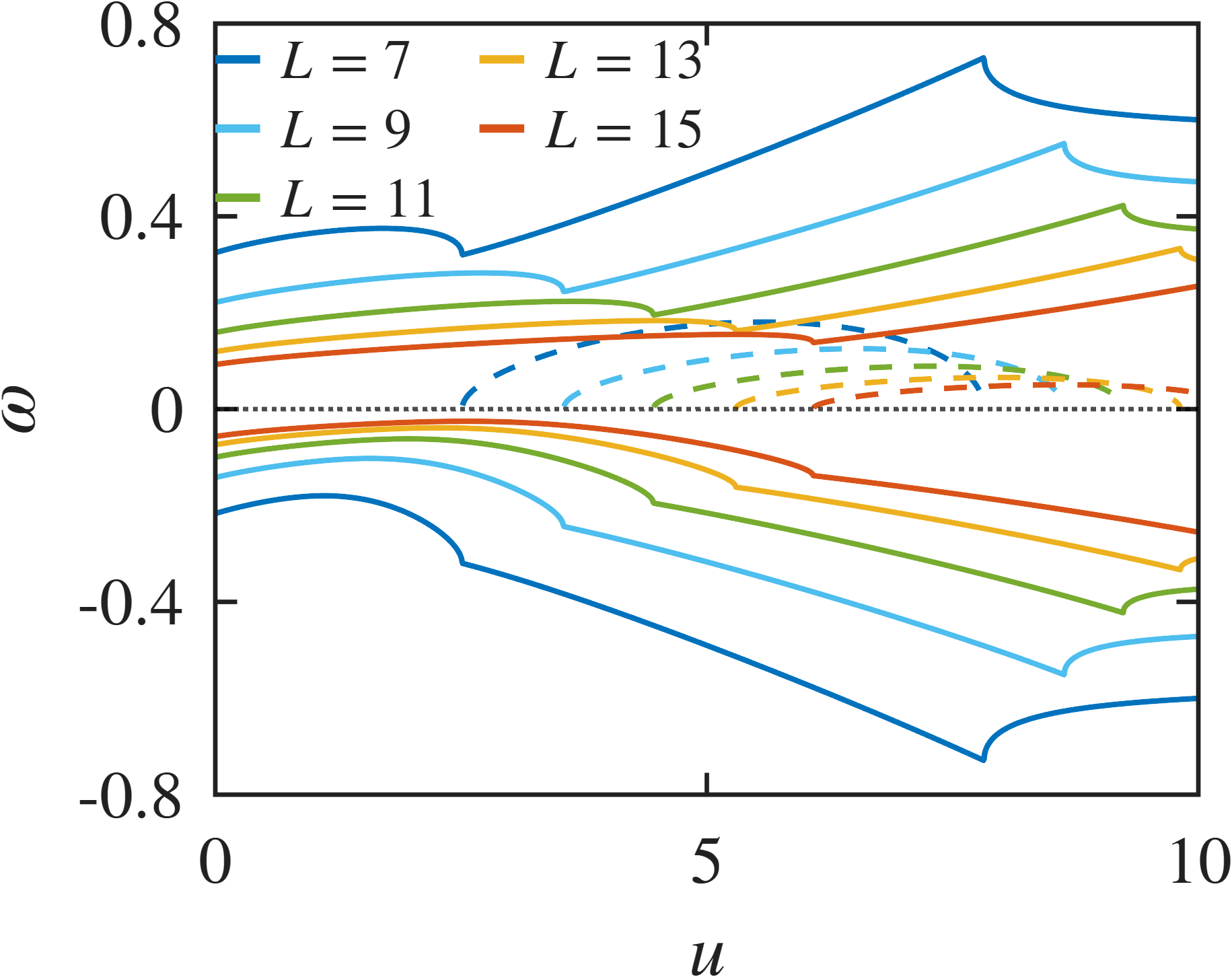}
\put(-3.2,75){(a)}
\end{overpic}
\begin{overpic}[width=7cm]{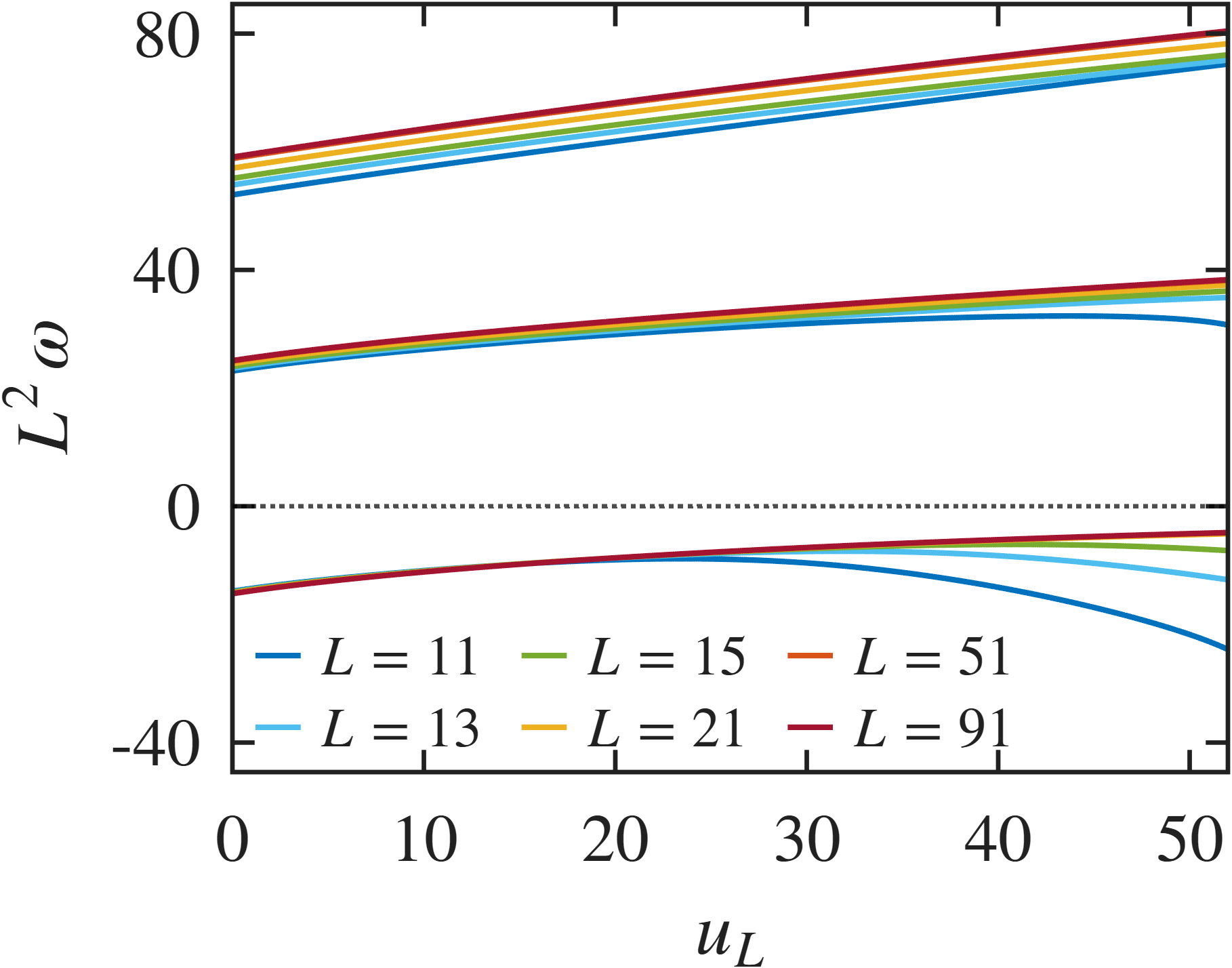}
\put(-3.2,75){(b)}
\end{overpic}

\caption{ 
{\bf Bogoliubov frequencies for short and long chains}. 
The lowest Bogoliubov frequencies for the ${m_o{=}2}$ stationary solution as a function of the interaction. The upper panel demonstrates how the complexity fades out for ${L_s{=}7,9,11,13,15}$. The dashed lines are for $\rm Im[\omega_q]$. 
The lower panel, with scaled axes $Lu$ and $L^2\omega$, 
demonstrates convergence to the GPE limit, with ${L_s{=}11,13,15,21,51,91}$.  
Note that for more than 20 sites, the curves cannot be resolved.
} 
\label{fBogoLs}
\end{figure}

The lowest Bogoliubov frequencies of the ${m_o=2}$ stationary solution are presented in \Fig{fBogoLs} for chains that have up to 100~sites. In the first panel, we show the first two frequencies for ${L_s{=}7,9,11,13,15}$.   
As $u$ is increased, the DNLSE instability is pushed to larger $u$ values, and the GPE regime is exposed, where the negative frequency approaches zero. 
In the second panel, we demonstrate that the results in the GPE regime scale very well for chains with more than 20 sites.   
In \Fig{fBogoManySites} we showed the Bogoliubov frequencies for the ${m_o{=}4}$ stationary solution of a chain that has $L_s{=}51$ sites. There, the results are compared with the outcome of a simple-minded approximation, which is further discussed below.

The $\bm{P}$ matrix is determined by the SP wavefunction. 
For example for $L_{s}=7$ we get for $m_o{=}2$ and zero interaction
\Eq{eSP7}. Consequently, in the orbital representation, we get 
\beq
\tilde{\bm{P}} \Big|_{u{=}0} = \frac{1}{16}
\begin{pmatrix}
2 & 0 & 1 & 0 & -1 & 0 & 0 \\
0 & 3 & 0 & 0 & 0 & -1 & 0 \\
1 & 0 & 2 & 0 & 0 & 0 & -1 \\
0 & 0 & 0 & 2 & 0 & 0 & 0 \\
-1 & 0 & 0 & 0 & 2 & 0 & 1 \\
0 & -1 & 0 & 0 & 0 & 3 & 0 \\
0 & 0 & -1 & 0 & 1 & 0 & 2
\end{pmatrix}. 
\eeq
An additional example that illustrates the general case has been provided in \Fig{fPmatrix} for the $m_o{=}4$ stationary state of an $L_s{=}51$ chain. As $u$ is increased, the off-diagonal elements get smaller, but on the other hand, a pattern appears, as discussed in \Sec{sec:SP}.
On the basis of this observation, one may speculate that a leading order approximation for the eigenvalues of $\bm{W}$ can be obtained via diagonalization of either $2\times2$ Blocks (high levels) or $4\times4$ Blocks (low levels):
\beq 
&&\begin{pmatrix}
\tilde{\mathcal{E}}_k & -\Delta_{k}  \\ 
\Delta_{k}   & -\varepsilon \\
\end{pmatrix},
\\
&&\begin{pmatrix}
\tilde{\mathcal{E}}_{-} & 2\Delta_{q}  & -\Delta_{k}  & -\Delta_{q}   \\ 
2\Delta_{q}   & \tilde{\mathcal{E}}_{+} & -\Delta_{q}   & -\Delta_{k}  \\ 
\Delta_{k}  & \Delta_{q}   & -\tilde{\mathcal{E}}_{-} & -2\Delta_{q}   \\
\Delta_{q}   & \Delta_{k}  & -2\Delta_{q}   & -\tilde{\mathcal{E}}_{+}
\end{pmatrix},
\eeq
where 
\beq
\Delta_{o} &=& NU \sum_j \frac{2}{L}[\sin(k_ox_j)]^2 p_j, \\
\Delta_{k} &=& NU \sum_j \frac{2}{L} [\sin(k x_j)]^2 p_j, \\
\Delta_{q} &=& NU \sum_j \frac{2}{L}\sin((k_o{+}q) x_j) \sin((k_o{-}q) x_j) p_j, \\
\tilde{\mathcal{E}}_k &=& \varepsilon_k +  2\Delta_{o} - \mu \ \ = \ \ 
\varepsilon_k {-} \varepsilon_o(u) + 2\Delta_{o} {-} \Delta.
\hspace{12mm} 
\eeq
The $\tilde{\mathcal{E}}_{\pm}$ are the $\tilde{\mathcal{E}}_k$ for the two respective orbitals. For ${u=0}$ one finds ${\Delta_{o}   = (3/2)\Delta}$, 
while for ${k\ne k_o}$ it is ${\Delta_{k} = \Delta}$, 
and for ${q\ne0}$ one finds ${\Delta_{q}   = (1/2)\Delta}$. 
For large interaction $\Delta_{o}$ approaches~$\Delta$,
while $\Delta_q$ approaches~$0$, see \Fig{fPmatrix}.
The characteristic polynomial of the $4\times4$ Block is a bi-quadratic equation ${\lambda^4 - \alpha \lambda^2 + \beta = 0}$, 
whose discriminant is positive if ${\tilde{\mathcal{E}} > \Delta}$,  
where ${\tilde{\mathcal{E}} =  (\tilde{\mathcal{E}}_{+} +  \tilde{\mathcal{E}}_{-})/2}$. Furthermore, in such a case, the roots come out real. 

In practice, we see in \Fig{fBogoManySites} that the $2\times2$ Blocks indeed provide a very good approximation for the high positive frequencies. But the $4\times4$ Blocks provide only a quantitatively bad approximation for the low positive frequencies, and false instability instead of the negative frequencies. As discussed in the main text, this shortcoming should have been expected because the solitonic comb has been ignored.

\section{Bogoliubov frequencies, derivation of exact results}
\label{sec:BogoExact}

For a chain with an odd number of sites, the middle orbital, aka the Dark state, is not affected by the interaction.
For example, for a 5-site chain, it is 
\beq
\psi^{(3)} = \frac{1}{\sqrt{3}} \{1, 0, -1, 0, 1\}.
\eeq
The Bogoliubov frequencies are indexed as ${q=-2,-1,0,1,2}$, and we get  
\beq
\omega_{+1} &=& +\left[1/4 + u^2/18 + u/18 \sqrt{u^2 - 9}\right]^{1/2}, 
\\
\omega_{-1} &=& -\left[1/4 + u^2/18 - u/18 \sqrt{u^2 - 9}\right]^{1/2}, 
\\
\omega_{+2} &=& +\left[3/4 + u^2/18 + u/18 \sqrt{u^2 - 27}\right]^{1/2},
\\
\omega_{-2} &=& -\left[3/4 + u^2/18 - u/18 \sqrt{u^2 - 27}\right]^{1/2}.
\eeq

Also, the SP that supports the first excited condensate of the 5-site chain is not affected by the interaction. The non-trivial frequencies are the solution of a cubic equation   
\begin{equation}
P(\lambda) = \lambda^3 - c_2\lambda^2 + c_1\lambda - c_0 = 0,
\end{equation}
with coefficients
%
\begin{align}
    c_0 &= \frac{1}{16} + \frac{u^3}{64}, \quad c_1 = \frac{3}{4} + \frac{u}{8} + \frac{u^3}{32}, \nonumber \\
    c_2 &= \frac{9}{4} + \frac{u}{4} + \frac{u^2}{16}. \nonumber
\end{align}
One defines
\begin{align}
\Delta_0 &= c_2^2 - 3c_1, \quad \Delta_1 = 2c_2^3 - 9c_1c_2 + 27c_0, \nonumber\\
C &= \left[\frac{1}{2}\left(\Delta_1 + \sqrt{\Delta_1^2 - 4\Delta_0^3}\right)\right]^{1/3}.
\nonumber 
\end{align}
And then the roots \Eq{eOmegaChain} are obtained from Cardano's formula. 
For small $u$, the $C$ comes out negative 
and consequently $|C|^2{=}\Delta_0$, and the frequencies come out real. 
The critical value is the value of $u$ 
for which the discriminant 
${\Delta_1^2 - 4\Delta_0^3}$ becomes positive. 
It comes out $u_c \approx 1.307$.



\clearpage 
\end{document}